\documentclass[amssymb,amsmath,prd,twocolumn,preprintnumbers,superscriptaddress,nofootinbib]{revtex4-2}

\usepackage[10pt]{type1ec}  
\usepackage[T1]{fontenc}
\usepackage{CJKutf8}
\usepackage[overlap, CJK]{ruby}
\usepackage{CJKulem}
\newenvironment{SChinese}{
  \CJKfamily{gbsn}
  \CJKtilde
  \CJKnospace
}{}

\usepackage{graphicx}

\usepackage{bm}
\usepackage{amssymb}
\usepackage{amsmath}
\usepackage{mathrsfs}
\usepackage{latexsym}

\usepackage{dcolumn}
\usepackage[colorlinks=true,citecolor=blue,urlcolor=blue]{hyperref}

\usepackage{color}
\usepackage[usenames,dvipsnames]{xcolor}
\usepackage{xspace}

\usepackage{multirow}
\usepackage{makecell}

\usepackage{xfp} 
\usepackage{nameref}

\newcommand{\rhonuc}{\ensuremath{\rho_{\mathrm{sat}}}}

\newcommand{\ep}{\ensuremath{\varepsilon}}

\newcommand{\Msolar}{\ensuremath{\mathrm{M}_\odot}}

\newcommand{\moifeature}{\ensuremath{\mathcal{D}^{I}_{M}}}

\newcommand{\latentenergy}{\ensuremath{\Delta(E/N)}}

\newcommand{\result}[1]{#1}

\newcommand{\externalresult}[1]{#1}

\newcommand{\LowMassBinMin}{\result{0.8}}
\newcommand{\LowMassBinMax}{\result{1.1}}

\newcommand{\MidMassBinMin}{\result{1.1}}
\newcommand{\MidMassBinMax}{\result{1.6}}

\newcommand{\HighMassBinMin}{\result{1.6}}
\newcommand{\HighMassBinMax}{\result{2.3}}

\newcommand{\MassUnit}{\result{\ensuremath{M_\odot}}}

\newcommand{\MidLatentEnergy}{\result{10}}
\newcommand{\HighLatentEnergy}{\result{50}}
\newcommand{\HugeLatentEnergy}{\result{100}}

\newcommand{\LatentEnergyUnit}{\result{\ensuremath{\mathrm{MeV}}}}

\newcommand{\MaxLikeRatioBranchLowMassPSR}{\result{1.00}} 
\newcommand{\MaxLikeRatioBranchMidMassPSR}{\result{1.00}} 
\newcommand{\MaxLikeRatioBranchHighMassPSR}{\result{1.00}} 

\newcommand{\MaxLikeRatioBranchLowMassGW}{\result{0.84}} 
\newcommand{\MaxLikeRatioBranchMidMassGW}{\result{0.81}} 
\newcommand{\MaxLikeRatioBranchHighMassGW}{\result{0.75}} 

\newcommand{\MaxLikeRatioBranchLowMassXray}{\result{0.45}} 
\newcommand{\MaxLikeRatioBranchMidMassXray}{\result{0.33}} 
\newcommand{\MaxLikeRatioBranchHighMassXray}{\result{0.68}} 

\newcommand{\MaxLikeRatioBranchLowMassPSRGW}{\result{0.79}} 
\newcommand{\MaxLikeRatioBranchMidMassPSRGW}{\result{0.23}} 
\newcommand{\MaxLikeRatioBranchHighMassPSRGW}{\result{0.69}} 

\newcommand{\MaxLikeRatioBranchLowMassPSRGWXray}{\result{0.47}} 
\newcommand{\MaxLikeRatioBranchMidMassPSRGWXray}{\result{0.14}} 
\newcommand{\MaxLikeRatioBranchHighMassPSRGWXray}{\result{0.20}} 

\newcommand{\MaxLikeRatioMoILowMassMidLatentEnergyPSR}{\result{1.00}} 
\newcommand{\MaxLikeRatioMoILowMassHighLatentEnergyPSR}{\result{1.00}} 
\newcommand{\MaxLikeRatioMoILowMassHugeLatentEnergyPSR}{\result{1.00}} 

\newcommand{\MaxLikeRatioMoIMidMassMidLatentEnergyPSR}{\result{1.00}} 
\newcommand{\MaxLikeRatioMoIMidMassHighLatentEnergyPSR}{\result{1.00}} 
\newcommand{\MaxLikeRatioMoIMidMassHugeLatentEnergyPSR}{\result{1.00}} 

\newcommand{\MaxLikeRatioMoIHighMassMidLatentEnergyPSR}{\result{1.00}} 
\newcommand{\MaxLikeRatioMoIHighMassHighLatentEnergyPSR}{\result{1.00}} 
\newcommand{\MaxLikeRatioMoIHighMassHugeLatentEnergyPSR}{\result{1.00}} 

\newcommand{\MaxLikeRatioMoILowMassMidLatentEnergyGW}{\result{1.01}} 
\newcommand{\MaxLikeRatioMoILowMassHighLatentEnergyGW}{\result{1.01}} 
\newcommand{\MaxLikeRatioMoILowMassHugeLatentEnergyGW}{\result{1.01}} 

\newcommand{\MaxLikeRatioMoIMidMassMidLatentEnergyGW}{\result{1.01}} 
\newcommand{\MaxLikeRatioMoIMidMassHighLatentEnergyGW}{\result{1.01}} 
\newcommand{\MaxLikeRatioMoIMidMassHugeLatentEnergyGW}{\result{1.01}} 

\newcommand{\MaxLikeRatioMoIHighMassMidLatentEnergyGW}{\result{0.91}} 
\newcommand{\MaxLikeRatioMoIHighMassHighLatentEnergyGW}{\result{0.91}} 
\newcommand{\MaxLikeRatioMoIHighMassHugeLatentEnergyGW}{\result{0.83}} 

\newcommand{\MaxLikeRatioMoILowMassMidLatentEnergyXray}{\result{0.95}} 
\newcommand{\MaxLikeRatioMoILowMassHighLatentEnergyXray}{\result{0.73}} 
\newcommand{\MaxLikeRatioMoILowMassHugeLatentEnergyXray}{\result{0.68}} 

\newcommand{\MaxLikeRatioMoIMidMassMidLatentEnergyXray}{\result{0.83}} 
\newcommand{\MaxLikeRatioMoIMidMassHighLatentEnergyXray}{\result{0.73}} 
\newcommand{\MaxLikeRatioMoIMidMassHugeLatentEnergyXray}{\result{0.68}} 

\newcommand{\MaxLikeRatioMoIHighMassMidLatentEnergyXray}{\result{0.83}} 
\newcommand{\MaxLikeRatioMoIHighMassHighLatentEnergyXray}{\result{0.73}} 
\newcommand{\MaxLikeRatioMoIHighMassHugeLatentEnergyXray}{\result{0.68}} 

\newcommand{\MaxLikeRatioMoILowMassMidLatentEnergyPSRGW}{\result{0.88}} 
\newcommand{\MaxLikeRatioMoILowMassHighLatentEnergyPSRGW}{\result{0.86}} 
\newcommand{\MaxLikeRatioMoILowMassHugeLatentEnergyPSRGW}{\result{0.31}} 

\newcommand{\MaxLikeRatioMoIMidMassMidLatentEnergyPSRGW}{\result{0.85}} 
\newcommand{\MaxLikeRatioMoIMidMassHighLatentEnergyPSRGW}{\result{0.78}} 
\newcommand{\MaxLikeRatioMoIMidMassHugeLatentEnergyPSRGW}{\result{0.31}} 

\newcommand{\MaxLikeRatioMoIHighMassMidLatentEnergyPSRGW}{\result{0.78}} 
\newcommand{\MaxLikeRatioMoIHighMassHighLatentEnergyPSRGW}{\result{0.78}} 
\newcommand{\MaxLikeRatioMoIHighMassHugeLatentEnergyPSRGW}{\result{0.31}} 

\newcommand{\MaxLikeRatioMoILowMassMidLatentEnergyPSRGWXray}{\result{0.57}} 
\newcommand{\MaxLikeRatioMoILowMassHighLatentEnergyPSRGWXray}{\result{0.49}} 
\newcommand{\MaxLikeRatioMoILowMassHugeLatentEnergyPSRGWXray}{\result{0.26}} 

\newcommand{\MaxLikeRatioMoIMidMassMidLatentEnergyPSRGWXray}{\result{0.57}} 
\newcommand{\MaxLikeRatioMoIMidMassHighLatentEnergyPSRGWXray}{\result{0.49}} 
\newcommand{\MaxLikeRatioMoIMidMassHugeLatentEnergyPSRGWXray}{\result{0.26}} 

\newcommand{\MaxLikeRatioMoIHighMassMidLatentEnergyPSRGWXray}{\result{0.52}} 
\newcommand{\MaxLikeRatioMoIHighMassHighLatentEnergyPSRGWXray}{\result{0.49}} 
\newcommand{\MaxLikeRatioMoIHighMassHugeLatentEnergyPSRGWXray}{\result{0.29}} 

\newcommand{\BayesBranchLowMassPSR}{\result{\ensuremath{0.169 \pm 0.012}}} 
\newcommand{\BayesBranchMidMassPSR}{\result{\ensuremath{0.102 \pm 0.009}}} 
\newcommand{\BayesBranchHighMassPSR}{\result{\ensuremath{1.007 \pm 0.043}}} 

\newcommand{\BayesBranchLowMassGW}{\result{\ensuremath{0.872 \pm 0.010}}} 
\newcommand{\BayesBranchMidMassGW}{\result{\ensuremath{1.369 \pm 0.014}}} 
\newcommand{\BayesBranchHighMassGW}{\result{\ensuremath{0.586 \pm 0.017}}} 

\newcommand{\BayesBranchLowMassXray}{\result{\ensuremath{0.115 \pm 0.010}}} 
\newcommand{\BayesBranchMidMassXray}{\result{\ensuremath{0.042 \pm 0.005}}} 
\newcommand{\BayesBranchHighMassXray}{\result{\ensuremath{0.384 \pm 0.028}}} 

\newcommand{\BayesBranchLowMassPSRGW}{\result{\ensuremath{0.421 \pm 0.043}}} 
\newcommand{\BayesBranchMidMassPSRGW}{\result{\ensuremath{0.029 \pm 0.005}}} 
\newcommand{\BayesBranchHighMassPSRGW}{\result{\ensuremath{0.088 \pm 0.027}}} 

\newcommand{\BayesBranchLowMassPSRGWXray}{\result{\ensuremath{0.362 \pm 0.036}}} 
\newcommand{\BayesBranchMidMassPSRGWXray}{\result{\ensuremath{0.030 \pm 0.006}}} 
\newcommand{\BayesBranchHighMassPSRGWXray}{\result{\ensuremath{0.147 \pm 0.028}}} 

\newcommand{\BayesBranchLowMassGWGivenPSR}{\result{\ensuremath{2.485 \pm 0.181}}} 
\newcommand{\BayesBranchMidMassGWGivenPSR}{\result{\ensuremath{0.282 \pm 0.064}}} 
\newcommand{\BayesBranchHighMassGWGivenPSR}{\result{\ensuremath{0.088 \pm 0.026}}} 

\newcommand{\BayesBranchLowMassGWXrayGivenPSR}{\result{\ensuremath{2.219 \pm 0.162}}} 
\newcommand{\BayesBranchMidMassGWXrayGivenPSR}{\result{\ensuremath{0.291 \pm 0.055}}} 
\newcommand{\BayesBranchHighMassGWXrayGivenPSR}{\result{\ensuremath{0.120 \pm 0.026}}} 

\newcommand{\BayesMoILowMassMidLatentEnergyPSR}{\result{\ensuremath{1.781 \pm 0.014}}} 
\newcommand{\BayesMoILowMassHighLatentEnergyPSR}{\result{\ensuremath{0.624 \pm 0.008}}} 
\newcommand{\BayesMoILowMassHugeLatentEnergyPSR}{\result{\ensuremath{0.373 \pm 0.010}}} 

\newcommand{\BayesMoIMidMassMidLatentEnergyPSR}{\result{\ensuremath{1.865 \pm 0.016}}} 
\newcommand{\BayesMoIMidMassHighLatentEnergyPSR}{\result{\ensuremath{0.950 \pm 0.012}}} 
\newcommand{\BayesMoIMidMassHugeLatentEnergyPSR}{\result{\ensuremath{0.516 \pm 0.011}}} 

\newcommand{\BayesMoIHighMassMidLatentEnergyPSR}{\result{\ensuremath{2.671 \pm 0.028}}} 
\newcommand{\BayesMoIHighMassHighLatentEnergyPSR}{\result{\ensuremath{2.265 \pm 0.029}}} 
\newcommand{\BayesMoIHighMassHugeLatentEnergyPSR}{\result{\ensuremath{1.366 \pm 0.027}}} 

\newcommand{\BayesMoILowMassMidLatentEnergyGW}{\result{\ensuremath{1.244 \pm 0.005}}} 
\newcommand{\BayesMoILowMassHighLatentEnergyGW}{\result{\ensuremath{1.379 \pm 0.007}}} 
\newcommand{\BayesMoILowMassHugeLatentEnergyGW}{\result{\ensuremath{1.393 \pm 0.010}}} 

\newcommand{\BayesMoIMidMassMidLatentEnergyGW}{\result{\ensuremath{1.250 \pm 0.006}}} 
\newcommand{\BayesMoIMidMassHighLatentEnergyGW}{\result{\ensuremath{1.426 \pm 0.008}}} 
\newcommand{\BayesMoIMidMassHugeLatentEnergyGW}{\result{\ensuremath{1.377 \pm 0.009}}} 

\newcommand{\BayesMoIHighMassMidLatentEnergyGW}{\result{\ensuremath{0.457 \pm 0.006}}} 
\newcommand{\BayesMoIHighMassHighLatentEnergyGW}{\result{\ensuremath{0.512 \pm 0.007}}} 
\newcommand{\BayesMoIHighMassHugeLatentEnergyGW}{\result{\ensuremath{0.604 \pm 0.009}}} 

\newcommand{\BayesMoILowMassMidLatentEnergyXray}{\result{\ensuremath{1.519 \pm 0.016}}} 
\newcommand{\BayesMoILowMassHighLatentEnergyXray}{\result{\ensuremath{0.451 \pm 0.008}}} 
\newcommand{\BayesMoILowMassHugeLatentEnergyXray}{\result{\ensuremath{0.254 \pm 0.009}}} 

\newcommand{\BayesMoIMidMassMidLatentEnergyXray}{\result{\ensuremath{1.420 \pm 0.016}}} 
\newcommand{\BayesMoIMidMassHighLatentEnergyXray}{\result{\ensuremath{0.682 \pm 0.011}}} 
\newcommand{\BayesMoIMidMassHugeLatentEnergyXray}{\result{\ensuremath{0.350 \pm 0.011}}} 

\newcommand{\BayesMoIHighMassMidLatentEnergyXray}{\result{\ensuremath{1.761 \pm 0.030}}} 
\newcommand{\BayesMoIHighMassHighLatentEnergyXray}{\result{\ensuremath{1.596 \pm 0.030}}} 
\newcommand{\BayesMoIHighMassHugeLatentEnergyXray}{\result{\ensuremath{0.914 \pm 0.026}}} 

\newcommand{\BayesMoILowMassMidLatentEnergyPSRGW}{\result{\ensuremath{0.897 \pm 0.017}}} 
\newcommand{\BayesMoILowMassHighLatentEnergyPSRGW}{\result{\ensuremath{0.355 \pm 0.011}}} 
\newcommand{\BayesMoILowMassHugeLatentEnergyPSRGW}{\result{\ensuremath{0.067 \pm 0.005}}} 

\newcommand{\BayesMoIMidMassMidLatentEnergyPSRGW}{\result{\ensuremath{0.778 \pm 0.018}}} 
\newcommand{\BayesMoIMidMassHighLatentEnergyPSRGW}{\result{\ensuremath{0.368 \pm 0.011}}} 
\newcommand{\BayesMoIMidMassHugeLatentEnergyPSRGW}{\result{\ensuremath{0.073 \pm 0.004}}} 

\newcommand{\BayesMoIHighMassMidLatentEnergyPSRGW}{\result{\ensuremath{0.512 \pm 0.020}}} 
\newcommand{\BayesMoIHighMassHighLatentEnergyPSRGW}{\result{\ensuremath{0.469 \pm 0.020}}} 
\newcommand{\BayesMoIHighMassHugeLatentEnergyPSRGW}{\result{\ensuremath{0.170 \pm 0.010}}} 

\newcommand{\BayesMoILowMassMidLatentEnergyPSRGWXray}{\result{\ensuremath{1.222 \pm 0.020}}} 
\newcommand{\BayesMoILowMassHighLatentEnergyPSRGWXray}{\result{\ensuremath{0.366 \pm 0.011}}} 
\newcommand{\BayesMoILowMassHugeLatentEnergyPSRGWXray}{\result{\ensuremath{0.117 \pm 0.008}}} 

\newcommand{\BayesMoIMidMassMidLatentEnergyPSRGWXray}{\result{\ensuremath{1.043 \pm 0.020}}} 
\newcommand{\BayesMoIMidMassHighLatentEnergyPSRGWXray}{\result{\ensuremath{0.463 \pm 0.013}}} 
\newcommand{\BayesMoIMidMassHugeLatentEnergyPSRGWXray}{\result{\ensuremath{0.152 \pm 0.009}}} 

\newcommand{\BayesMoIHighMassMidLatentEnergyPSRGWXray}{\result{\ensuremath{1.012 \pm 0.035}}} 
\newcommand{\BayesMoIHighMassHighLatentEnergyPSRGWXray}{\result{\ensuremath{0.898 \pm 0.034}}} 
\newcommand{\BayesMoIHighMassHugeLatentEnergyPSRGWXray}{\result{\ensuremath{0.383 \pm 0.023}}} 

\newcommand{\BayesMoILowMassMidLatentEnergyGWGivenPSR}{\result{\ensuremath{0.504 \pm 0.009}}} 
\newcommand{\BayesMoILowMassHighLatentEnergyGWGivenPSR}{\result{\ensuremath{0.570 \pm 0.017}}} 
\newcommand{\BayesMoILowMassHugeLatentEnergyGWGivenPSR}{\result{\ensuremath{0.180 \pm 0.013}}} 

\newcommand{\BayesMoIMidMassMidLatentEnergyGWGivenPSR}{\result{\ensuremath{0.417 \pm 0.009}}} 
\newcommand{\BayesMoIMidMassHighLatentEnergyGWGivenPSR}{\result{\ensuremath{0.388 \pm 0.012}}} 
\newcommand{\BayesMoIMidMassHugeLatentEnergyGWGivenPSR}{\result{\ensuremath{0.142 \pm 0.009}}} 

\newcommand{\BayesMoIHighMassMidLatentEnergyGWGivenPSR}{\result{\ensuremath{0.192 \pm 0.007}}} 
\newcommand{\BayesMoIHighMassHighLatentEnergyGWGivenPSR}{\result{\ensuremath{0.207 \pm 0.009}}} 
\newcommand{\BayesMoIHighMassHugeLatentEnergyGWGivenPSR}{\result{\ensuremath{0.124 \pm 0.008}}} 

\newcommand{\BayesMoILowMassMidLatentEnergyGWXrayGivenPSR}{\result{\ensuremath{0.684 \pm 0.011}}} 
\newcommand{\BayesMoILowMassHighLatentEnergyGWXrayGivenPSR}{\result{\ensuremath{0.588 \pm 0.016}}} 
\newcommand{\BayesMoILowMassHugeLatentEnergyGWXrayGivenPSR}{\result{\ensuremath{0.292 \pm 0.021}}} 

\newcommand{\BayesMoIMidMassLowLatentEnergyGWXrayGivenPSR}{\result{\ensuremath{0.552 \pm 0.010}}} 
\newcommand{\BayesMoIMidMassMidLatentEnergyGWXrayGivenPSR}{\result{\ensuremath{0.563 \pm 0.010}}} 
\newcommand{\BayesMoIMidMassHighLatentEnergyGWXrayGivenPSR}{\result{\ensuremath{0.481 \pm 0.013}}} 
\newcommand{\BayesMoIMidMassHugeLatentEnergyGWXrayGivenPSR}{\result{\ensuremath{0.267 \pm 0.017}}} 

\newcommand{\BayesMoIHighMassLowLatentEnergyGWXrayGivenPSR}{\result{\ensuremath{0.385 \pm 0.013}}} 
\newcommand{\BayesMoIHighMassMidLatentEnergyGWXrayGivenPSR}{\result{\ensuremath{0.387 \pm 0.013}}} 
\newcommand{\BayesMoIHighMassHighLatentEnergyGWXrayGivenPSR}{\result{\ensuremath{0.399 \pm 0.015}}} 
\newcommand{\BayesMoIHighMassHugeLatentEnergyGWXrayGivenPSR}{\result{\ensuremath{0.256 \pm 0.016}}} 

\newcommand{\numpriorsamples}{310,000}
\newcommand{\astroneff}{19,300}

\newcommand{\psrneffpercent}{80}

\newcommand{\CIT}{\affiliation{TAPIR, California Institute of Technology, Pasadena, California 91125, USA}}
\newcommand{\CITLab}{\affiliation{LIGO Laboratory, California Institute of Technology, Pasadena, CA 91125, USA}}

\begin{document}

\preprint{N3AS-22-021, INT-PUB-22-025}

\title{
Phase Transition Phenomenology with Nonparametric Representations of the Neutron Star Equation of State
}

\author{Reed Essick}
\email{essick@cita.utoronto.ca}
\affiliation{Canadian Institute for Theoretical Astrophysics, University of Toronto, Toronto, ON M5S 3H8}
\affiliation{Department of Physics, University of Toronto, Toronto, ON M5S 1A7}
\affiliation{David A. Dunlap Department of Astronomy, University of Toronto, Toronto, ON M5S 3H4}
\affiliation{Perimeter Institute for Theoretical Physics, Waterloo, Ontario, Canada, N2L 2Y5}

\author{Isaac Legred}
\email{ilegred@caltech.edu}\CIT\CITLab

\author{Katerina Chatziioannou}
\email{kchatziioannou@caltech.edu}\CIT\CITLab

\author{
Sophia Han
(\begin{CJK}{UTF8}{}\begin{SChinese}韩 君\end{SChinese}\end{CJK})
}
\email{sjhan@sjtu.edu.cn}
\affiliation{Tsung-Dao Lee Institute and School of Physics and Astronomy, Shanghai Jiao Tong University, Shanghai 200240, China}
\affiliation{Institute for Nuclear Theory, University of Washington, Seattle, WA~98195, USA}
\affiliation{Department of Physics, University of California, Berkeley, CA~94720, USA}

\author{Philippe Landry}
\email{plandry@cita.utoronto.ca}
\affiliation{Canadian Institute for Theoretical Astrophysics, University of Toronto, Toronto, ON M5S 3H8}

\date{\today}

\begin{abstract}
    Astrophysical observations of neutron stars probe the structure of dense nuclear matter and have the potential to reveal phase transitions at high densities.
    Most recent analyses are based on parametrized models of the equation of state with a finite number of parameters and occasionally  include extra parameters intended to capture phase transition phenomenology.
    However, such models restrict the types of behavior allowed and may not match the true equation of state.
    We introduce a complementary approach that extracts phase transitions directly from the equation of state without relying on, and thus being restricted by, an underlying parametrization. 
    We then constrain the presence of phase transitions in neutron stars with astrophysical data.
    Current pulsar mass, tidal deformability, and mass-radius measurements disfavor only the strongest of possible phase transitions (latent energy per particle \result{$\gtrsim 100\,\mathrm{MeV}$}).
    Weaker phase transitions are consistent with observations.
    We further investigate the prospects for measuring phase transitions with future gravitational-wave observations and find that catalogs of \result{$O(100)$} events will (at best) yield Bayes factors of \result{$\sim 10:1$} in favor of phase transitions even when the true equation of state contains very strong phase transitions.
    Our results reinforce the idea that neutron star observations will primarily constrain trends in macroscopic properties rather than detailed microscopic behavior.
    Fine-tuned equation of state models will likely remain unconstrained in the near future.
\end{abstract}

\maketitle


\section{Introduction}
\label{sec:introduction}

Recent astronomical data, such as gravitational waves (GWs) from coalescing neutron star (NS) binaries~\cite{GW170817, GW190425} observed by LIGO~\cite{LIGO} and Virgo~\cite{Virgo}, X-ray pulse profiles from hotspots on rotating NSs observed by NICER~\cite{J0030-Miller, J0030-Riley, J0740-Miller, J0740-Riley}, and mass measurements for heavy radio pulsars~\cite{J0348-Antoniadis, J0740-Cromartie, J0740-Fonseca}, have advanced our understanding of matter at supranuclear densities~\cite{GW170817-Mass-Radius,Landry:2020vaw,Pang:2021jta,J0030-Raaijmakers,J0740-Raaijmakers,Biswas:2021yge,Jiang:2019rcw,Dietrich:2020efo,Legred:2021hdx}.
Nonetheless, there is still considerable uncertainty in the equation of state (EoS) of cold, dense matter, which relates the pressure $p$ to the energy density $\ep$, or rest-mass density $\rho$.
The data favor a sound speed $c_s = \sqrt{dp/d\ep}$ that exceeds the conjectured conformal bound of $\sqrt{1/3}$ expected for weakly interacting ultra-relativistic particles~\cite{Bedaque:2014sqa,McLerran:2018hbz,Landry:2020vaw,Legred:2021hdx}.
The potential violation of this bound at high densities may point to a state of matter with strongly coupled interactions.

Such strong couplings call into question the accuracy of perturbative expansions of interactions between neutrons, protons, and pions at high densities, and raise the possibility that other degrees of freedom may be a more natural description.
Theoretical studies have investigated whether the smooth crossover from hadron resonance gas to quark-gluon plasma observed with lattice quantum chromodynamics (QCD) at low baryon chemical potential and high temperature implies the existence of a critical endpoint in the QCD phase diagram~\cite{Baym:2017whm} and how EoS calculations at low density and temperature connect to perturbative QCD (pQCD) calculations at high densities ($\sim 40$ times nuclear saturation $\rhonuc$)~\cite{Komoltsev:2021jzg, Gorda:2022jvk, Somasundaram:2022ztm}.
Other work predicts a variety of phase transitions stemming from a range of microphysical descriptions for dense matter~\cite{Schertler:2000xq,Glendenning:2001pe,Schaffner-Bielich:2000nft,Alford:2004pf,Zdunik:2012dj,Hempel:2013tfa,Fukushima:2015bda,Baym:2017whm,McLerran:2018hbz,Alford:2019oge}.

Many theorized phase transitions in NS matter are characterized by a softening of the EoS, i.e., a decrease in $c_s$.
This occurs because the NS is supported by degeneracy pressure, and additional degrees of freedom (e.g., hyperons or quarks) initially do not contribute significantly to the pressure due to their low number density $n$.
This manifests as an interval of nearly constant pressure (small $c_s$) over a density range in which the new degrees of freedom first appear.
A decrease in pressure support relative to an EoS without a phase transition leads to more compact NSs.
Such compactification can lead to bends or kinks in the relation between macroscopic observables, such as the gravitational mass $M$, radius $R$, tidal deformability $\Lambda$, and moment of inertia $I$.
The strongest phase transitions can even give rise to disconnected sequences of stable NSs separated by a range of central densities for which no stable NSs exist.
This manifests as, e.g., two or more disconnected branches in the $M$-$R$ relation and twin stars with the same mass but different radii~\cite{Lindblom:1998dp,Schaeffer:1983,Seidov:1971,Schertler:2000xq,Alford:2013aca,Alford:2017qgh,Han:2018mtj,Montana:2018bkb}.
Moreover, the relative loss of pressure support from the phase transition often reduces the maximum mass ($M_\mathrm{TOV}$) for cold, non-rotating NSs. 

Current observational evidence for a sudden softening in the EoS is inconclusive.
Both the PREX neutron skin measurement~\cite{PREX} and the existence of $2\,\Msolar$ pulsars~\cite{J0740-Fonseca} suggest a relatively stiff EoS (near $\rhonuc$ and above $\sim 3\rhonuc$, respectively).
In contrast, the relatively small tidal deformability of GW170817 points to a moderately soft EoS around $\sim 2\rhonuc$~\cite{GW170817-Mass-Radius, Legred:2021hdx}.
While this stiff--soft--stiff sequence resembles the morphology of a phase transition, the actual statistical evidence for or against this scenario remains inconclusive~\cite{Legred:2021hdx,Pang:2021jta,Gorda:2022lsk}.
Furthermore, while observations favor a violation of the conformal bound around $\sim3\rhonuc$, they do not strictly rule out EoSs with $c_s \leq \sqrt{1/3}$ at higher densities~\cite{Legred:2021hdx}.
Additionally, the CREX collaboration's neutron skin measurement favors lower pressures near $\rhonuc$ than PREX~\cite{CREX}.
At present, consistency between \textit{ab initio} theoretical models, laboratory experiments, and astrophysical data within statistical uncertainties does not require a phase transition~\cite{Essick:2021kjb, Essick:2021ezp}.

Several features of NSs' macroscopic properties have been proposed as a way to identify a phase transition in NS matter with forthcoming GW observations.
During a compact binary's inspiral (before the objects touch), the relevant observable is the (adiabatic or static) tidal deformability~\cite{Flanagan:2007ix,Wade:2014vqa,Chatziioannou:2020pqz}, which is strongly correlated with the radius.
Both are expected to be smaller for NSs with exotic cores than their nucleonic counterparts.
\citet{Chen:2019rja} leveraged this fact to search for phase transitions via a change in the slope of the inferred $M$--$R$ relation, parametrized as a piecewise linear function.
\citet{Chatziioannou:2019yko} pursued a related method, modeling the detected binary merger population hierarchically and searching for a subpopulation with smaller radii.
Parametrizing the $M$--$\Lambda$ relation itself, \citet{LandryChakravarti2022} sought to identify twin stars in the binary NS population based on gaps in the joint distribution of masses and binary tidal deformabilities.
Proposals for identifying phase transitions based on the presence of disconnected stable branches in the $M$--$R$ or $M$--$\Lambda$ relation, independently of a parametrization, have also been investigated~\cite{Essick:2019ldf,Pang:2021jta,Legred:2021hdx}.
However, approaches that directly model macroscopic observables cannot easily enforce physical precepts like causality and thermodynamic stability, nor do they offer an obvious pathway to microscopic EoS properties.
At best, one can constrain proxies for microphysical phase transitions, such as the difference between radii at different masses, e.g., $\Delta R \equiv R_{1.4} - R_{2.0}$~\cite{Drischler:2020fvz,Pang:2021jta,J0740-Raaijmakers,Legred:2021hdx}.
Moreover, macroscopic signatures test a sufficient, but not necessary, condition for exotic phases.
A phase transition may not be strong enough to leave a measurable imprint on NS observables. This ambiguity is known as the masquerade problem~\cite{Alford:2004pf}.

An alternative approach is to directly model the EoS and connect it to macroscopic NS observables by solving the Tolman-Oppenheimer-Volkoff (TOV) equations~\cite{Tolman:1939, Oppenheimer:1939}.
A plethora of phenomenological EoS parametrizations adapted to phase transitions have been proposed~\cite{Alford:2013aca,Gorda:2022lsk,Tan:2021ahl}.
For example,~\citet{Pang:2020ilf} modeled the EoS as a piecewise polytrope, including a segment with vanishing adiabatic index ($c_s = 0$) to represent the phase transition.
They performed model selection on a catalog of simulated GW observations to test whether they favored the presence of a phase transition.
\citet{Tan:2021ahl} performed a similar analysis with a more complex parametric EoS model, which nonetheless retained the characteristic morphology of regions of large $c_s$ bracketing a range of densities with small $c_s$.
We discuss these and other approaches at length in Sec.~\ref{sec:discussion}.

However, it is also possible to model the EoS directly without introducing a parametrization.
Flexible nonparametric models, such as the Gaussian process (GP) representation introduced in Refs.~\cite{Landry:2019, Essick:2019ldf, Landry:2020vaw}, avoid the \textit{ad hoc} correlations across density scales that are inevitable in parametric representations with a finite number of parameters~\cite{Legred:2022pyp}.
While some interdensity correlations are desirable (e.g., those dictated by causality, thermodynamic stability, or predictions from nuclear theory), phenomenological parametric models implicitly impose much stronger prior assumptions by virtue of their chosen functional form.
Nonparametric models need not impose such correlations.
They can also provide a faithful representation of theoretical uncertainty at low densities without sacrificing model flexibility at high densities~\cite{Essick:2020, Essick:2021kjb, Essick:2021ezp}.
However, the lack of phenomenological parameters can make it difficult to map features in the EoS to underlying microphysics.
In order to address this, a generic mapping from the EoS to a set of physically interpretable microscopic parameters is needed.

We develop such a mapping: a phenomenological approach to identifying physically meaningful properties of phase transitions via softening in the EoS.
We show that a nonparametric model's lack of obvious physically interpretable parameters does not fundamentally limit its utility for inferences about phase transitions in NSs.
We propose and test model-independent features that characterize a broad range of phase transition phenomenology.
Our procedure goes beyond existing nonparametric tests based on the number of distinct stable NS sequences in the $M$--$R$ (or $M$--$\Lambda$) relation~\cite{Essick:2019ldf, Landry:2020vaw, Legred:2021hdx} and enables us to directly extract information about the onset and strength of both large and weak phase transitions that respectively do and do not create multiple stable branches.
As such, it provides an alternative to parametric phase transition inferences, whose inflexible parametrizations may introduce systematic biases if they do not closely match the true EoS~\cite{Lindblom:2010, Greif:2018njt, Carney:2018sdv, Legred:2022pyp}.

\begin{figure*}
    \begin{minipage}{0.52\textwidth}
        \includegraphics[width=1.0\textwidth, clip=True, trim=0.0cm 0.1cm 0.3cm 0.3cm]{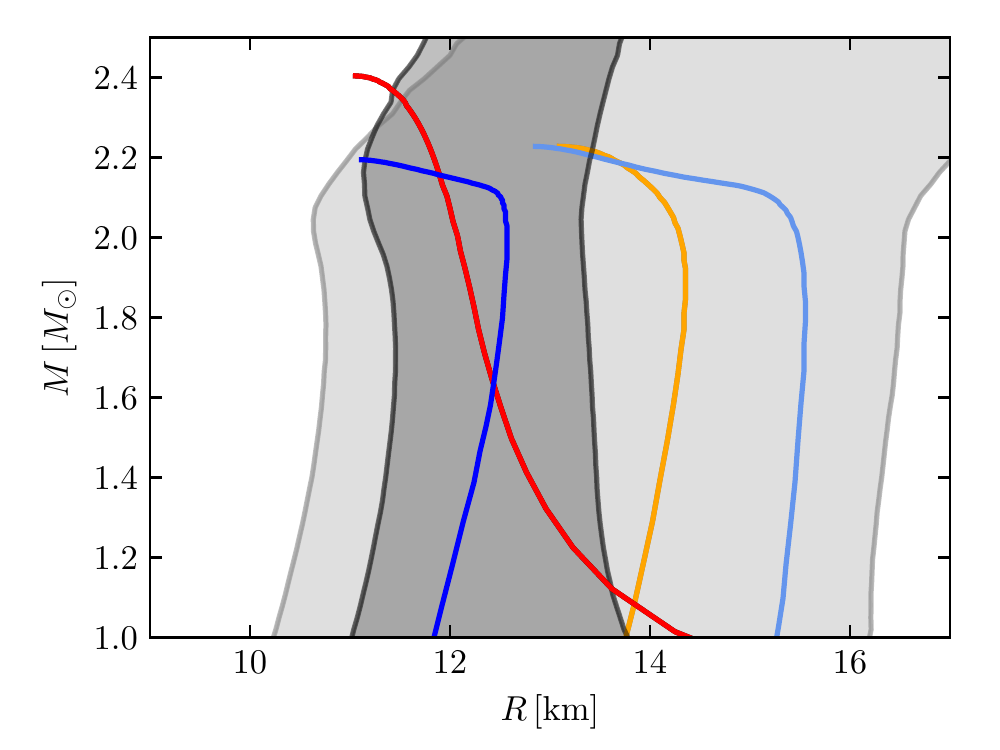} 
    \end{minipage}
    \begin{minipage}{0.47\textwidth}
        \includegraphics[width=1.0\textwidth, clip=True, trim=0.75cm 0.9cm 0.5cm 0.5cm]{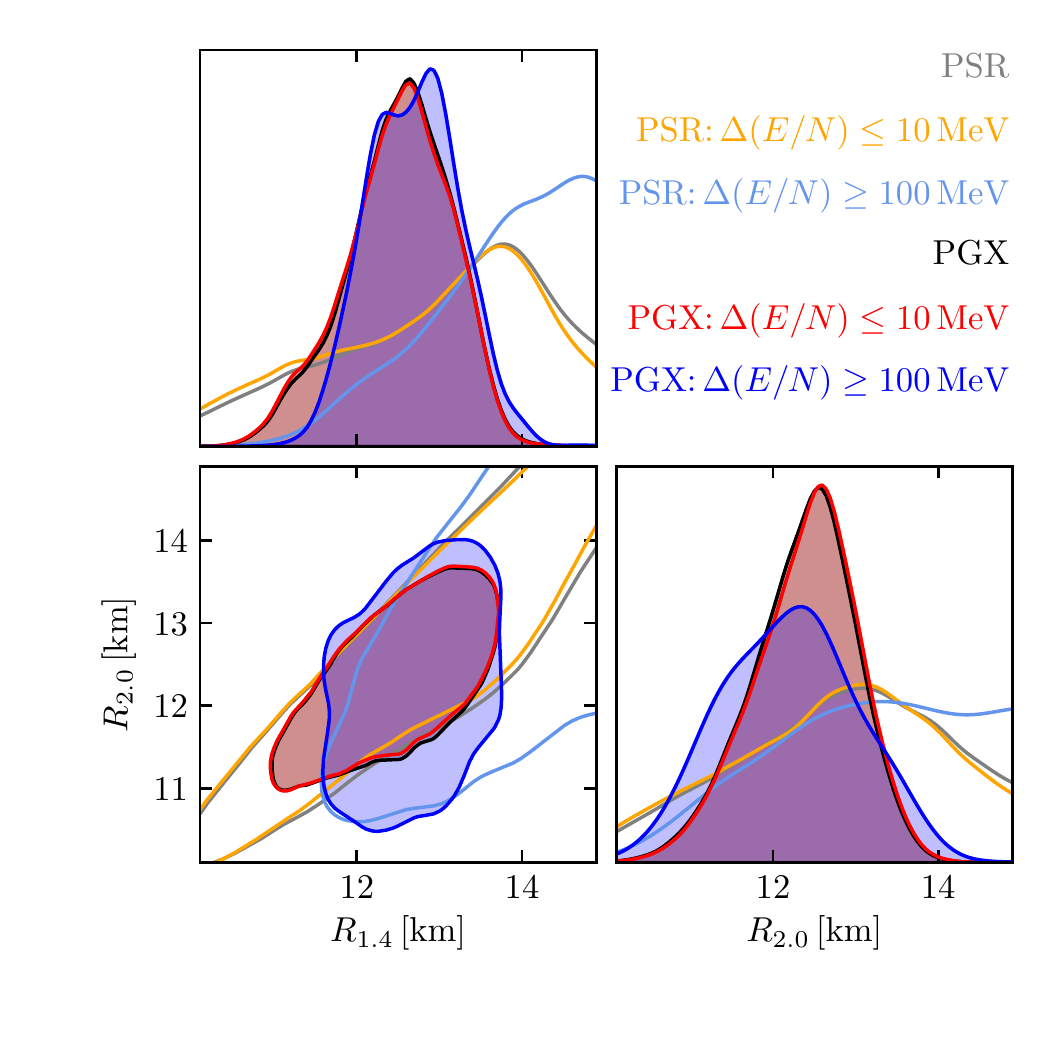}
    \end{minipage}
    \caption{
        (\textit{left}) one-dimensional 90\% symmetric marginal posterior credible regions for the radius as a function of mass conditioned on current data. We show results with only pulsar masses (denoted PSR) and pulsar masses, GW observations, and NICER X-ray pulse profiling (denoted PGX).
        We additionally show maximum-likelihood EoSs from subsets of the prior conditioned on the size of the latent energy per particle \latentenergy~of phase transitions that overlap with the central densities of NSs between \result{1.1--2.3$\,\Msolar$} (\textit{small}: \result{$\latentenergy \leq 10\,\mathrm{MeV}$} and \textit{large}: \result{$\latentenergy \geq 100\,\mathrm{MeV}$}).
        (\textit{right}) Correlations between the radius at two reference masses: $M=1.4$ and $2.0\,\Msolar$.
        While the one-dimensional marginal distributions are similar, EoSs with small \latentenergy~show stronger correlations between $R_{1.4}$ and $R_{2.0}$ than EoSs with large \latentenergy.
        This is because the radius can change rapidly when \latentenergy~is large, as is evident in the maximum-likelihood EoS.
    }
    \label{fig:current constraints M-R}
\end{figure*}

We introduce our methodology in Sec.~\ref{sec:observables}.
Section~\ref{sec:phenomenology} reviews the basic phenomenology of phase transitions and, motivated by these considerations, Sec.~\ref{sec:new stats} proposes novel features that can be used to identify the presence of a phase transition and extract physically relevant properties without the need for a direct parametrization.
Our new features are based on the mass dependence of the moment of inertia ($I$) and the density dependence of the speed of sound, although similar features can also be derived from other macroscopic observables.
We apply our methodology to current astrophysical data in Sec.~\ref{sec:constraints}.
Current astrophysical data (Fig.~\ref{fig:current constraints M-R}) disfavor the strongest of possible phase transitions, but only when those transitions occur within NSs between \result{$\sim 1$--$2\,\Msolar$}.
Even the presence of multiple stable branches cannot be unambiguously ruled out, although they are disfavored compared to EoS with a single branch and smaller phase transitions.
Section~\ref{sec:prospects} examines the prospects for detecting and characterizing phase transitions with large catalogs of simulated GW detections.
We obtain Bayes factors of \result{$\sim 10:1$} in favor of phase transitions with \result{$O(10^2)$} events, a larger catalog than is likely~\cite{Observing-Scenarios} within the lifetime of advanced LIGO~\cite{LIGO} and Virgo~\cite{Virgo}.
We discuss our conclusions in the context of previous studies in the literature as well as possible future research in Sec.~\ref{sec:discussion}.


\section{Phenomenological identification of phase transitions}
\label{sec:observables}

\begin{figure*}
    \begin{center}
        \large{Weak Maxwell CSS Phase Transition}
    \end{center}
    \vspace{-0.30cm}
    \begin{minipage}{0.66\textwidth}
        \begin{center}
            \includegraphics[width=1.0\textwidth, clip=True, trim=0.0cm 0.00cm 0.0cm 0.25cm]{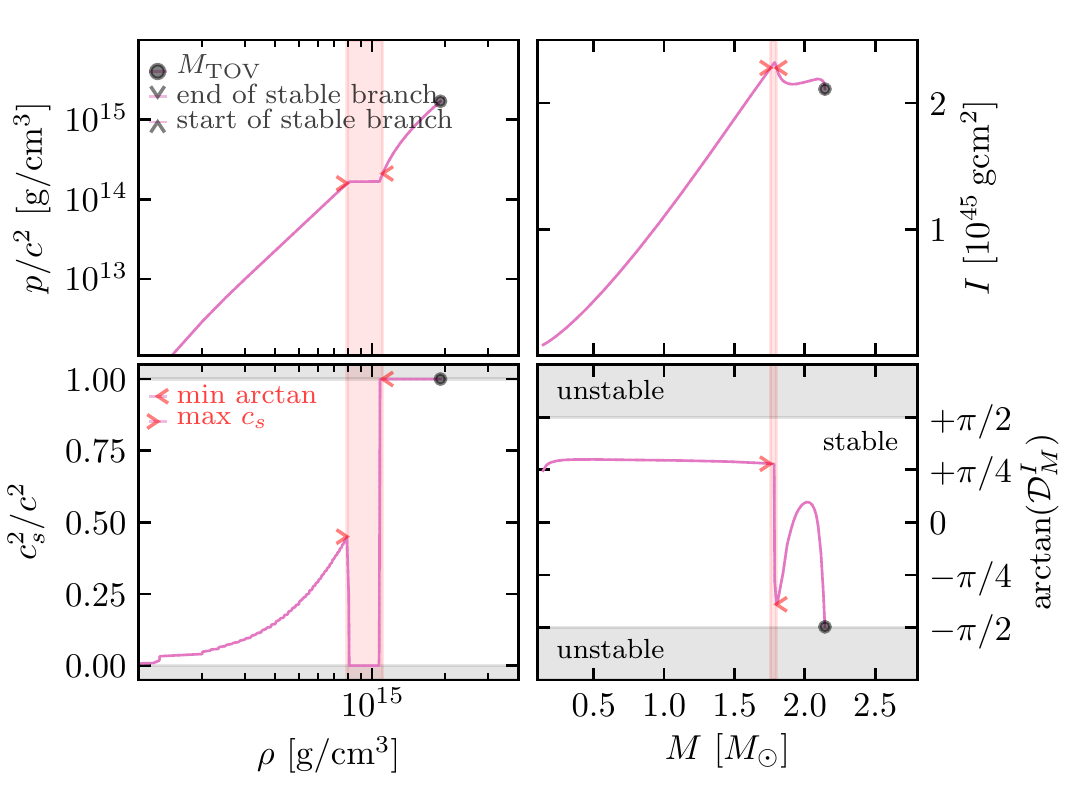}
        \end{center}
    \end{minipage}
    \hfill
    \begin{minipage}{0.33\textwidth}
        \begin{center}
            \includegraphics[width=1.0\textwidth, clip=True, trim=5.4cm 0.00cm 0.0cm 0.25cm]{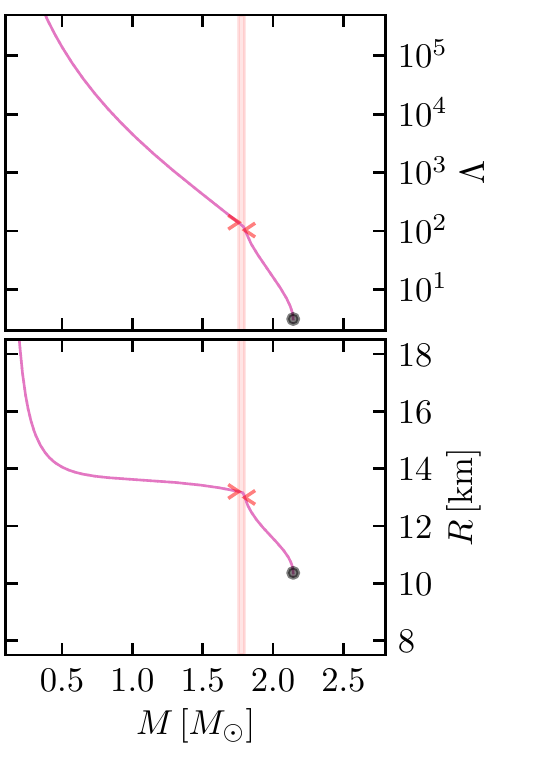}
        \end{center}
    \end{minipage}
    \begin{center}
        \large{Strong Maxwell CSS Phase Transition}
    \end{center}
    \vspace{-0.30cm}
    \begin{minipage}{0.66\textwidth}
        \begin{center}
            \includegraphics[width=1.0\textwidth, clip=True, trim=0.0cm 0.00cm 0.0cm 0.25cm]{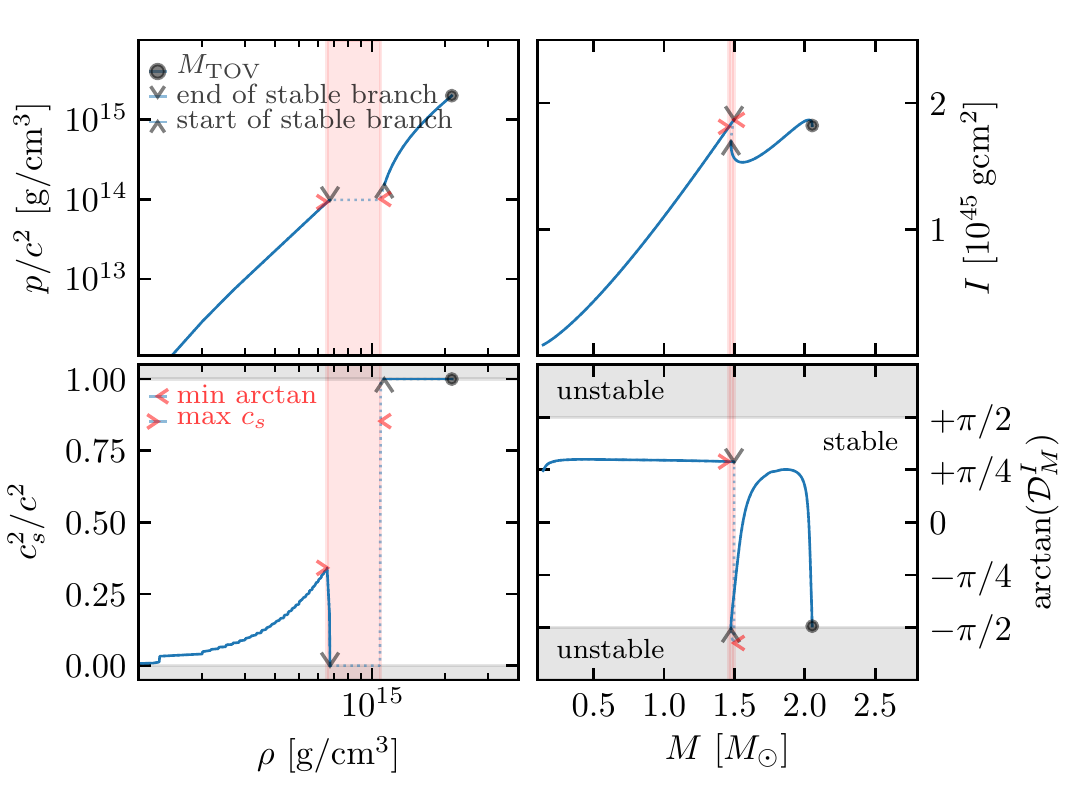}
        \end{center}
    \end{minipage}
    \hfill
    \begin{minipage}{0.33\textwidth}
        \begin{center}
            \includegraphics[width=1.0\textwidth, clip=True, trim=5.4cm 0.00cm 0.0cm 0.25cm]{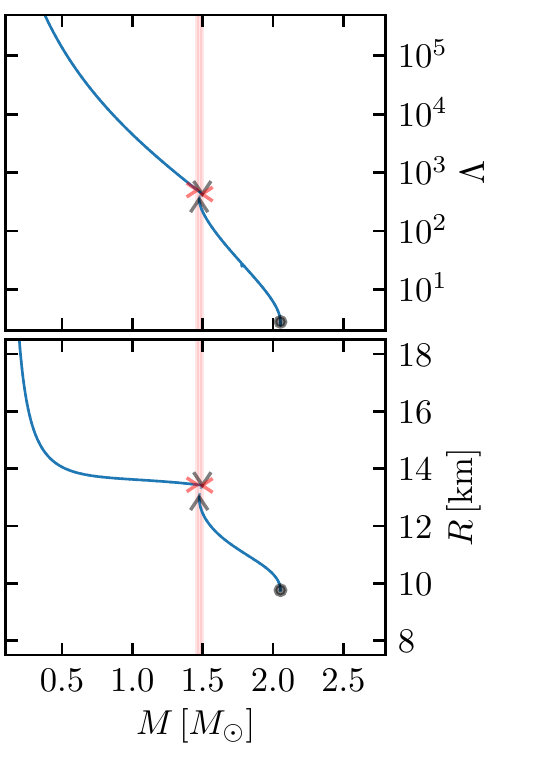}
        \end{center}
    \end{minipage}
    \caption{
        Examples of CSS EoSs based on DBHF~\cite{Gross-Boelting:1998xsk} with a causal extension ($c_s=c$) beyond the end of the phase transition.
        We show examples with (\textit{top}) weak and (\textit{bottom}) strong phase transitions, defined by whether there are multiple stable branches.
        For each EoS, we show (\textit{top left}) the pressure and (\textit{bottom left}) the sound-speed as a function of baryon density, (\textit{top center}) the moment of inertia and (\textit{bottom center}) the novel feature introduced in Sec.~\ref{sec:new stats} 
        (Eq.~\eqref{eq:moi feature}) as a function of gravitational mass, and (\textit{top right}) the $M$--$\Lambda$ and the (\textit{bottom right }) $M$--$R$ relations.
        Stable (unstable) branches are shown with dark solid (light dashed) lines.
        Each curve is labeled with connections between macroscopic phenomenology and microphysical features.
        (\textit{black annotations}) The maximum mass of cold, non-rotating stars ($M_\mathrm{TOV}$) and, where relevant, the beginning and end of stable branches.
        (\textit{red annotations}) The beginning and end of features as identified by the procedure in Sec.~\ref{sec:new stats}.
        (\textit{red shading}) The extent of the identified features.
    }
    \label{fig:DBHF examples}
\end{figure*}

\begin{figure*}
    \begin{center}
        \large{Gibbs Phase Transition}
    \end{center}
    \vspace{-0.30cm}
    \begin{minipage}{0.66\textwidth}
        \begin{center}
            \includegraphics[width=1.0\textwidth, clip=True, trim=0.0cm 0.00cm 0.0cm 0.25cm]{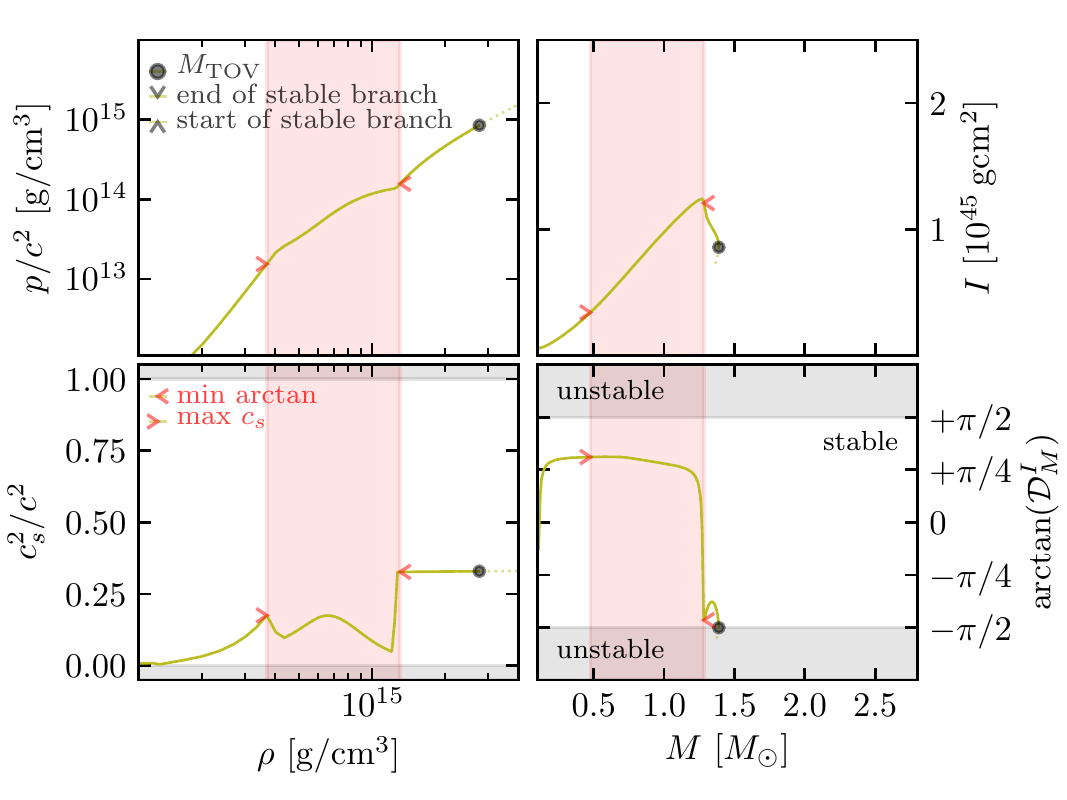}
        \end{center}
    \end{minipage}
    \hfill
    \begin{minipage}{0.33\textwidth}
        \begin{center}
            \includegraphics[width=1.0\textwidth, clip=True, trim=5.4cm 0.00cm 0.0cm 0.25cm]{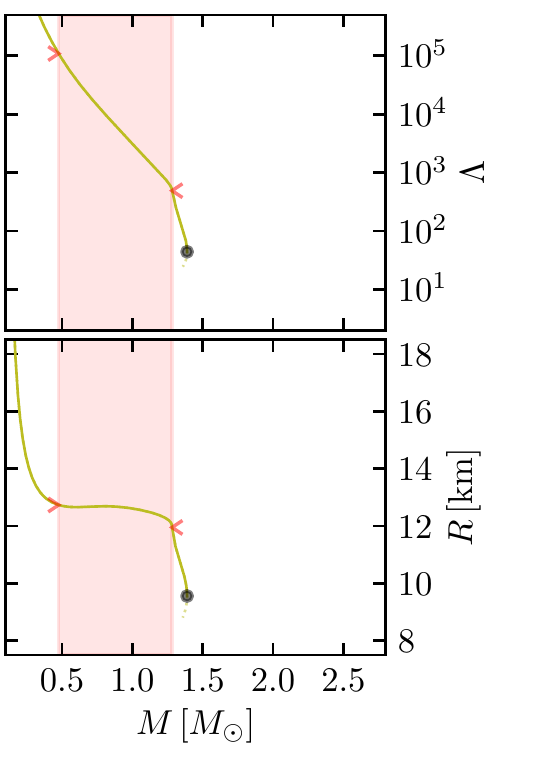}
        \end{center}
    \end{minipage}
    \caption{
        Analogous to Fig.~\ref{fig:DBHF examples} but for more complicated phase transition phenomenology associated with mixed phases (Gibbs construction) from~\citet{Han:2019}, obtained by implementing specific hadronic and quark models..
        Again, the features introduced in Sec.~\ref{sec:new stats} correctly identify the beginning and end of the phase transition even though there is no discontinuity in $c_s$ at the onset and the phase transition corresponds to a wide range of masses.
        The broad extent of the phase transition is not readily apparent from the macroscopic properties alone, which show a sharp feature only at the end of the phase transition.
    }
    \label{fig:Gibbs examples}
\end{figure*}

We begin by reviewing the basic phenomenology of phase transitions from microscopic and macroscopic perspectives in Sec.~\ref{sec:phenomenology} and then introduce our novel model-independent features in Sec.~\ref{sec:new stats}.
We discuss our ability to identify phase transitions in the context of the masquerade problem in Sec.~\ref{sec:masquerade problem}.


\subsection{Phase Transition Morphology}
\label{sec:phenomenology}

The basic phenomenology associated with the phase transitions we consider is a softening of the EoS over some density range.
The following microscopic picture is often invoked.
Consider two species of degenerate, noninteracting fermions with light ($m_l$) and heavy ($m_h > m_l$) rest masses, respectively.
At zero temperature, the system will fill all states up to the Fermi energy ($E_F$) choosing between light and heavy fermions to balance their chemical potentials.
The partial pressure contributed by each fermion will be determined by their respective number densities.
The relation between $E_F$ and the fermion rest masses then determines the system's composition.

If $E_F < m_h$, only light fermions exist.
As the density increases, the pressure must increase as additional light fermions are added to high-momentum states.
However, if $E_F \geq m_h$, heavy fermions in low-momentum states can become energetically favorable. 
These heavy fermions contribute to the rest-mass (and energy) density but have a much lower partial pressure due to their relatively low number density.
The total pressure, then, remains nearly constant at the pressure set by the light fermions at $E_F$.
This will continue until enough heavy fermions appear that a significant fraction of additional particles are light fermions (to balance the chemical potential of heavy fermions) or the partial pressure of the heavy fermions becomes comparable to that of the light fermions.
At that point, the pressure will once again increase with density.

The actual microphysics in a NS is complicated by interactions between particles, but the expected softening based on this heuristic picture is often present in more complicated models. 
Fig.~\ref{fig:DBHF examples} shows the typical behavior of a first-order phase transition with examples constructed from a hadronic model (DBHF~\cite{Gross-Boelting:1998xsk}) at low densities and a constant sound-speed (CSS) extension~\cite{Alford:2013aca} to higher densities.
These EoSs have a sharp boundary separating the two different phases (Maxwell construction); $\ep$ is discontinuous across the boundary and $c_s$ vanishes within the transition.
The EoS in Fig.~\ref{fig:Gibbs examples} employs a mixed phase (Gibbs) construction that exhibits more complicated sound-speed behavior~\cite{Han:2019}, taking into account global charge neutrality (valid for small surface tension between the two phases~\cite{Glendenning:1992vb}) when hadronic and quark matter coexist.
The sound-speed decreases across the phase transition, but does not necessarily drop all the way to zero.
The EoS also shows an approximately density-independent sound speed  towards high densities (due to the specific vMIT model for the pure quark phase), which can be well represented by the generic CSS parametrization.
In both figures, $c_s$ initially increases at low densities, then suddenly decreases across the density range corresponding to the phase transition before recovering and plateauing at a value set by the CSS extension (Maxwell case) or by the microscopic model describing the high-density pure phase (Gibbs case).

While the microscopic details of the phases and their interface may vary, the phase transitions can be characterized phenomenologically by a few parameters, such as the onset density (or pressure) at which the phase transition begins, the density at which it ends, and the latent energy of the transition.
We consider the difference in energy per particle across the phase transition
\begin{equation}\label{eq:latent energy}
    \latentenergy \equiv \left(\frac{\ep}{n}\right)_\mathrm{end} - \left(\frac{\ep}{n}\right)_\mathrm{onset}
\end{equation}
We compute the energy per particle from the energy density $\ep$ and rest-mass density $\rho$ assuming a typical nucleonic mass of \result{$m_n = 938.5\,\mathrm{MeV}$} via $E/N = m_n (\ep/\rho)$.

We wish to associate these microscopic properties of the phase transition with the behavior of macroscopic observables (such as the masses and radii of NSs) that can be probed astronomically.
Strong phase transitions can produce sharp features, such as bends or kinks, in the $M$--$R$ relation.
Figs.~\ref{fig:current constraints M-R} and~\ref{fig:DBHF examples} show examples.
However, EoSs with less abrupt phase transitions, such as the example in Fig.~\ref{fig:Gibbs examples}, may not have a perceptible impact on NS properties.
Moreover, even if a bend or kink is readily apparent in, e.g., the $M$--$R$ relation, it is not immediately clear how to best extract the relevant microphysical parameters of the phase transition.


\subsection{Phase Transition Feature Extraction}
\label{sec:new stats}

We now introduce a set of statistics to identify phase-transition-like behavior in nonparametric EoS realizations.
These statistics are motivated by common features observed in EoSs with phase transitions, such as the ones in Figs.~\ref{fig:DBHF examples} and~\ref{fig:Gibbs examples}, and nonparametric EoS realizations with multiple stable branches.
Our statistics comprise both macroscopic and microscopic features of the EoS and are not tied to an underlying parametrization.
A key macroscopic feature associated with phase transitions is the presence of bends or kinks in the $M$-$R$, $M$-$\Lambda$, and $M$-$I$ relations.\footnote{A feature in one of these relations is accompanied by a similar feature in the others.}
We consider the $M$--$I$ relation, but our procedure also works with other NS observables.

\begin{figure*}
    \begin{center}
        \large{Novel Phase Transition Identification Algorithm}
    \end{center}
    \begin{minipage}{0.64\textwidth}
        \begin{center}
            \includegraphics[width=1.0\textwidth, clip=True, trim=0.0cm 1.15cm 0.0cm 3.7cm]{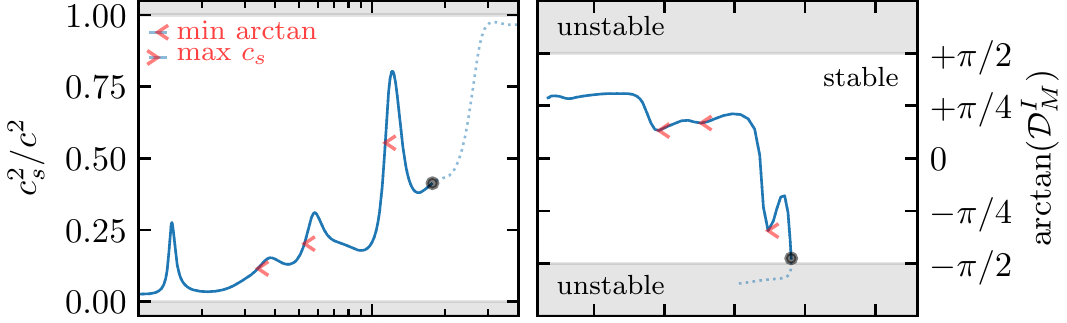} \\
            \vspace{0.05cm}
            \includegraphics[width=1.0\textwidth, clip=True, trim=0.0cm 1.15cm 0.0cm 3.7cm]{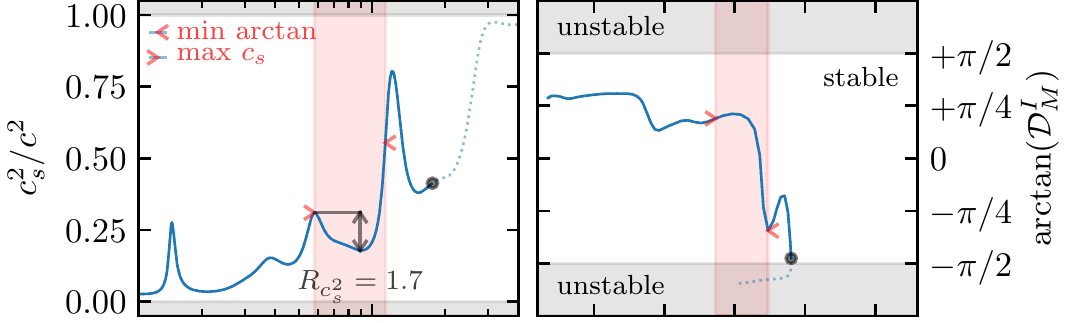} \\
            \vspace{0.05cm}
            \includegraphics[width=1.0\textwidth, clip=True, trim=0.0cm 1.15cm 0.0cm 3.7cm]{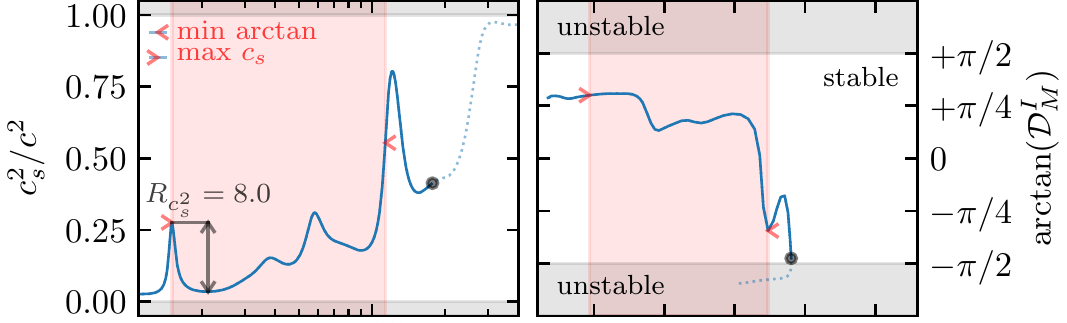} \\
            \vspace{0.05cm}
            \includegraphics[width=1.0\textwidth, clip=True, trim=0.0cm 0.00cm 0.0cm 3.7cm]{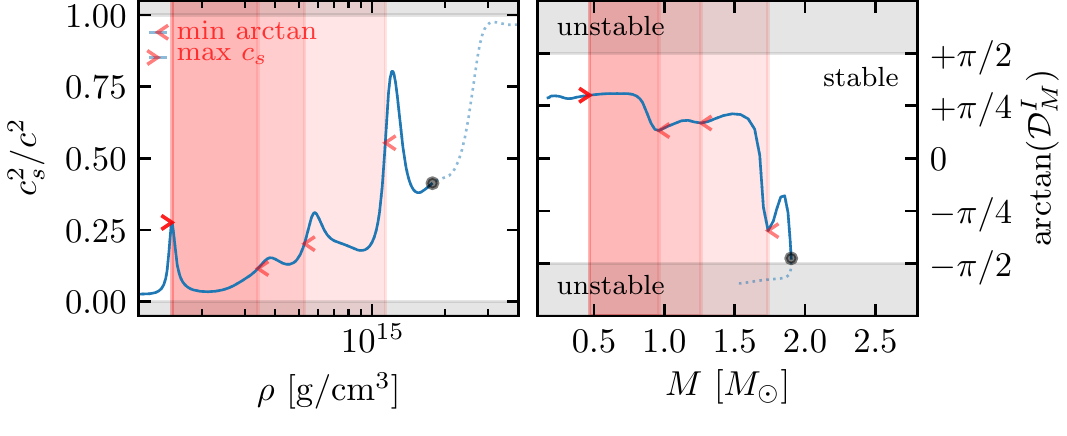}
        \end{center}
    \end{minipage}
    \hfill
    \begin{minipage}{0.32\textwidth}
        Identify all local minima in $\arctan(\moifeature)$.
        In this example there are three with $M \gtrsim 1 \Msolar$. Each local minimum is associated with the end of a candidate phase transition. \\
        \vspace{0.50cm}
        For each local minimum, find the preceding running local maximum in $c_s$.
        This is the start of the candidate phase transition.
        Compute the fraction by which $c_s^2$ decreases from the running local maximum to the smallest $c_s^2$ observed within the candidate phase transition ($R_{c_s^2}$). \\
        \vspace{0.50cm}
        If $R_{c_s^2}$ is sufficiently large, accept the candidate onset density. 
        Proceed to the next local minimum in $\arctan(\moifeature)$. \\
        \vspace{0.50cm}
        Otherwise, reject the candidate's running local maximum $c_s$ and proceed to the next largest running local maximum.
        Compute the new $R_{c_s^2}$ and compare to the threshold.
        Repeat until $R_{c_s^2}$ is large enough or there are no remaining running local maxima in $c_s$.
        If $R_{c_s^2}$ never passes the threshold, reject this local minimum in $\arctan(\moifeature)$ entirely. \\
        \vspace{0.50cm}
        Repeat for remaining local minima.
        This EoS has three local minima that pair with the same running local maximum to produce $R_{c_s^2} \geq 2$ (larger than the threshold used in our main results). \\
        \vspace{1.00cm}
    \end{minipage}
    \caption{
        The feature extraction algorithm: (\textit{left}) the sound-speed as a function of baryon density and (\textit{right}) $\arctan(\moifeature)$~(Eq.~\ref{eq:moi feature}) as a function of the gravitational mass.
        The algorithm progresses from top to bottom, first with the identification of local minima in $\arctan(\moifeature)$ and then pairing each with a corresponding running local maximum in $c_s$.
        The \textit{number of features} reported corresponds to the number of unique running local maxima in $c_s$ selected; in this case 1.
        The \textit{multiplicity of each feature} corresponds to the number of local minima in $\arctan(\moifeature)$~that are paired with the same running local max in $c_s$; in this case 3.
        For demonstration purposes, we show how the algorithm would progress if we had $R_{c_s^2} > 1.7$.
        If the threshold on the drop in the sound-speed $R_{c_s^2}$ was $\leq 1.7$, the algorithm would accept the first pairing (second row) and instead report two features: one at lower densities with multiplicity two and one at higher densities with multiplicity one.
        This would be the case for the main results presented in Secs.~\ref{sec:constraints} and~\ref{sec:prospects}, which use a threshold \result{$R_{c_s^2} > 1.1$}.
    }
    \label{fig:flowchart}
\end{figure*}

We identify phase transitions by looking for characteristic behavior in the derivative of the moment of inertia along a NS sequence.
Specifically, we examine the logarithmic derivative
\begin{equation}\label{eq:moi feature}
    \moifeature \equiv \frac{d\log I/d\log p_c}{d\log M/d\log p_c}\,,
\end{equation}
where $p_c$ is the central pressure.
To aid in categorization, we map the logarithmic derivative to a finite interval by considering its arctangent.\footnote{Technically, we consider \texttt{arctan2}$(d\log I/d\log p_c,\, d\log M/d\log p_c)$ which preserves information about the relative signs of the numerator and denominator within Eq.~\eqref{eq:moi feature}.}
For example, if $|\arctan(\moifeature)| > \pi/2$, then $dM/dp_c < 0$ and the NS is unstable.
If $|\arctan(\moifeature)| < \pi/2$, then $dM/dp_c > 0$ and the NS is stable.
Importantly, the logarithmic derivative is typically constant for EoSs not undergoing a phase transition, but it varies rapidly across the density interval associated with rapid changes in compactness.
Sudden changes in compactness can be caused by a phase transition or the final collapse to a black hole (BH) near $M_\mathrm{TOV}$.
Appendix~\ref{sec:newtonian stars} provides a simple example of this behavior with an incompressible Newtonian star.

A phase transition is identified by a sharp decrease in $\arctan(\moifeature)$.
The change can be discontinuous, but need not be.
Similarly, $\arctan(\moifeature)$ may decrease enough that the star loses stability, but it does not have to.
One can often identify a feature in $\arctan(\moifeature)$ regardless of the exact behavior of $c_s$ or whether there are multiple stable branches.
Thus, it can identify both weak or strong phase transitions, including those with mixed phases.

More concretely, Fig.~\ref{fig:flowchart} demonstrates our algorithm for one EoS drawn from our nonparametric prior process.
We implement the following scheme for identifying phase transitions in arbitrary EoS realizations:

\textbf{(1) Identify candidate ends of phase transitions as local minima in $\arctan(\moifeature)$}.
We first search for local minima in $\arctan(\moifeature)$ bracketed by stable NSs.
This excludes the sudden decrease in $\arctan(\moifeature)$ associated with the collapse to a BH above $M_{\rm TOV}$.
Each such feature is associated with a phase transition, and the density at which this $\moifeature$ feature occurs is taken to be the end of the phase transition ($\ep_e$).
In the absence of a suitable local minimum, we deem the EoS to have no phase transition.

\textbf{(2) Identify a candidate onset density for an end point}.
We then associate each local minimum in $\arctan(\moifeature)$ with the largest local maximum in $c_s$ that precedes it (i.e., occurs at lower densities).
Specifically, we select a running maximum in $c_s$, defined as the local maximum that is larger than all preceding local maxima.
The density at which this $c_s$ feature occurs becomes the candidate for the onset density $\ep_t$.
If there is no preceding local maximum in $c_s$, then we deem the EoS to have no phase transition.

\textbf{(3) Repeat step (2) until an acceptable onset density is found}. 
We require the minimum $c_s^2$ between the candidate onset and end densities to be at least \result{10\%} smaller than $c_s^2$ at the onset.
If this threshold on the fractional change ($R_{c_s^2}$) is not met, the candidate onset density is rejected, and the preceding running local maximum is considered in its place.
This procedure is repeated until $R_{c_s^2}$ is large enough (candidate is accepted) or there are no more local maxima in $c_s^2$ (candidate phase transition is rejected).
See Appendix~\ref{sec:thresholds} for more discussion of thresholds within the feature selection process.

\textbf{(4) Repeat steps (2-3) for remaining local minima in $\arctan(\moifeature)$}.
We identify exactly one onset density for each end density.

If there is more than one local minimum in $\arctan(\moifeature)$, several of them may be associated with the same onset density.
In that case, we define the \textit{multiplicity} of the phase transition as the number of local minima in $\arctan(\moifeature)$ associated with the same running local maximum in $c_s$.
We use the multiplicity of the phase transition as a proxy for the complexity of the phase transition morphology.
For example, the complexity of the sound speed's behavior within the phase transition could indicate the (dis)appearance of (new) species of particles within the system or be related to inflection points in the particle fractions.
See, e.g., examples of the equilibrium sound speed profiles in \citet{Constantinou:2021hba,Constantinou:2023ged} exploring various conditions.
Complementarily, the number of selected running local maxima in $c_s^2$ defines the number of \moifeature~features within the EoS.
These basic counting exercises provide a classification scheme for simple (multiplicity 1) and complex (multiplicity $>1$) $c_s$ structure within the phase transition along with the number of transitions.

After this procedure, each phase transition is characterized by an onset density (or pressure or stellar mass) and an end density (largest density of all local minima in $\arctan(\moifeature)$ associated with the onset).
Based on these points, we define various properties of the phase transition.
We focus on \latentenergy~in Secs.~\ref{sec:constraints} and~\ref{sec:prospects}.

Of course, the points identified by the above procedure are only proxies for the true onset and end of the phase transition.
While the correspondence is excellent for Maxwell constructions (Fig.~\ref{fig:DBHF examples}), it may not be perfect for more complicated models.
See, e.g., Fig.~\ref{fig:Gibbs examples extra}.
Moreover, because the feature identification hinges on the presence of local minima in $\arctan(\moifeature)$, we sometimes cannot identify phase transitions that occur near $M_\mathrm{TOV}$, i.e., that terminate in collapse to a BH.
As such, it may be difficult to determine whether NSs collapse to BHs because of a sudden decrease in $c_s$ at high densities or whether $c_s$ remains large and the NS's self-gravity wins without assistance.
Empirically, we find a correlation between the sharpness of the bend in $\arctan(\moifeature)$ near the collapse to a BH and the existence of a phase transition at those densities, but we leave further investigations of this to future work.

Additionally, the specific onset, end, and latent energy values we extract for the phase transition are sensitive to the threshold on $R_{c_s^2}$.
A lower threshold would favor the identification of a greater number of weaker phase transitions at the risk of selecting small upward fluctuations in $c_s$ (unconstrained by current data) as the onset even if more plausible features in $c_s$ exist at lower densities.
A higher threshold would retain only the strongest phase transitions.
In what follows, we choose to ignore phase-transition-like features with \result{$R_{c_s^2} < 1.1$} as an attempt to balance these extremes, but the exact choice is \textit{ad hoc}.
See Appendix~\ref{sec:thresholds} for more discussion.


\subsection{Connections between Macroscopic and Microphysical Behavior: the Masquerade Problem}
\label{sec:masquerade problem}

\begin{figure}
    \includegraphics[width=.49\textwidth]{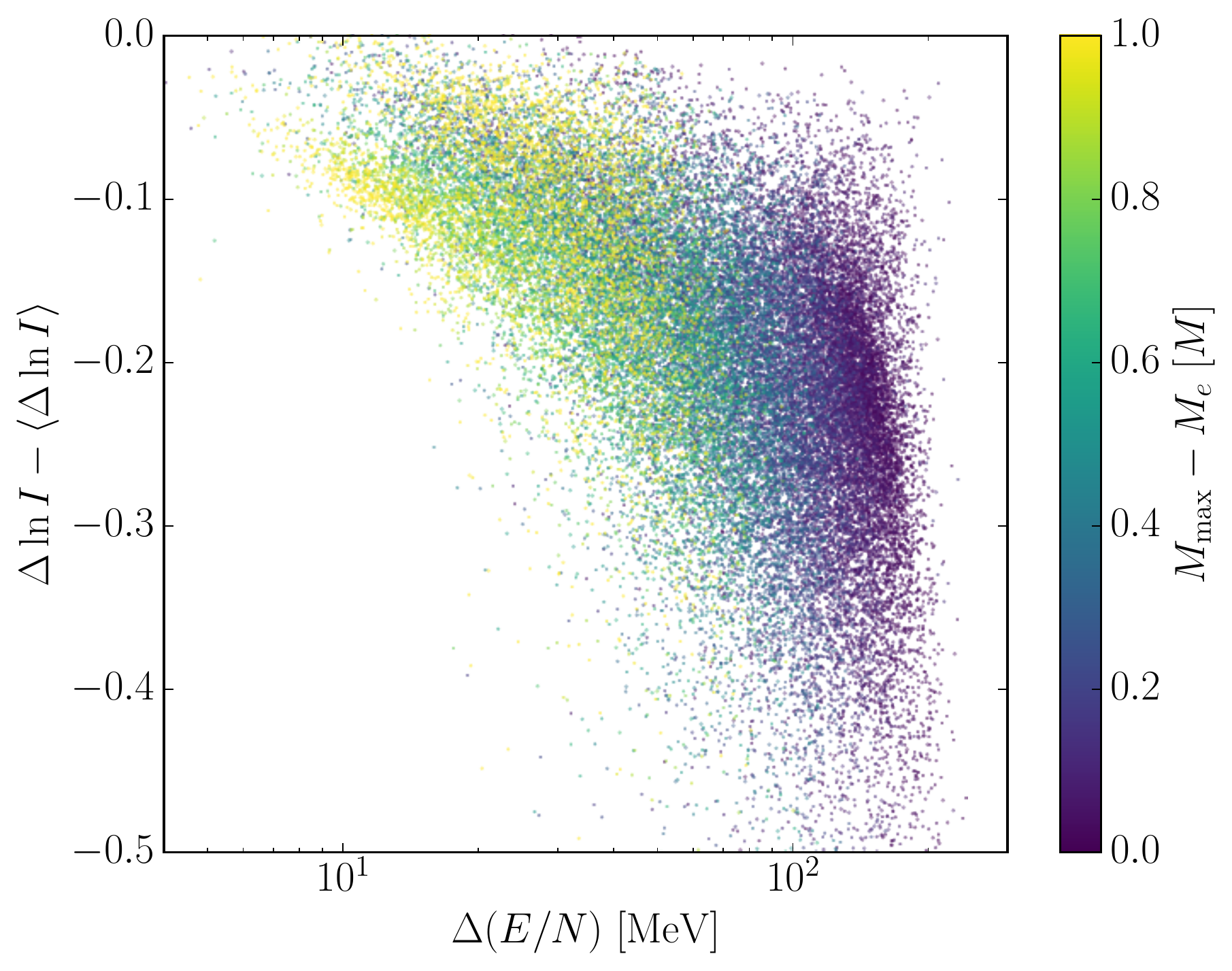}
    \caption{
        Correlations between the divergence between macroscopic properties caused by a phase transition $\Delta \ln I - \left<\Delta \ln I\right>$ and the latent energy per particle of the associated phase transition \latentenergy\ for all transitions that begin at masses greater than \result{$0.7 \,\Msolar$}.
        Color indicates the proximity of the phase transition's end to $M_\mathrm{TOV}$.
        Large divergences in macroscopic properties can only be caused by phase transitions with large \latentenergy, but not all phase transitions with large \latentenergy~cause large divergences in macroscopic properties.
    }
    \label{fig:delta moi feature}
\end{figure}

We expect \latentenergy~to be related to phase transition's impact on macroscopic properties.
However, this mapping is complicated because the same \latentenergy~can lead to very different changes in NS properties depending on the onset density and pressure. In order to explore this relation, we consider how much the phase transition causes the macroscopic properties to diverge from what they would have been without it.
This provides a natural interpretation to the masquerade problem, as it will be difficult to distinguish between two nearby $M$--$I$ curves that never diverge from each other without extremely precise observations.

While it is not trivial to construct such a divergence without an underlying parametrization (one cannot just ``turn off'' the phase transition), Fig.~\ref{fig:delta moi feature} shows an example: the difference between the change in the (logarithm of the) moment of inertia across the phase transition and what it would have been if the transition was not present.
We measure the actual $\Delta \ln I$ directly from the identified onset and end of a transition, and approximate what it would have been without a phase transition via the following observation.
In the absence of phase-transition-like behavior, \moifeature~is roughly constant: $\left<\moifeature\right>$.
Appendix~\ref{sec:newtonian stars} shows that $\left<\moifeature\right>=5/3$ for incompressible Newtonian stars, and we empirically find values near $\left<\moifeature\right> \sim 1.7$ for general EoSs in full General Relativity.
Therefore, we approximate the change in the moment of inertia that would have occurred without the phase transition as $\left<\Delta \ln I\right> = \left<\moifeature\right> \Delta \ln M$, where $\Delta \ln M$ is again defined by the onset and end of the transition.

Fig.~\ref{fig:delta moi feature} shows $\Delta \ln I - \left<\Delta \ln I\right>$ as a function of the phase transition's latent energy per particle.
We see that large $|\Delta \ln I - \left<\Delta \ln I\right>|$ are only possible with large \latentenergy, but large \latentenergy~do not always lead to large divergences.
Again, this demonstrates the masquerade problem: large microphysical changes may not always manifest as observable features within macroscopic NS observables.
Additionally, large \latentenergy~tend to produce end masses (NS mass with central density at the end of the phase transition) close to $M_\mathrm{TOV}$.
This is because large phase transitions imply very compact stellar cores (due to relatively low pressures at high densities), which are likely to collapse to BHs if even a small amount of additional matter is added.
Similarly, transitions with very large \latentenergy~may lead to direct collapse to a BH.
Because our identification algorithm (Sec.~\ref{sec:new stats}) struggles to detect features that cause the stellar sequence to collapse to a BH, this may cause a selection in the maximum \latentenergy~for which we can identify \moifeature~features in Fig.~\ref{fig:delta moi feature}.
Empirically, we only identify $\latentenergy \lesssim 300\,\mathrm{MeV}$.


 \section{Constraints with Current Astrophysical Observations}
\label{sec:constraints}

Equipped with the procedure defined in Sec.~\ref{sec:new stats}, we now turn to current astrophysical observations.
Following~\citet{Legred:2021hdx}, we consider GW observations (GW170817~\cite{GW170817, GW170817-properties} and GW190425~\cite{GW190425}) assuming that all objects below (above) $M_\mathrm{TOV}$ are NSs (BHs), NICER observations of pulsar hotspots (J0030+0451~\cite{J0030-Miller} and J0740+6620\footnote{We use the headline results from~\citet{J0740-Miller} rather than~\citet{J0740-Riley} because the former implements the measured cross-calibration between NICER and XMM. Note that~\citet{J0740-Riley} also presents posteriors conditioned on the nominal published cross-calibration, although their headline results implement looser priors. See also~\citet{Salmi:2022cgy}.}~\cite{J0740-Miller}), and radio-based mass measurements of pulsars (J0348+0432~\cite{J0348-Antoniadis} and J0740+6620~\cite{J0740-Cromartie, J0740-Fonseca}).

We use a model-agnostic nonparametric EoS prior, which by construction includes little information from either nuclear theory or experiment at any density beyond the requirements of thermodynamic stability and causality.
See e.g.,~\citet{Essick:2019ldf}.
This prior allows us to isolate the impact of astrophysical observations on the high-density EoS  ($\gtrsim \rhonuc$) without introducing modeling artifacts, as are common in phenomenological parametric models~\cite{Legred:2022pyp}.
Compared to some nonparametric efforts (e.g.,~\citet{J0740-Miller}), our nonparametric prior was constructed with the goal of maximizing model freedom.
It therefore already contains many EoS realizations that exhibit characteristics of phase transition phenomenology, including EoSs with multiple stable branches.
While additional theoretical and/or experimental low-density information could be considered, see e.g., Refs.~\cite{Essick:2020, Essick:2021kjb, Essick:2021ezp}, we leave those to future work and focus on astrophysical observations. Similarly, we do not incorporate pQCD calculations at high densities~\cite{Komoltsev:2021jzg, Gorda:2022jvk} as initial explorations indicated that these constraints are model-dependent.\footnote{Specifically, when evaluating the pQCD likelihood at $10 \rhonuc$ we find that pQCD results influence NS near $M_\mathrm{TOV}$ in agreement with~\citet{Gorda:2022jvk}. However, those constraints are weaker when we use the central density of stars with $M=M_\mathrm{TOV}$, in agreement with~\citet{Somasundaram:2022ztm}. Therefore, the exact impact of pQCD constraints on the inference of the EoS at lower densities is still somewhat uncertain because it changes with the choice of where the integral constraints are applied.
}

\begin{figure*}
    \centering
    \includegraphics[width=.99\textwidth]{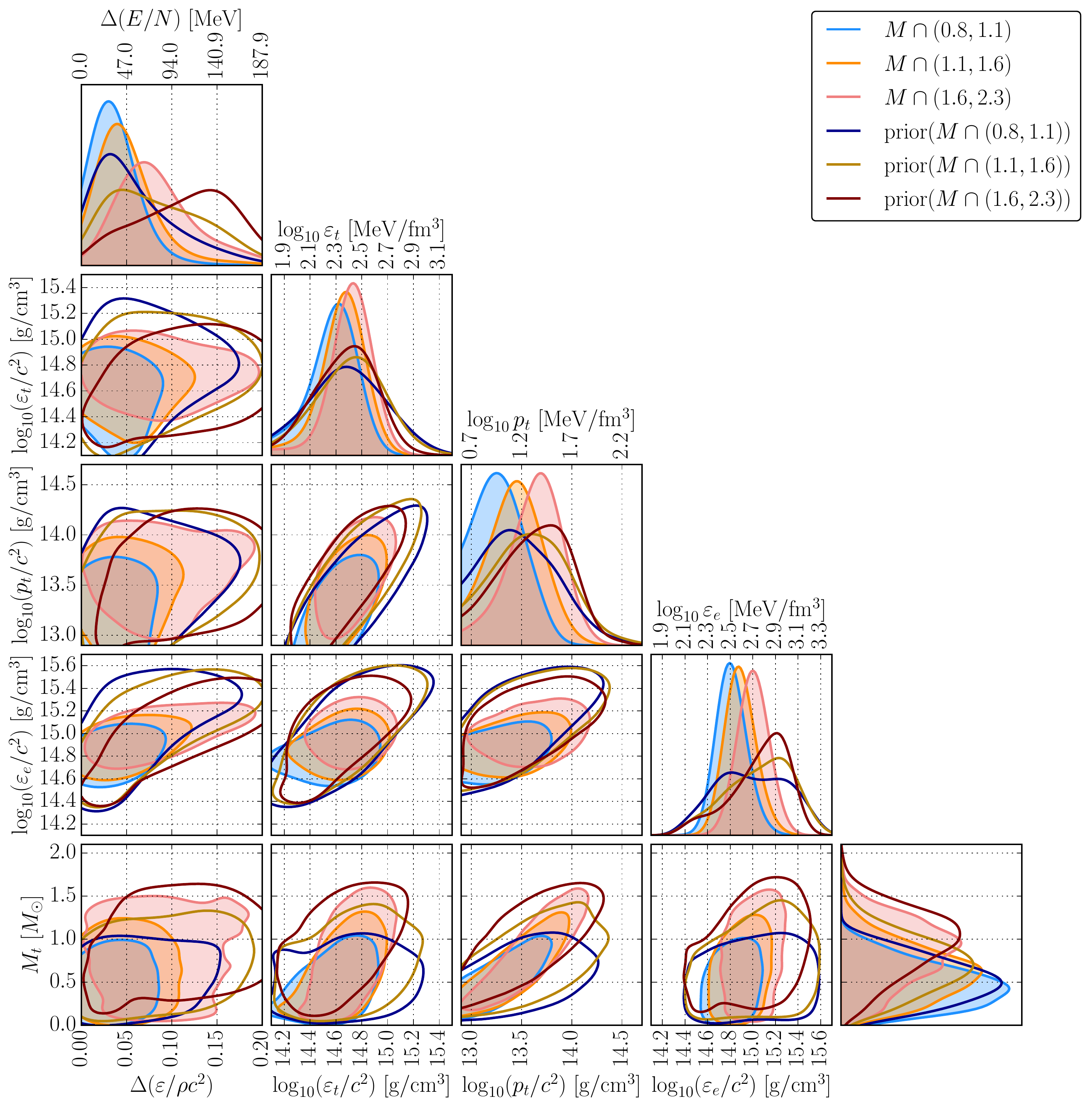}
    \caption{
        Marginalized (\textit{unshaded}) priors and (\textit{shaded}) posteriors for parameters that characterize phase transitions based on current astrophysical data from pulsar masses, GWs, and X-ray mass-radius measurements.
        For each EoS we report the properties of the transition with the largest \latentenergy~that overlaps with each mass interval.
        We report (\textit{left to right}), the latent energy (\latentenergy), the onset energy density ($\ep_t$), the onset pressure ($p_t$), the energy density at the end of the transition ($\ep_e$), and the onset mass scale ($M_t$) for three mass-overlap regions: 0.8--1.1$\,\Msolar$, 1.1--1.6$\,\Msolar$, and 1.6--2.3$\,\Msolar$.
    }
    \label{fig:current constraints corner}
\end{figure*}

\begin{table*}
    \centering
    \caption{
        Ratios of maximized and marginalized likelihoods for different types of features based on current astrophysical observations: (P) pulsar masses, (G) GW observations from LIGO/Virgo, and (X) X-ray timing from NICER.
        See Eqs.~\eqref{eq:maxL nomenclature} and~\eqref{eq:bayes factor notation} for an explicit definition of this notation.
        We consider multiple mass ranges (features must span stellar masses that overlap with the specified range) and latent energies (where appropriate, there must be at least one feature with latent energy larger than the threshold).
        We show the statistics for both the number of stable branches and $\moifeature$~features.
        Error estimates for Bayes factors ($\mathcal{B}$) approximate 1-$\sigma$ uncertainty from the finite Monte Carlo sample size.
        See Tables in Appendix~\ref{sec:additional representations} for additional combinations of subsets of astrophysical data.
    }
    \label{tab:current constraints}
    {\renewcommand{\arraystretch}{1.5}
    \begin{tabular}{@{\extracolsep{0.2cm}} c ccc c ccc}
        \hline\hline
        \multirow{2}{*}{\thead{$M$\\$[\MassUnit]$}} & \multicolumn{3}{c}{Stable Branches}
          & \multirow{2}{*}{\thead{$\min\latentenergy$\\$[\LatentEnergyUnit]$}} & \multicolumn{3}{c}{\moifeature~Features} \\
         & $\max\mathcal{L}^{n>1}_{n=1}(\mathrm{PGX})$ & $\mathcal{B}^{n>1}_{n=1}(\mathrm{PGX})$ & $\mathcal{B}^{n>1}_{n=1}(\mathrm{GX|P})$
         & & $\max\mathcal{L}^{n>0}_{n=0}(\mathrm{PGX})$ & $\mathcal{B}^{n>0}_{n=0}(\mathrm{PGX})$ & $\mathcal{B}^{n>0}_{n=0}(\mathrm{GX|P})$ \\
        \hline\hline
        \multirow{3}{*}{\LowMassBinMin--\LowMassBinMax}
         & \multirow{3}{*}{\MaxLikeRatioBranchLowMassPSRGWXray} & \multirow{3}{*}{\BayesBranchLowMassPSRGWXray} & \multirow{3}{*}{\BayesBranchLowMassGWXrayGivenPSR}
           & \MidLatentEnergy & \MaxLikeRatioMoILowMassMidLatentEnergyPSRGWXray & \BayesMoILowMassMidLatentEnergyPSRGWXray & \BayesMoILowMassMidLatentEnergyGWXrayGivenPSR \\
         & & &
           & \HighLatentEnergy & \MaxLikeRatioMoILowMassHighLatentEnergyPSRGWXray & \BayesMoILowMassHighLatentEnergyPSRGWXray & \BayesMoILowMassHighLatentEnergyGWXrayGivenPSR\\
         & & &
           & \HugeLatentEnergy & \MaxLikeRatioMoILowMassHugeLatentEnergyPSRGWXray & \BayesMoILowMassHugeLatentEnergyPSRGWXray & \BayesMoILowMassHugeLatentEnergyGWXrayGivenPSR \\
        \cline{2-8}
        \multirow{3}{*}{\MidMassBinMin--\MidMassBinMax}
         & \multirow{3}{*}{\MaxLikeRatioBranchMidMassPSRGWXray} & \multirow{3}{*}{\BayesBranchMidMassPSRGWXray} & \multirow{3}{*}{\BayesBranchMidMassGWXrayGivenPSR}
           & \MidLatentEnergy & \MaxLikeRatioMoIMidMassMidLatentEnergyPSRGWXray & \BayesMoIMidMassMidLatentEnergyPSRGWXray & \BayesMoIMidMassLowLatentEnergyGWXrayGivenPSR \\
         & & &
           & \HighLatentEnergy & \MaxLikeRatioMoIMidMassHighLatentEnergyPSRGWXray & \BayesMoIMidMassHighLatentEnergyPSRGWXray & \BayesMoIMidMassLowLatentEnergyGWXrayGivenPSR \\
         & & &
           & \HugeLatentEnergy & \MaxLikeRatioMoIMidMassHugeLatentEnergyPSRGWXray & \BayesMoIMidMassHugeLatentEnergyPSRGWXray & \BayesMoIMidMassHugeLatentEnergyGWXrayGivenPSR \\
        \cline{2-8}
        \multirow{3}{*}{\HighMassBinMin--\HighMassBinMax}
         & \multirow{3}{*}{\MaxLikeRatioBranchHighMassPSRGWXray} & \multirow{3}{*}{\BayesBranchHighMassPSRGWXray} & \multirow{3}{*}{\BayesBranchHighMassGWXrayGivenPSR}
           & \MidLatentEnergy & \MaxLikeRatioMoIHighMassMidLatentEnergyPSRGWXray & \BayesMoIHighMassMidLatentEnergyPSRGWXray & \BayesMoIHighMassLowLatentEnergyGWXrayGivenPSR \\
         & & &
           & \HighLatentEnergy & \MaxLikeRatioMoIHighMassHighLatentEnergyPSRGWXray & \BayesMoIHighMassHighLatentEnergyPSRGWXray & \BayesMoIHighMassLowLatentEnergyGWXrayGivenPSR \\
         & & &
           & \HugeLatentEnergy & \MaxLikeRatioMoIHighMassHugeLatentEnergyPSRGWXray & \BayesMoIHighMassHugeLatentEnergyPSRGWXray & \BayesMoIHighMassHugeLatentEnergyGWXrayGivenPSR \\
        \hline
    \end{tabular}
    }
\end{table*}

Current observations span masses roughly between \externalresult{1.2-2.1 $\Msolar$}.\footnote{The smallest observed mass we consider is likely the secondary in GW190425~\cite{GW190425}, although there is considerable uncertainty in the event's mass ratio. The largest observed mass is J0740+6620~\cite{J0740-Fonseca}.}
What is more, the answer to questions such as, ``how many phase transitions does the EoS have?'' depends on the mass or density range considered, and we do not wish to confound our inference with the presence of \moifeature~features that occur at masses below the smallest observed NS.
As such, we divide the prior into multiple sets defined by whether or not the EoS has a \moifeature~feature that overlaps with a specific mass range.
That is, whether the range of densities spanning the feature overlaps with the range of central densities for stellar models within a specified mass interval.
We consider three mass ranges:
\begin{itemize}
    \item $M \in [0.8,\,1.1)\,\Msolar$: features that occur below the current observed set of NSs.
    \item $M \in [1.1,\,1.6)\,\Msolar$: features that could influence observed NSs, particularly in the peak of the distribution of known galactic pulsars~\cite{Alsing:2017bbc,Farr:2019}.
    \item $M \in [1.6,\,2.3)\,\Msolar$: features that may influence observed NSs, but at high enough masses that individual GW systems are unlikely to confidently bound the tidal deformability away from zero.
\end{itemize}
Individual EoSs may belong to multiple sets if they have multiple or large~\moifeature~features or just happen to straddle a boundary.

Table~\ref{tab:current constraints} presents ratios of maximized and marginal likelihoods conditioned on different datasets.
The ratio of maximized likelihoods for all astrophysical data (pulsars (P), GWs (G), and X-ray observations (X)) for different subsets of our prior ($A$ and $B$) is
\begin{equation}\label{eq:maxL nomenclature}
    \max\mathcal{L}^A_B(\mathrm{PGX}) = \frac{\max\limits_{\ep\in A}\, p(\mathrm{PGX}|\ep)}{\max\limits_{\ep \in B}\, p(\mathrm{PGX}|\ep)}\,,
\end{equation}
where the maximization is over different EoSs $\ep$.
The Bayes factor is the ratio of marginal likelihoods
\begin{equation}\label{eq:bayes factor notation}
    \mathcal{B}^A_B(\mathrm{GX}|\mathrm{P}) = \frac{p(\mathrm{GX}|\mathrm{P}; A)}{p(\mathrm{GX}|\mathrm{P}; B)}\,,
\end{equation}
where, for example, 
\begin{equation}
    p(\mathrm{GX}|\mathrm{P}; A) = \int \mathcal{D}\ep\, p(\mathrm{GX}|\ep) p(\ep|\mathrm{P}, A)\,,
\end{equation}
and
\begin{equation}
    p(\ep|\mathrm{P}, A) = \frac{p(\mathrm{P}|\ep) p(\ep|A)}{\int \mathcal{D} \ep\, p(\mathrm{P}|\ep) p(\ep|A)}\,.
\end{equation}
We report these statistics for both the number of stable branches and the number of \moifeature~features, conditioned on several minimum \latentenergy~thresholds.
We present both statistics because each has its relative strengths and weaknesses.
While Occam factors may be important for  Bayes factors, they do not affect the ratio of maximized likelihoods.
At the same time, the maximized likelihoods may correspond to an extremely rare EoS, whereas the Bayes factors provide an average over typical EoS behavior.
We therefore should trust statements about which both statistics broadly agree.

\begin{figure*}
    {\large $\max \mathcal{L}(\mathrm{PGX}|N) / \max \mathcal{L}(\mathrm{PGX})$}
    \hfill
    {\large \textcolor{red}{$p(N|\mathrm{PGX}) / p(N)$}}
    \hfill
    {\large \textcolor{blue}{$p(N|\mathrm{PGX}) / p(N|\mathrm{PSR})$}}
    \\
    \vspace{0.25cm}
    \includegraphics[width=0.264\textwidth, clip=True, trim=0.00cm 1.10cm 0.0cm 0.0cm]{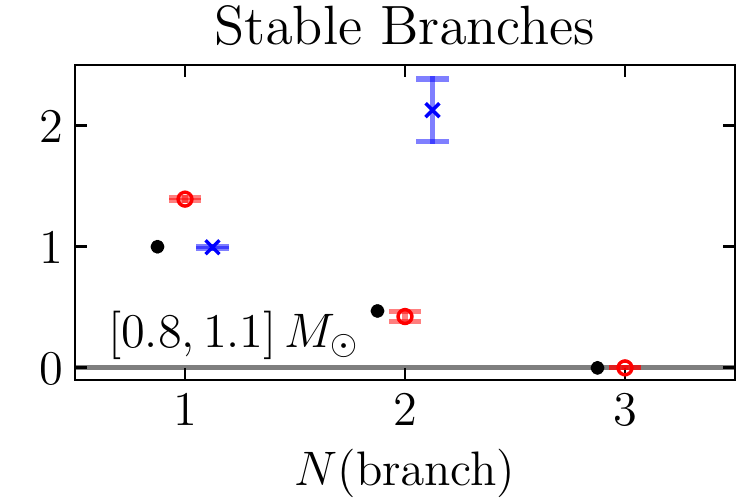}
    \includegraphics[width=0.239\textwidth, clip=True, trim=0.75cm 1.10cm 0.0cm 0.0cm]{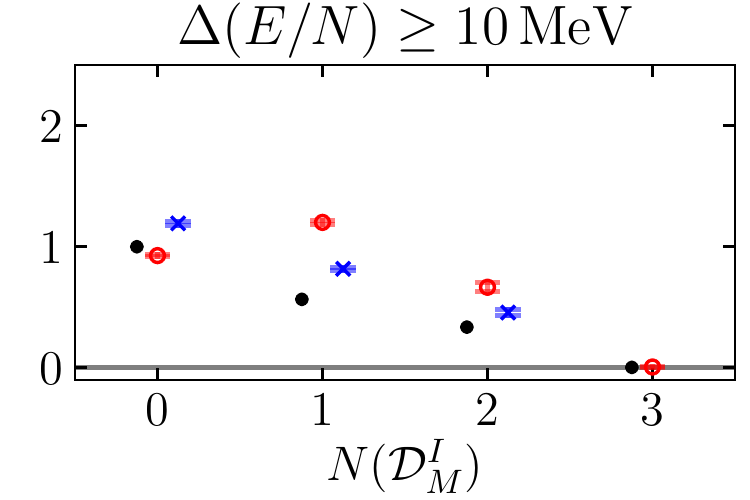}
    \includegraphics[width=0.239\textwidth, clip=True, trim=0.75cm 1.10cm 0.0cm 0.0cm]{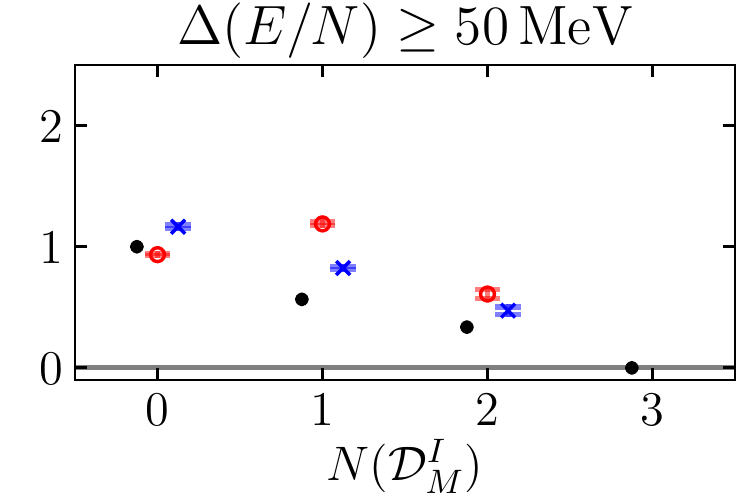}
    \includegraphics[width=0.239\textwidth, clip=True, trim=0.75cm 1.10cm 0.0cm 0.0cm]{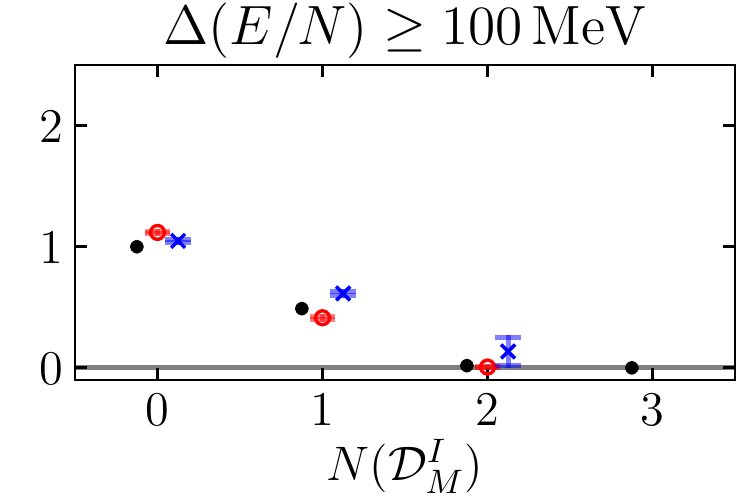} \\
    \includegraphics[width=0.264\textwidth, clip=True, trim=0.00cm 1.10cm 0.0cm 0.60cm]{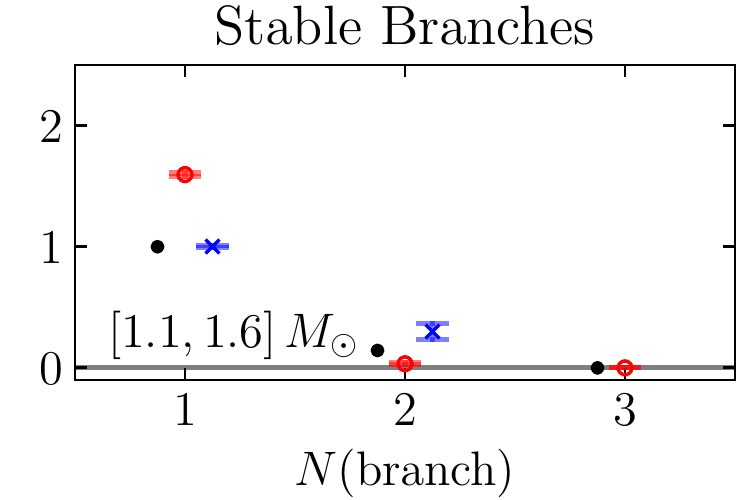}
    \includegraphics[width=0.239\textwidth, clip=True, trim=0.75cm 1.10cm 0.0cm 0.60cm]{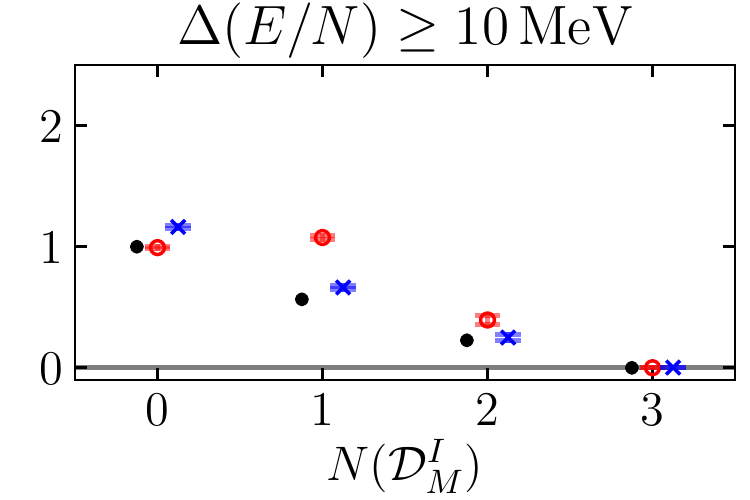}
    \includegraphics[width=0.239\textwidth, clip=True, trim=0.75cm 1.10cm 0.0cm 0.60cm]{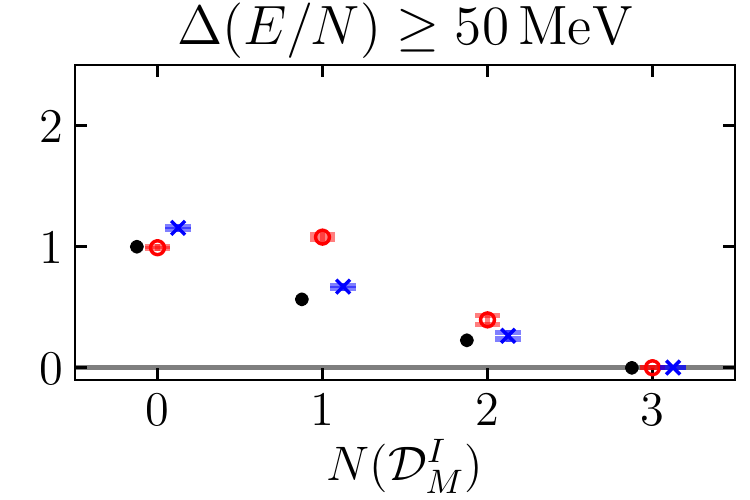}
    \includegraphics[width=0.239\textwidth, clip=True, trim=0.75cm 1.10cm 0.0cm 0.60cm]{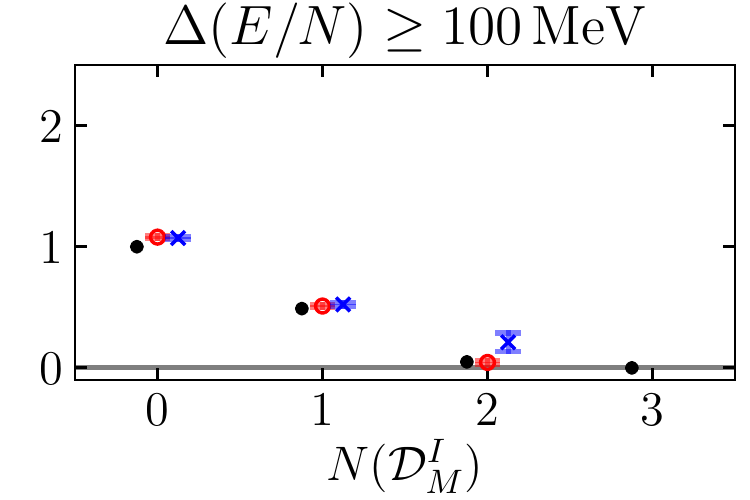} \\
    \includegraphics[width=0.264\textwidth, clip=True, trim=0.00cm 0.00cm 0.0cm 0.60cm]{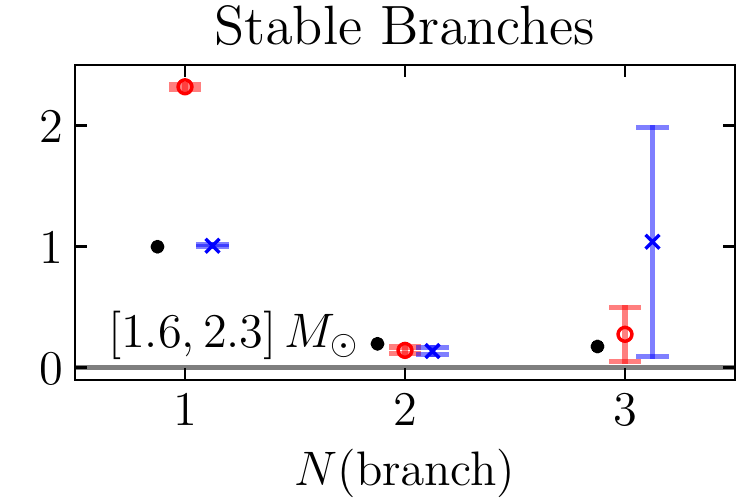}
    \includegraphics[width=0.239\textwidth, clip=True, trim=0.75cm 0.00cm 0.0cm 0.60cm]{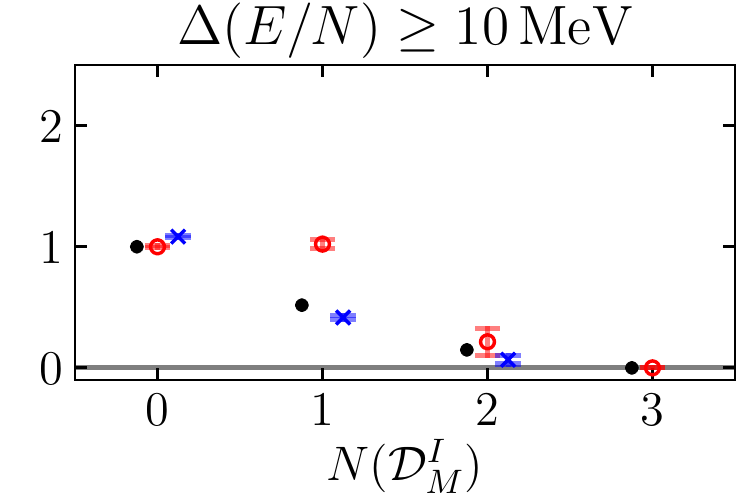}
    \includegraphics[width=0.239\textwidth, clip=True, trim=0.75cm 0.00cm 0.0cm 0.60cm]{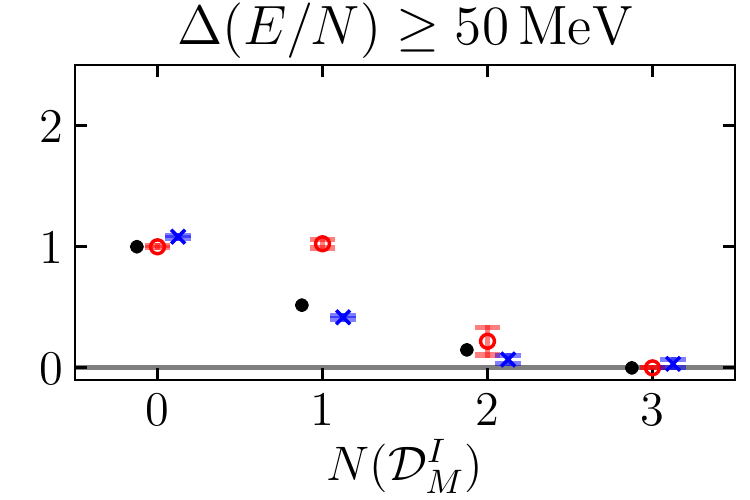}
    \includegraphics[width=0.239\textwidth, clip=True, trim=0.75cm 0.00cm 0.0cm 0.60cm]{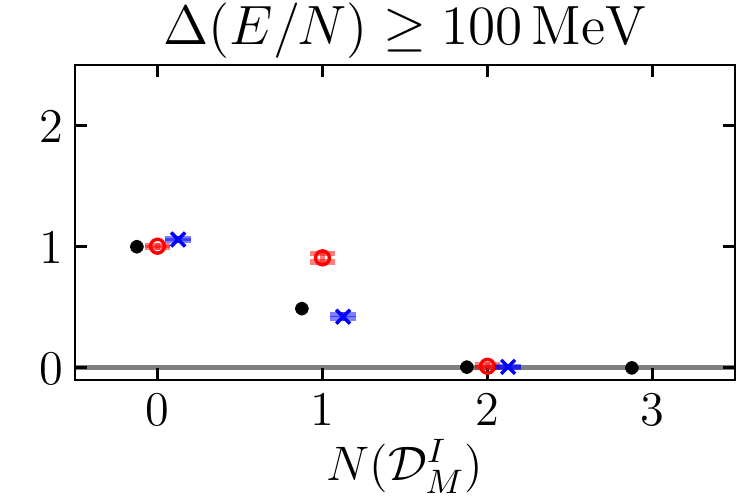}
    \caption{
        Ratios of probabilities conditioned on different numbers of features.
        Compare to Table~\ref{tab:current constraints}; see Eqs.~\eqref{eq:maxL nomenclature} and~\eqref{eq:bayes factor notation} for an explicit definitions of our notation.
        (\textit{left}) Distributions over the number of stable branches and (\textit{right}) distributions over the number of \moifeature~features for EoSs with $\latentenergy \geq$ 10, 50, and 100 MeV, respectively for different mass-overlap regions: (\textit{top}) 0.8--1.1$\,\Msolar$, (\textit{middle}) 1.1--1.6$\,\Msolar$, and (\textit{bottom}) 1.6--2.3$\,\Msolar$.
        We show the ratio of maximum likelihoods (\textit{black dots}) and the posterior divided by the prior (\textit{circles and x's}).
        As in Table~\ref{tab:current constraints}, we consider (PGX, \textit{red circles}) the ratio of the posterior conditioned on PSR masses, GW coalescences, and X-ray timing and compare it to our nonparametric prior as well as (\textit{blue x's}) the posterior conditioned on only PSR masses.
        Error bars approximate 1-$\sigma$ uncertainties from the finite size of our prior sample.
        In general, a single stable branch without strong \moifeature~features is preferred.
    }
    \label{fig:current constraints counts}
\end{figure*}

Overall, we expect stronger constraints on features that overlap with the observed mass range.
In Figs.~\ref{fig:current constraints corner},~\ref{fig:current constraints counts}, and Table~\ref{tab:current constraints}, we indeed find the strongest constraints on phase transitions that occur in NSs less massive than $1.6 \,\Msolar$, although constraints for $M \in [0.8, 1.1) \,\Msolar$ and $M \in [1.1, 1.6) \,\Msolar$ are comparable.
Indeed, in Fig.~\ref{fig:current constraints corner} the posterior for the latent energy is more constrained with respect to the prior for masses below $1.6 \,\Msolar$.
\result{Furthermore, Table~\ref{tab:current constraints} shows that the Bayes factor using all astrophysical data disfavors the presence of large \moifeature~features ($\latentenergy \geq 100\,\mathrm{MeV}$) at low and medium masses (0.8--1.1 and 1.1--1.6$\,\Msolar$) approximately three times as strongly as at high masses (1.6--2.3$\,\Msolar$).}

As shown in~\citet{Legred:2021hdx}, all NS observations are consistent with a single radius near $\sim 12.5\,\mathrm{km}$.
We therefore expect the data to disfavor the existence of strong phase transitions and place an upper limit on $\latentenergy$.
Fig.~\ref{fig:current constraints corner} bears this out.
It shows posterior distributions on the properties of the \moifeature~feature with the largest \latentenergy~that overlaps with the specified mass range (i.e., features with larger \latentenergy~may exist in the EoS, but they do not overlap with the mass range).
Astrophysical data place an upper limit on the largest phase transition within an EoS, but are less informative about weaker phase transitions.

Figure~\ref{fig:current constraints corner} shows the onset energy density and pressure as well as the energy density at the end of the phase transition.
Beyond limiting the possible size of \moifeature~features, astrophysical data also disfavor phase transitions with large onset densities and pressures.
This likely corresponds to the observation that the sound-speed must increase rapidly around $3\rhonuc$ in order to support $\sim 2\,\Msolar$ pulsars against gravitational collapse while remaining compatible with observations at lower densities, primarily from GW170817~\cite{Legred:2021hdx}.
The peak in the posteriors for the onset parameters is likely due to a combination of the (peaked) prior and these upper limits. This trend is also encountered in the behavior of the $p$--$\ep$ bounds for EoSs with multiple stable branches.
That is, Fig. 8 in~\citet{Legred:2021hdx} suggests it is more likely for phase transitions to begin below $\rhonuc$ than above it when the EoS supports multiple stable branches.

Figure~\ref{fig:current constraints M-R} provides an additional perspective on current constraints by showing one-dimensional symmetric credible regions for the radius as a function of the gravitational mass.
While current astrophysical data generally disfavor EoSs with large \latentenergy, Fig.~\ref{fig:current constraints M-R} nevertheless shows that there are EoSs with large \latentenergy~that are consistent with observations.
In particular, the maximum-likelihood draw from the full PGX posterior conditioned on $\latentenergy \geq 100\,\mathrm{MeV}$ places a sharp feature in the $M$--$R$ curve at high masses, just above J0740+6620's observed mass.
Such behavior maximizes the likelihood from the PSR masses due to the assumption that the EoS itself is what limits the largest observed NS mass.
See discussions in~\cite{Landry:2020vaw,Miller:2019nzo}.
Furthermore, the maximum-likelihood EoS favors smaller radii at low masses (in line with GW170817) and larger radii at high masses (in line with J0740+6620).
Notably, the model-agnostic nonparametric prior was not designed to favor this specific behavior, which instead emerges from the data without direct supervision or fine-tuning.

We quantify the degree to which data prefer EoSs with different numbers and types of features in Table~\ref{tab:current constraints} and Fig.~\ref{fig:current constraints counts}.
Table~\ref{tab:current constraints} shows the ratio of maximized likelihoods as well as the ratio of marginal likelihoods for EoSs with different numbers of features.
We compare EoSs with a single stable branch against EoSs with multiple stable branches, as well as EoSs with and without at least one \moifeature~feature above a certain \latentenergy.
Generally, these statistics are consistent with Fig.~\ref{fig:current constraints corner}: the astrophysical data disfavor large phase transitions (multiple stable branches or large \latentenergy) more strongly than weaker ones.
However, the statistical evidence is still weak, and further observations are required to definitively rule out even the presence of multiple stable branches. 

Figure~\ref{fig:current constraints counts} expands on Table~\ref{tab:current constraints} by examining the preference for different numbers of features, rather than just their absence or presence.
That is, Table~\ref{tab:current constraints} in effect provides a summary of Fig.~\ref{fig:current constraints counts} by marginalizing over all EoS with more than one stable branch or at least one \moifeature~feature.
Overall, although current astrophysical observations cannot rule out the presence of a phase transition, they more strongly disfavor the presence of multiple features.
The astrophysical posterior strongly disfavors EoSs with more than two stable branches and less strongly disfavor EoSs with more than one large \moifeature~feature.
This suggests that one may not need to consider arbitrarily complicated EoS in order to model the observed population of NSs, or at least that there  is a limit to how exotic astrophysical NSs are.

Finally, current astrophysical data carries little information about the multiplicity of any phase transitions, should they exist.
Conditioning on the presence of a phase transition, we find Bayes factors between \result{$\sim$0.8--1.5 in favor of multiplicity $>1$ compared to multiplicity 1} for the feature with the largest \latentenergy~within each EoS, even for the strongest phase transitions.
This should be expected.
We cannot yet confidently determine whether a phase transition exists, and it would therefore be surprising if we could already identify even basic features of the phase transition.


\section{Future Prospects with Gravitational Wave Observations}
\label{sec:prospects}

Building upon current data, we now consider future prospects from GW observations of inspiraling compact binaries.
Section~\ref{sec:prospects for detection} explores the prospects for detecting the presence of phase transitions, and Sec.~\ref{sec:prospects for characterization} considers our ability to characterize them.
In brief, \result{we find that we will not be able to confidently detect the presence of even relatively extreme phase transitions with catalogs of $100$ events.
Rather, we will need at least 200 events or more.}
\result{However, we will be able to rule out the presence of multiple stable branches at low mass scales with 100 GW events.}
\result{Nevertheless, we will be able to infer the correct $\Lambda(M)$ for all $M$ simultaneously regardless of what the true EoS is, and obtain $\sim 6\%\ (50\%)$ relative uncertainty in $\Lambda_{1.2}\ (\Lambda_{2.0})$ after 100 GW detections.}

To explore a range of potential behavior, we simulate catalogs of GW events assuming a few representative CSS EoSs based on DBHF~\cite{Gross-Boelting:1998xsk}.
We consider
\begin{itemize}
    \item DBHF~\cite{Gross-Boelting:1998xsk}: a hadronic EoS without phase transitions.
    \item DBHF\_3504: a modification to DBHF with a weak phase transition at $\sim1.9\,\Msolar$ and a causal CSS extension at higher densities.
    \item DBHF\_2507: a modification to DBHF with a strong phase transition at $\sim1.5\,\Msolar$ and a causal CSS extension at higher densities. This is the Strong Maxwell CSS example in Fig.~\ref{fig:DBHF examples}.
\end{itemize}
These EoSs are not drawn from our nonparametric prior, and in fact their sharp features are relatively extreme examples of possible EoS behavior.
As such, we expect them to be rigorous tests of the inference framework.

The simulated catalogs assume a network signal-to-noise ratio ($\mathcal{S/N}$) detection threshold of \result{12}, and they approximate measurement uncertainty in the masses and tidal parameters according to the procedure described in~\citet{Landry:2020vaw}.
We inject a population of \result{non-spinning NSs uniform in component masses between 1.0 $\Msolar$ and $M_{\mathrm{TOV}}$}.
Injections are drawn assuming $p(\mathcal{S/N}) \sim (\mathcal{S/N})^{-4}$, consistent with a uniform rate per comoving volume at low redshift.
We assume the mass, spin, and redshift distributions are known exactly and therefore ignore selection effects.
For more details, see Refs.~\cite{Landry:2020vaw, Legred:2021hdx}.

For computational expediency, we consider the ability of GW observations alone to constrain phase transition phenomenology.
That is, we do not impose lower bounds on $M_\mathrm{TOV}$ from pulsar masses in order to retain a large effective sample size within the Monte Carlo integrals.
We do assume, however, that all objects below $M_\mathrm{TOV}$ are NSs, and, therefore, placing a lower limit on $\Lambda(M)$ from GW observations will \textit{de facto} place a lower limit on $M_\mathrm{TOV}$.
See Appendix~\ref{sec:sampling headaches} for more discussion.


\subsection{Prospects for Detecting Phase Transitions}
\label{sec:prospects for detection}

\begin{figure*}
    \begin{center}
        \includegraphics[width=1.0\textwidth]{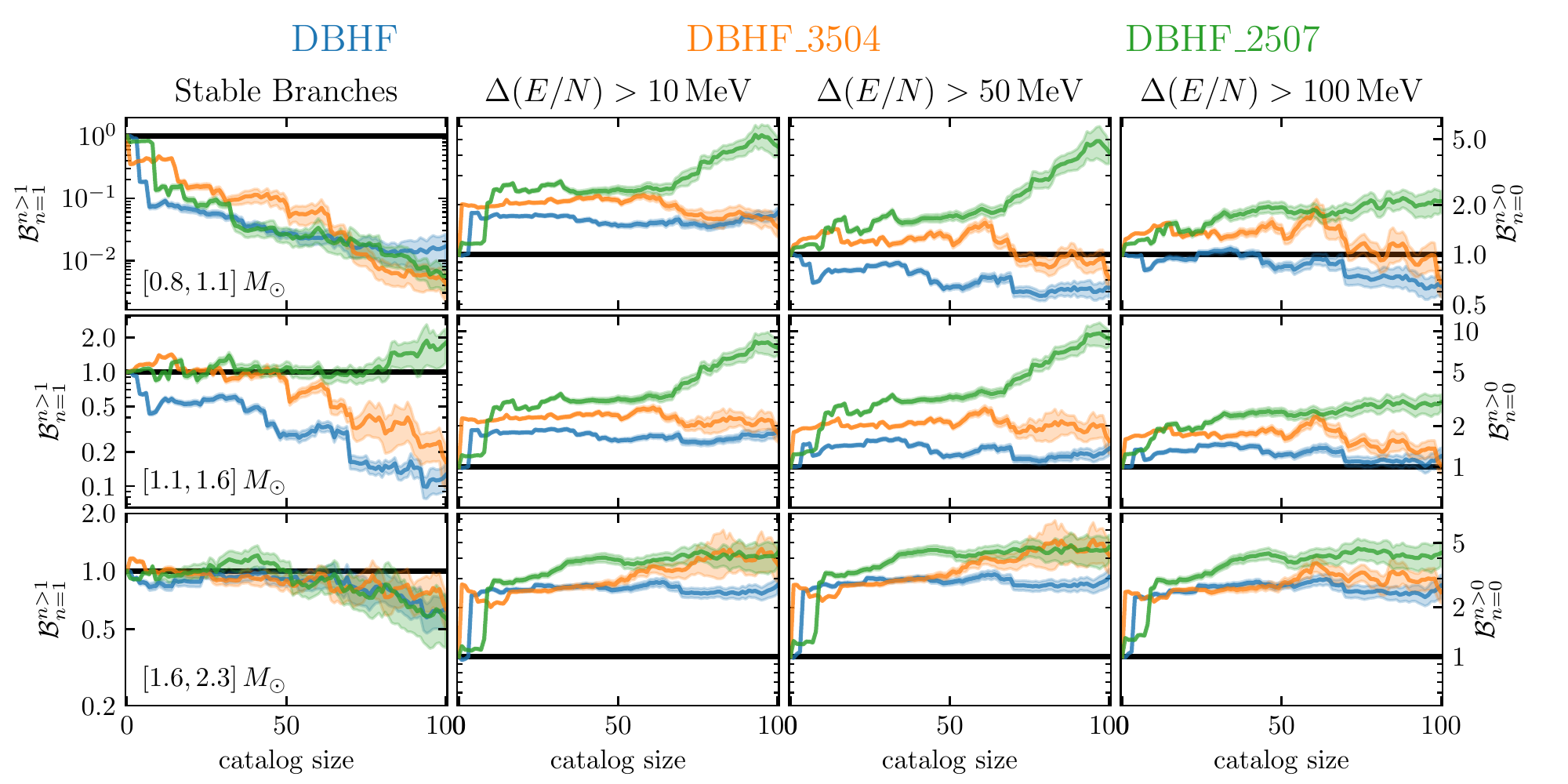}
    \end{center}
    \caption{
        Bayes factors vs. catalog size comparing (\textit{left-most column}) multiple stable branches vs. a single stable branch and (\textit{right three columns}) at least one \moifeature~feature vs. no \moifeature~features.
        We consider features that overlap with three mass ranges: (\textit{top row}) 0.8--1.1$\,\Msolar$, (\textit{middle row}) 1.1--1.6$\,\Msolar$, and (\textit{bottom row}) 1.6--2.3$\,\Msolar$.
        We also show three different injected EoSs: (\textit{blue}, no phase transition) DBHF, (\textit{orange}, weak phase transition at $\sim1.9\,\Msolar$) DBHF\_3504, and (\textit{green}, strong phase transition at $\sim1.5\,\Msolar$) DBHF\_2507.
        Shaded regions denote 1-$\sigma$ uncertainties from the finite size of our Monte Carlo sample sets.
        Different realizations of catalogs will also produce different trajectories; these should only be taken as representative.
    }
    \label{fig:prospects for detection}
\end{figure*}

We first consider detection of a phase transition with a catalog of GW events.
Fig.~\ref{fig:prospects for detection} shows the statistics from Table~\ref{tab:current constraints} for various simulated catalog sizes for injected EoSs both with and without a phase transition. Generally speaking, we recover the expected behavior: confidence in the presence (or absence) of a phase transition grows as the catalog increases.
Moreover, when a phase transition is present, evidence grows the most in the mass range where the phase transition occurs. 

\subsubsection{The Number of Stable Branches}
\label{sec:prospects for detection: branches}

We begin by considering the number of stable branches, with the left panels of Fig.~\ref{fig:prospects for detection} showing Bayes factors for multiple stable branches ($n>1$) vs. a single stable branch ($n=1$).
As none of the injected EoSs have a phase transition at low masses and GW observations should be able to confidently bound $\Lambda \gg 0$ at low masses, we quickly obtain relatively high confidence that there is only a single stable branch within 0.8--1.1$\,\Msolar$.
We find Bayes factors as large as \result{$\sim 100:1$ in favor of a single branch after 100 events}.

For moderate masses (1.1--1.6$\,\Msolar$), we again see the expected evidence in favor of a single stable branch for both DBHF (no phase transition) and DBHF\_3504 (phase transition at $\sim 1.9\,\Msolar$).
The Bayes factors are only \result{$\sim 10:1$ after 100 events}, but nonetheless the trend is clear.
In contrast, DBHF\_2507 (phase transition at $\sim 1.5\,\Msolar$ and multiple stable branches) exhibits a notably different pattern.
Although a strong preference is not developed either way, Bayes factors begin to (correctly) favor multiple stable branches after 100 events.

Finally,  we are not able to confidently distinguish between EoSs with a single stable branch or multiple stable branches in the mass range 1.6--2.3$\,\Msolar$.
This is because the individual events' uncertainties on $\Lambda$ are much larger than the true $\Lambda$ in this mass range.\footnote{$\Lambda$ typically scales as $\Lambda \propto M^{-5}$ and rapidly decreases at high masses.}
It will therefore take the combination of many GW events to be able to precisely resolve the true value of $\Lambda$ at high masses.


\subsubsection{The Number and Properties of \moifeature Features}
\label{sec:prospects for detection: features}

\begin{figure*}
    \includegraphics[width=1.0\textwidth]{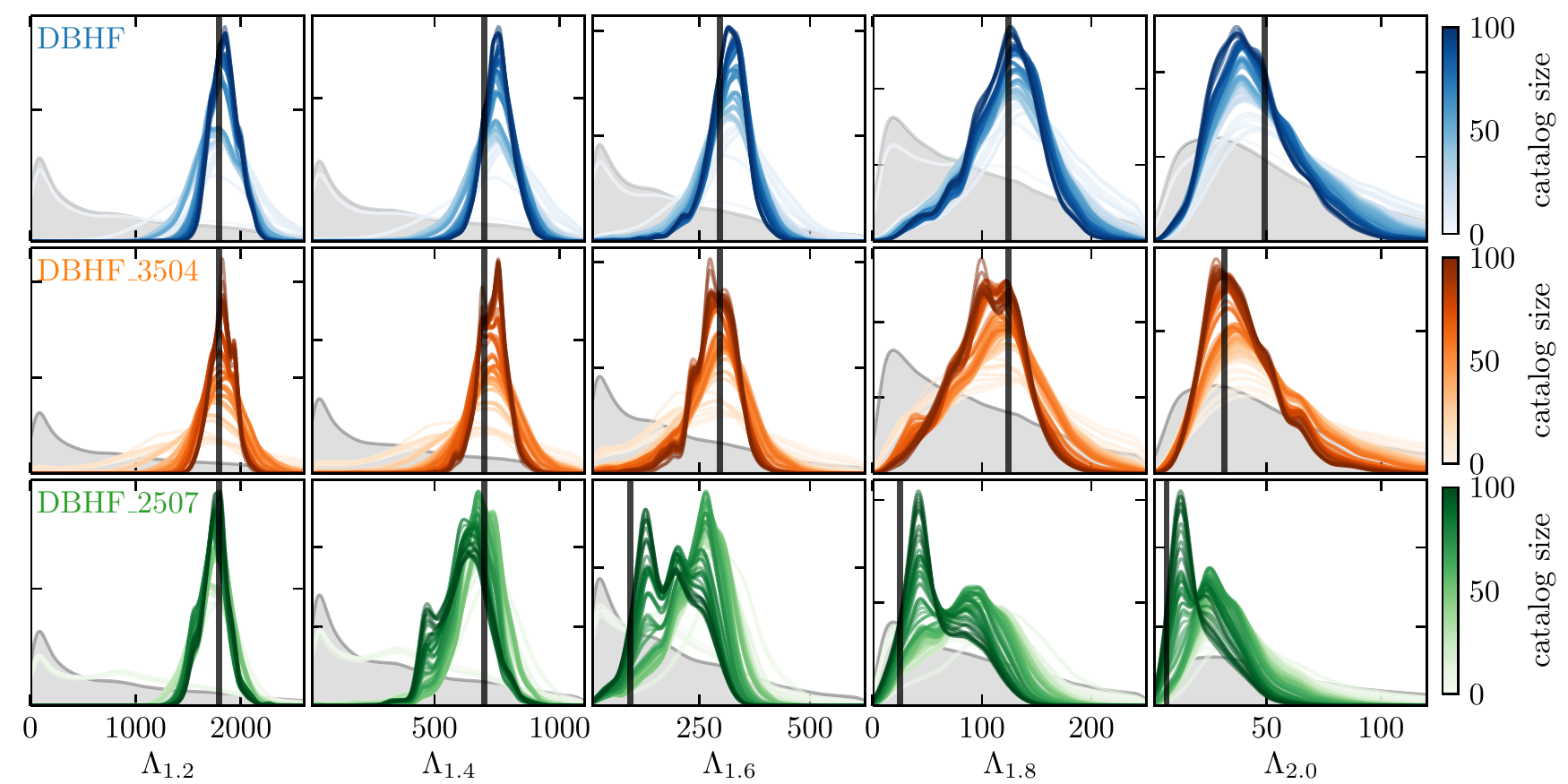}
    \caption{
        Sequences of one-dimensional marginal posteriors for $\Lambda(M)$ at (\textit{left to right}) $1.2$, $1.4$, $1.6$, $1.8$, and $2.0\,\Msolar$ for different simulated EoSs: (\textit{top, blue}) DBHF, (\textit{middle, orange}) DBHF\_3504 (phase transition at $\sim 1.9\,\Msolar$) and (\textit{bottom, green}) DBHF\_2507 (phase transition at $\sim1.5\,\Msolar$).
        These posteriors show the distributions of $\Lambda(M) > 0$ (i.e., they only consider EoSs with $M_\mathrm{TOV} \geq M$).
        These posteriors are conditioned only on simulated GW events (no real observations), and a line's color denotes the number of simulated GW events within the catalog (\textit{light to dark : fewer to more events}) along with the true injected values (\textit{vertical black lines}).
        The prior is shown for reference (\textit{grey shaded distributions}).
        For very small $\Lambda$, primarily associated with DBHF\_2507 at high masses, the true value falls near the lower bound in the prior.
        The primary effect of additional observations is to reduce support for larger values of $\Lambda$.
        While significant uncertainty in $\Lambda(M)$ remains after 100 events, the nonparametric prior is able to correctly infer $\Lambda(M)$ at all $M$ simultaneously, including sharp changes in $\Lambda(M)$ over relatively small mass ranges.
    }
    \label{fig:prospects for characterization envelops}
\end{figure*}

The remaining panels of Fig.~\ref{fig:prospects for detection} show similar trends for \moifeature~features.
We show Bayes factors for at least one \moifeature~feature ($n>0$) vs. no \moifeature~features ($n=0$).
In general, the strongest preference for a \moifeature~feature is for DBHF\_2507, which has the largest phase transition among the three EoSs we consider.
The evidence in favor of at least one \moifeature~feature is nevertheless smaller for the largest \latentenergy~($\geq 100\,\mathrm{MeV}$) compared to more moderate values ($\geq 50\,\mathrm{MeV}$).
This is true for all mass ranges, suggesting that we will be able to constrain a feature's \latentenergy~more easily than we may be able to constrain the mass range over which it occurs.
Additionally, we will need very large catalogs to confidently detect the presence of a \moifeature~feature.
At best, we find Bayes factors of \result{$\sim 10:1$ after 100 events}.
This matches previous estimates, which place the required number of events between \externalresult{200-400}~\cite{Chatziioannou:2019yko,Pang:2020ilf,LandryChakravarti2022}.
See Sec.~\ref{sec:discussion} for more discussion.
Furthermore, while there will not be unambiguous statistical evidence in favor of a \moifeature~feature at high masses (1.6--2.3$\,\Msolar$), we do see an upward trend for DBHF\_3504.
This suggests that, even though our individual-event uncertainties on tidal parameters are large at these masses, we will nevertheless eventually be able to detect small phase transitions at high masses given enough events.

Occam factors are readily apparent in these results, causing systematic shifts of comparable magnitude for all three injected EoSs.
These tend to favor the presence of \moifeature~features, as it is likely that very stiff EoSs at intermediate densities (unlikely to have \moifeature~features) are quickly ruled out by GW observations.
As such, some fraction of the prior is ruled out after only a few detections reducing the evidence even though there are still many EoSs without \moifeature~features that match the data well.
Furthermore, selecting EoSs with at least one feature at high masses requires $M_\mathrm{TOV}$ to be at least as high as the lower-edge of this mass range because of how our \moifeature~feature extraction algorithm works.
Such EoSs are better matches to the data for all the true EoSs considered.
Even a few detections can quickly rule out $M_\mathrm{TOV} \ll 1.6\,\Msolar$, which penalizes EoSs for which our algorithm did not detect a \moifeature~feature above $1.6\,\Msolar$ because the EoS's $M_\mathrm{TOV}$ was below $1.6\,\Msolar$.
Nevertheless, these Ocaam factors are typically \result{$\lesssim 2$}, implying that large Bayes factors can still be interpreted at face value.

Finally, it may be difficult to completely rule out the presence of \moifeature~features even if the true EoS does not have any phase transitions.
Fig.~\ref{fig:prospects for detection} shows a possible exception at the lowest masses considered, but even there the Bayes factors are only {$\sim 0.5$ after 100 events}.
This is yet another manifestation of the masquerade problem:
EoSs with and without \moifeature~features can produce similar $M$--$I$ relations, even for relatively large \latentenergy.


\subsection{Prospects for Characterizing Phase Transitions}
\label{sec:prospects for characterization}

\begin{figure*}
    \includegraphics[width=.49\textwidth]{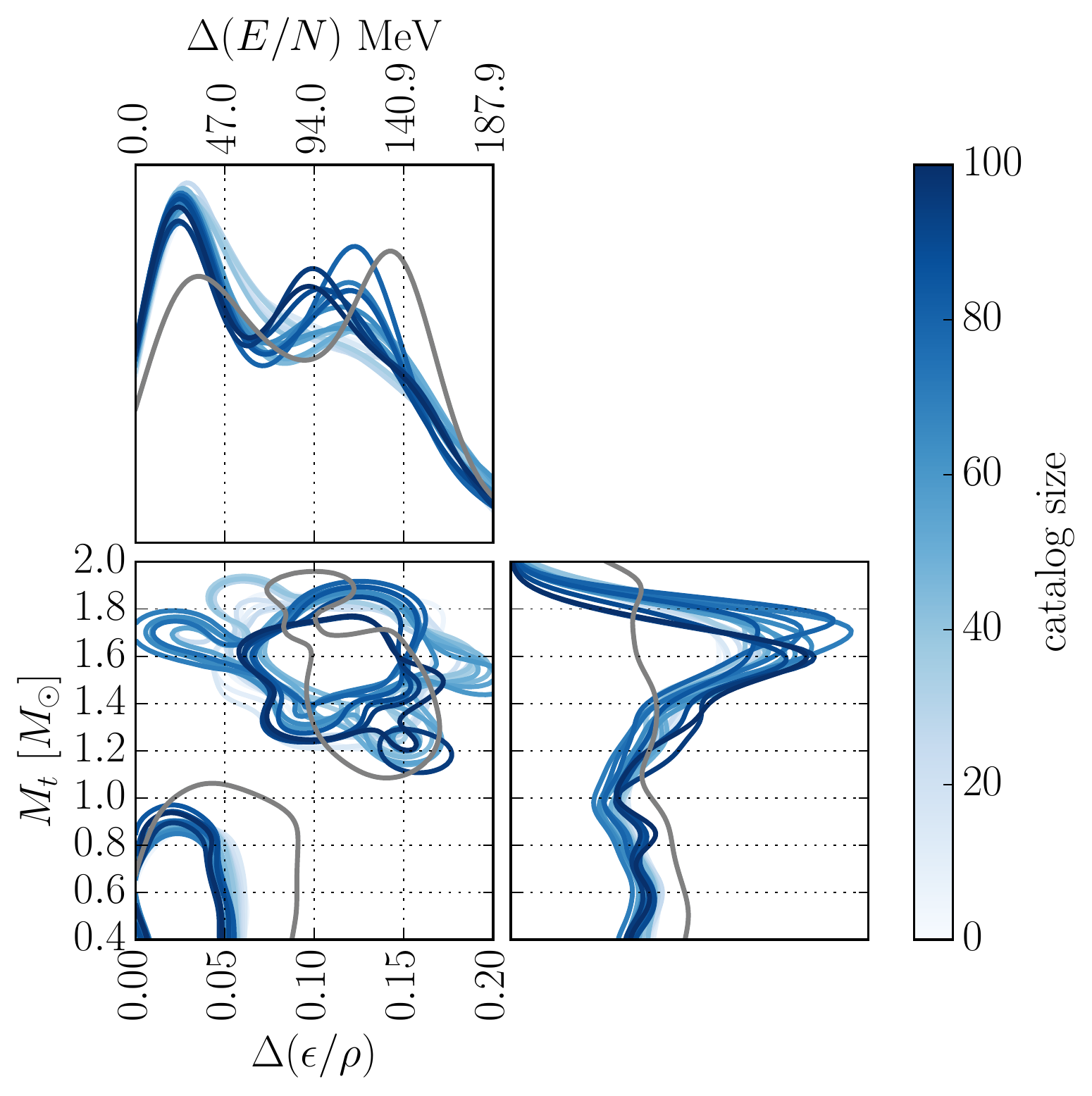}
    \includegraphics[width=.49\textwidth]{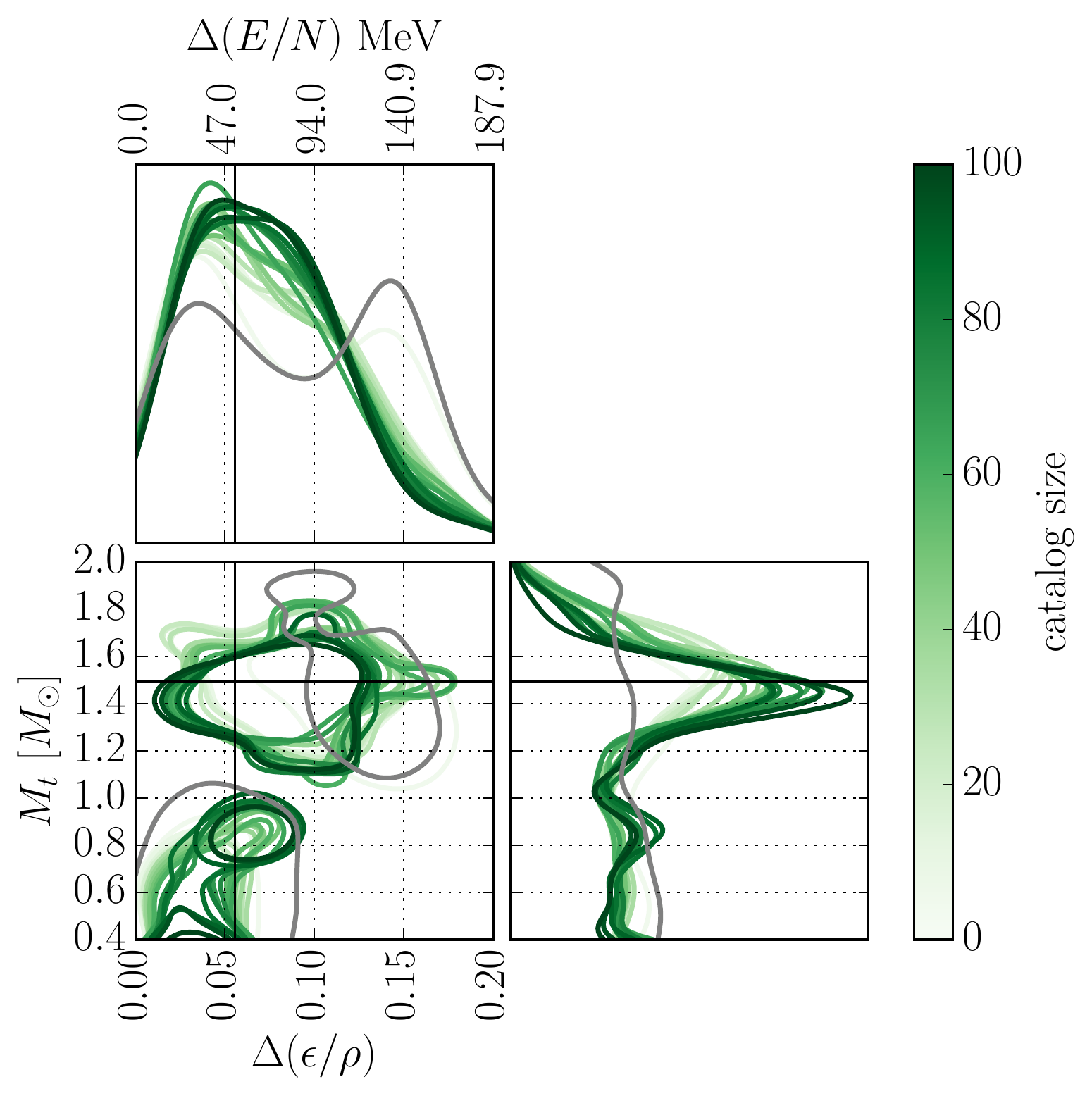}
    \caption{
        Joint posteriors for \latentenergy~and transition onset mass ($M_t$) inferred from simulated GW catalogs for (\textit{left, blue}) DBHF and (\textit{right, green}) DBHF\_2507.
        Grey curves denote the (reweighed) prior, color denotes the size of the catalog, and contours in the joint distribution are 50\% highest-probability-density credible regions.
        Solid lines denote the true parameters for DBHF\_2507; there are no such lines for DBHF because it does not contain a phase transition.
        As in Fig.~\ref{fig:current constraints corner}, extracted parameters correspond to the feature with the largest \latentenergy, but here we only require features to overlap the broad range 0.8--2.3$\,\Msolar$.
    }
    \label{fig:prospects for characterization}
\end{figure*}

In addition to detecting the presence of a phase transition, we wish to determine its properties should it exist.
Fundamental to this is the ability to infer the correct $M$--$\Lambda$ relation.
That is, to infer the correct $\Lambda(M)$ for all $M$ simultaneously.
Fig.~\ref{fig:prospects for characterization envelops} demonstrates that our nonparametric inference is capable of this, regardless of the true EoS used to generate injections.
This is often not the case for parametric models of the EoS (see~\cite{Pang:2020ilf,LandryChakravarti2022} and discussion in Sec.~\ref{sec:discussion}).
Fig.~\ref{fig:prospects for characterization envelops} shows one-dimensional marginal posteriors for $\Lambda(M)$ at $M=1.2$, 1.4, 1.6, 1.8, and 2.0$\,\Msolar$ for different catalog sizes and each of the three injected EoSs.
We find that the low-density (low-mass) EoS is relatively well measured.
\result{$\Lambda_{1.2}$ will have a relative uncertainty (standard deviation divided by the mean) between 6\% (DBHF\_3504) and 7\% (DBHF\_2507) at $M=1.2\,\Msolar$ after 100 detections}.
However, it will generally take more events before we can confidently resolve features at higher masses, even without the presence of a phase transition.
With catalogs of 100 events, we are only able to constrain \result{$\Lambda_{2.0}$ to between 40\% (DBHF\_3504) and 55\% (DBHF\_2507)}.
In agreement with Fig.~\ref{fig:prospects for detection}, it is likely to take \result{more than 100 events} to unambiguously distinguish between EoSs with and without phase transitions.
For example, the $\Lambda_{2.0}$ posterior for DBHF\_2507 still has nontrivial support at the location of the DBHF's $\Lambda_{2.0}$, and vice versa, even with the full catalog of 100 events.

Even though we identify phase transition features from macroscopic relations, we expect the inferred microscopic properties to be robust given the one-to-one mapping between $p$--$\ep$ and, e.g., $M$--$R$~\cite{Lindblom:1992}.
Fig.~\ref{fig:prospects for characterization} shows how constraints on the onset mass ($M_t$) and \latentenergy~evolve with the catalog size for DBHF (no phase transition) and DBHF\_2507 (strong phase transition).
In order to highlight constraints on the transition mass, Fig.~\ref{fig:prospects for characterization} additionally reweighs the posterior so that it corresponds to a (as much as possible) uniform prior in the transition mass.
It only shows EoSs that have at least one identified \moifeature~feature that overlaps with 0.8--2.3$\,\Msolar$.

Characterizing onset properties is challenging because of the wide variability in softening behavior during the course of the phase transition.
That is, the onset density as identified by a running local maximum in $c_s$ may not correspond to any immediately obvious features in macroscopic relations, as is the case in Fig.~\ref{fig:Gibbs examples}.
Therefore, we may expect a long tail towards low onset masses even if the end of the transition is well determined. 

Additionally, we sometimes observe unintuitive behavior when we condition on the presence of features that do not exist (left panel).
For example, the marginal posterior for $M_t$ (conditioned on the existence of at least one feature) peaks at $M_t \gtrsim 1.6\,\Msolar$ for DBHF.
Transitions that begin at these masses are difficult to detect with GW observations alone, see Figs.~\ref{fig:prospects for detection} and~\ref{fig:prospects for characterization envelops}.
Therefore, these EoSs are not strongly constrained by observations, particularly compared to EoSs that have transitions that begin at lower masses.
This explains why the posterior tends to disfavor low $M_t$, and the peak at higher masses should be interpreted primarily as a lower limit.

However, transitions that begin at very high masses ($M_t \gtrsim 1.8\,\Msolar$) are also disfavored by the data.
This is unintuitive, as we expect very weaker tidal constraints for high mass systems.
However, by conditioning on the presence of at least one identified \moifeature~feature, which in turn are only identified by our algorithm if the EoS does not collapse to a BH as part of the transition, we \textit{de facto} require EoSs with large onset masses to be rather stiff.
That is, only the stiffest EoS can have an \moifeature~feature begin at high mass and not collapse directly to a BH.
At the same time, these EoSs are ruled out by observations at smaller masses, which favor more compact stars and soft EoSs.
Therefore, a high $M_t$ is disfavored by low-mass observations and the correlation induced within the prior by requiring at least one identified \moifeature~feature at high mass.

We contrast this with DBHF\_2507, in which there is a phase transition near $1.5\,\Msolar$ (right panel).
Here, we find a similar peak in the one-dimensional marginal posterior for $M_t$, but there is additional information in the joint posterior for $M_t$ and \latentenergy.
The joint posterior for DBHF mostly follows the prior, particularly for $M_t\sim1.6\,\Msolar$, whereas for DBHF\_2507 it is shifted relative to the prior towards the injected values and disfavors large \latentenergy.
These considerations highlight the fact that low-dimensional marginal posteriors conditioned on specific, sometimes \textit{ad hoc}, features will require care to interpret correctly.
It may be better, then, to consider sets of marginal distributions for macroscopic observables, such as Fig.~\ref{fig:prospects for characterization envelops}, at the same time.
At the very least, the latter can provide context for inferred constraints on proxies for microphysical properties.


\section{Discussion}
\label{sec:discussion}

We summarize our main conclusions in Sec.~\ref{sec:discussion summary} before comparing them to existing work in the literature in Sec.~\ref{sec:discussion comparison to other work}.
We conclude by discussing possible extensions to our study in Sec.~\ref{sec:discussion future work}.


\subsection{Summary}
\label{sec:discussion summary}

We introduced a new algorithm to identify phase transitions within the EoS of dense matter based on NS properties and the underlying $c_s$ behavior.
This algorithm does not rely on a parametrization, and as such works for both parametric and nonparametric representation of the EoS.
Our approach improves upon previous studies by demonstrating that physically meaningful density scales can be extracted directly from NS observables. We further demonstrated that nonparametric EoS inference can recover the correct macroscopic properties, such as $\Lambda(M)$, at all masses simultaneously.
As such, we suggest that extracting physical quantities from nonparametric EoS draws is preferable to directly modeling of the $p$--$\ep$ relation with \textit{ad hoc} parametric functional forms, as different choices for the parametrization can introduce strong model-dependence on the conclusions~\cite{Legred:2022pyp}.

This approach is similar in spirit to efforts to constrain the nuclear symmetry energy and its derivatives (slope parameter: $L$) with nonparametric EoSs~\cite{Essick:2021kjb, Essick:2021ezp}.
Studies based on parametric EoS models described in terms of $L$ have suggested tension between terrestrial experiments and astrophysical observations~\cite{Reed:2021, Biswas:2021, Biswas:2021yge}. 
Refs.~\cite{Essick:2021kjb, Essick:2021ezp} instead extracted $L$ from nonparametric EoS realizations by imposing $\beta$-equilibrium at $\rhonuc$ without relying on an explicit parametrization far from $\rhonuc$.
They demonstrated that any apparent tension was due to model assumptions rather than the data, as nonparametric models were able to accommodate both terrestrial constraints on $L$ and astrophysical observations of NSs.

Returning to this work, we showed that current astrophysical data disfavor only the strongest phase transitions and the presence of multiple phase transitions.
However, the data are still consistent with two stable branches and/or one moderate phase transition.
We also showed that we will not be able to confidently detect the presence of a phase transition with catalogs of $\leq 100$ GW events.
Although we do not directly estimate how many events will be needed for computational reasons, extrapolating Fig.~\ref{fig:prospects for detection} suggests that we may need several hundred events to reach Bayes factors $\gtrsim 100$, often taken as a rule-of-thumb for confident detections~\cite{Kass:1995}.
We can, however, expect to confidently rule out the presence of multiple stable branches at low masses after 100 events.
While the exact rates of NS coalescences and future GW-detector sensitivities are still uncertain, it is unlikely that we will obtain a catalog of this size within the lifetime of the advanced LIGO and Virgo detectors~\cite{Observing-Scenarios}.


\subsection{Comparison to other work}
\label{sec:discussion comparison to other work}

As discussed briefly in Sec.~\ref{sec:introduction}, several authors have proposed tests based on features in the distribution of macroscopic observables.
\citet{Chen:2019rja} investigated a piecewise linear fit of the $M$--$R$ relation with two segments that captures phase transitions through a change in the slope.
However, beyond possible systematics associated with the simplicity of the piecewise linear model, quantitative conclusions hinge on the assumption that the measurement uncertainty on $R$ from GW events is roughly the same for all masses.
This is unrealistic for massive systems in which the relative uncertainty in the tidal deformability grows quickly.
\citet{Chatziioannou:2019yko} pursued a related method that models the population of detections hierarchically and searches for a second population with significantly different radii at high masses.\footnote{\citet{Chen:2019aiw} proposed a similar technique to distinguish between binary NS and NS-BH systems. In this case, a reduced inferred radius is attributed to the presence of a BH in the binary (which does not exhibit tidal effects) rather than a softening in the EoS.}
They found that phase transitions could be identified with ${\cal{O}}(100)$ events if hybrid stars emerge at $\sim1.4\,\Msolar$. \citet{LandryChakravarti2022} introduced a method for identifying the presence of twin stars, which can arise due to strong first-order phase transitions, in the population of merging binary NSs based on gaps in the joint distribution of masses and binary tidal deformabilities.
However, these and related approaches that directly model the $M$-$\Lambda$ relation~\cite{DelPozzo:2013ala, Agathos:2015uaa} offer no obvious pathway to microscopic EoS properties nor the ability to enforce physical precepts such as causality and thermodynamic stability.
What is more, not all microscopic models that contain phase transitions produce macroscopic observables with this phenomenology (the masquerade problem), and this phenomenology might be caused by other effects, such as a mix of binary NS and NS-BH binaries at the same masses~\cite{Chen:2019aiw} or even dark matter~\cite{Rutherford:2022xeb}.

Alternative approaches involve modeling the $p$--$\ep$ relation directly.
Several authors have attempted this with parametric models of varying complexity.
\citet{Pang:2020ilf} introduced a piecewise-polytropic model for first-order phase transitions and carried out model selection between models that do and do not support phase transitions, respectively.
They concluded that a strong phase transition could be identified with \externalresult{$12$} GW events, each with signal-to-noise ratio \externalresult{$\mathcal{S/N}>30$}.\footnote{Assuming merging binaries are uniformly distributed in volume within a Euclidean universe, the $\mathcal{S/N}$ is distributed as $p(\mathcal{S/N}) \propto (\mathcal{S/N})^{-4}$. This means that to observe $12$ events with $\mathcal{S/N}>30$ requires a total of \result{$> 187$} events above the detection threshold used in Sec.~\ref{sec:prospects} ($\mathcal{S/N} = 12$) and \result{324} events above the more realistic detection threshold $\mathcal{S/N} = 10$~\cite{GWTC-3-O3-Injections, GWTC-3-O1+O2+O3-Injections}.}
However, in addition to technical issues associated with their Bayes factor calculation, their results appear to be affected by model systematics within their EoS parametrization.
They arrive at counterintuitive conclusions: weaker phase transitions are detected more easily than stronger ones (their Fig. 5), and the inference precision is largely unaffected by the observation of more events (their Fig. 9).\footnote{For most parameters, statistical uncertainty roughly scales as $N^{-1/2}$, where $N$ is the number of detections. Systematic uncertainty is independent of $N$.}
We speculate that the cause is the fact that their parametric EoS model does not closely reproduce either of their injected EoSs, leading to model systematics~\cite{Legred:2022pyp}.
If systematic issues are less severe for the injected EoS with a weak phase transition than the one with a strong transition, the former could be more easily distinguished from EoSs without phase transitions.

Two other recent studies have looked at the astrophysical evidence for or against the presence of phase transitions.
Both~\citet{Tan:2021ahl} and~\citet{Mroczek:2023eff} constructed EoS models by adding features to the speed of sound such as spikes, dips, and plateaus.
As explained in~\citet{Tan:2021ahl}, these features are motivated by specific theoretical expectations of phase transition phenomenology.
\citet{Mroczek:2023eff} employs underlying EoS realizations drawn from a few simple GP priors, resulting in what they call a modified Gaussian Process.
In comparison, our nonparametric prior inherently generates broad ranges of phase transition morphology without the need to modify realizations \textit{post hoc}.
\citet{Mroczek:2023eff} must add features by hand because their original GP was constructed with long correlation lengths and small variances.
As such, it only produces smooth EoSs without phase-transition-like features by itself.
Additionally, \citet{Mroczek:2023eff} report a Bayes factor for models with or without such features, finding no strong evidence either way.
Though this generally agrees with our conclusions, the quantitative comparison might be affected by the fact that their prior is first ``pruned'' by rejecting EoSs that do not fall within broad boundaries that represent realistic EoS.
Inevitably, these boundaries carry information about current astrophysical observations.
Therefore, it may not be surprising that subsets of different priors (each chosen to resemble current astrophysical data) predict the current observed data with comparable frequency, which is what is implied by a Bayes factor $\sim 1$.

Several other authors have investigated models intended to test specifically for the presence of deconfined quarks in NS cores, e.g.~\cite{Takatsy:2023xzf, Annala:2019puf, Annala:2023cwx}.
Many of these studies base the evidence for the presence of quark matter on the behavior of the polytropic index ($\gamma = d\log p/d\log \ep$) in addition to using various parametric and nonparametric representations of the EoS and approximations to astrophysical likelihoods.
For example,~\citet{Annala:2023cwx} present approximate ranges for $\gamma$, $c_s$, and other statistics and propose that massive NS cores likely contain matter displaying approximate conformal symmetry, which may be indicative of a transition to deconfined quarks.
These studies typically focus on the composition of matter at the highest densities possible within NSs (near $M_\mathrm{TOV}$).
Some studies have even claimed evidence for the presence of deconfined quark matter based on $\gamma$ at high densities.
Our \moifeature~features are more agnostic about the composition of new matter and are sensitive over a broad range of masses.
They should therefore provide a complementary approach to direct modeling based on assumptions about NS composition and microphysical interactions.

Finally, several other authors have introduced EoS models with many parameters and increased model freedom, some of which are implemented as neural networks of varying complexity~\cite{Fujimoto:2018, Fujimoto:2020, Fujimoto:2021zas, Han:2021kjx, Han:2022rug}.
Our conclusions based on current observations are broadly consistent with these other approaches, and therefore we only remark that our \moifeature~feature could be extracted from any EoS, regardless of the underlying model (or lack thereof).
It should be straightforward to investigate phase transition phenomenology with realizations from any EoS prior in the literature, although this is beyond the scope of our current study.


\subsection{Future work}
\label{sec:discussion future work}

Finally, we discuss possible extensions and the impact that additional assumptions may have on our analysis.

As mentioned in Sec.~\ref{sec:constraints}, we intentionally condition our nonparametric prior on very little information from nuclear theory or experiment beyond causality and thermodynamic stability.
It would be of interest to better understand how terrestrial experiments or \textit{ab initio} theoretical calculations such as chiral EFT at low densities may impact our conclusions.
For example, Fig. 3 from \citet{Essick:2020} shows that improved constraints at very low densities ($\lesssim \rhonuc/2$) can improve uncertainty in the pressure at higher densities ($\sim 3\rhonuc$) when combined with astrophysical data.
Furthermore, theoretical calculations suggest a moderate value of $L$, which would remove even the hint that a phase transition may occur at low densities found in~\citet{Essick:2021ezp} when they assumed $L$ was large.

At the other extreme, it is worth discussing the impact of pQCD calculations further.
Several conflicting reports exist in the literature, suggesting that the pressures at very high densities ($\sim 40\rhonuc$) limit the pressures achieved in the highest-mass NS~\cite{Gorda:2022jvk, Gorda:2023usm}, while other studies point out that these conclusions depend on the details of how the densities relevant for NSs are extrapolated to the pQCD regime~\cite{Somasundaram:2022ztm}.
Indeed, the current proposal for mapping pQCD calculations to lower densities~\cite{Komoltsev:2021jzg} implements something similar to a maximization over the extrapolation rather than marginalizing over the EoS within the extrapolation region (candidate EoS are given equal weight as long as they can be connected to high-density pQCD preditions regardless of how many such connections exist or the relatively (prior) probability of any of the possible connections), although \citet{Gorda:2023usm} marginalize over a nonparametric extrapolation based on GPs for at least part of the extrapolation region (up to $\sim 10\rhonuc$ but not all the way to $\sim 40\rhonuc$).
Alternatively, Refs.~\cite{Annala:2023cwx, Annala:2019puf, Annala:2021gom} implement parametric models spanning the entire extrapolated region, although their parametric models contain a relatively small number of segments (4 segments for a piecewise polytrope in~\citet{Annala:2019puf} and 3-5 stitching points for a piecewise linear $c_s$-construction in~\citet{Annala:2021gom}).
The precise impact of constraints from such parametric inferences with small numbers of parameters can depend strongly on the exact functional forms assumed~\cite{Legred:2022pyp}.
The fact that the impact of pQCD constraints depends on the choice of how the extrapolation is performed and/or where the extrapolation begins suggests that they depend on the prior assumptions for EoS behavior within the (unobserved and unobservable) extrapolation region between the central density of $M_\mathrm{TOV}$ stars and the pQCD regime. 
Further study is needed to disentangle the impact of such prior choices from the physical limits imposed by thermodynamic consistency between densities relevant for NSs and the pQCD calculations.

Additional information about the EoS will be imprinted in post-merger signals from coalescing NS systems.
An extensive literature exists (e.g., Refs.~\cite{Most:2018eaw,Bauswein:2018bma}) mostly focusing on the ability to resolve the dominant frequency of the post-merger emission thought to be associated with the fundamental 2-2 mode of the massive remnant.
Additional work will be needed to connect our nonparametric inference based on tides observed during the GW inspiral to the complicated physics at work during the post-merger.
See, e.g.,~\citet{Wijngaarden:2022sah} for a way to model the full GW signal.
This may include extending our nonparametric EoS representation to include finite-temperature effects~\cite{Blacker:2023onp}.

In addition to incorporating more information within the inference, we may be able to dig deeper into features of the current data.
As mentioned in Sec.~\ref{sec:new stats}, our procedure does not identify phase transitions that results in the direct collapse to a BH, although we do find that the sharpness of the final decrease in $\arctan(\moifeature)$ may correlate with whether the collapse was due to only self-gravity or assisted by a sudden decrease in $c_s$.
Future work may develop additional features targeting this phenomenology, as it could have implications for the behavior of merger remnants that may or may not power electromagnetic counterparts depending on how long the remnant survives~\cite{Margalit:2017dij, Shibata:2019ctb, Koppel:2019pys}.

Assuming a phase transition is identified, an open challenge is to extend the inference to determine the order of the phase transition (e.g., first- vs. second-order). A smooth crossover from hadronic to quark matter may, for example, be mimicked by either a weak first-order phase transition or a second-order one~\cite{FujimotoFukushima2023}. Condensation of pions or kaons may also give rise to a second-order phase transition~\cite{PereiraBejger2022}. 
Our feature is able to detect a variety of possible morphologies, but additional statistics will need to be developed to further categorize the $c_s$ behavior within the phase transition's extent.

Finally, we would also be remiss if we did not remind the reader that our feature specifically targets phenomenology associated with decreases in $c_s$ and associated increase of compactness.
If, instead, a 
smooth crossover as realized in, e.g., quarkyonic matter~\cite{Fukushima:2015bda,Baym:2017whm,McLerran:2018hbz} only manifests as a sudden increase in the speed of sound, the features introduced here will not detect it.
Additional features targeting such behavior would need to be developed.
To that end, it may be of general interest to more carefully study the types of correlations between $c_s$ at different densities that are preferred by astrophysical data.
In the future, we will interrogate our nonparametric posteriors to not only constrain $c_s$ but also how quickly $c_s$ can vary.
For example, we do not expect periodic, extremely rapid oscillations in $c_s$ to have a significant impact on NS properties, and therefore they may only be very weakly constrained by the data.
See, e.g.,~\citet{Tan:2021ahl} for more discussion.
However, this will likely require more advanced sampling techniques to efficiently draw representative sets from our nonparametric processes.
See Appendix~\ref{sec:sampling headaches}.


\acknowledgements

The authors thank Aditya Vijaykumar for reviewing this manuscript within the LIGO Scientific Collaboration.

R.E. and P.L. are supported by the Natural Sciences \& Engineering Research Council of Canada (NSERC).

Research at Perimeter Institute is supported in part by the Government of Canada through the Department of Innovation, Science and Economic Development Canada and by the Province of Ontario through the Ministry of Colleges and Universities.
R.E. also thanks the Canadian Institute for Advanced Research (CIFAR) for support. 

The work of S.H. was supported by Startup Funds from the T.D. Lee Institute and Shanghai Jiao Tong University. S.H. also acknowledges support from the Network for Neutrinos, Nuclear Astrophysics, and Symmetries (N3AS) during the early stages of this project, funded by the National Science Foundation under cooperative agreements 2020275 and 1630782 and by the Heising-Simons Foundation under award 00F1C7. 

IL and KC acknowledge support from the Department of Energy under award number DE-SC0023101 and the Sloan Foundation.

The authors gratefully acknowledge the program ``Neutron Rich Matter on Heaven and Earth'' (INT-22-2a) held at the Institute for Nuclear Theory, University of Washington for useful discussion.
They also thank the LIGO laboratory for providing computational resources supported by National Science Foundation Grants PHY-0757058 and PHY-0823459.
This material is based upon work supported by NSF's LIGO Laboratory which is a major facility fully funded by the National Science Foundation.


\bibliography{references}


\newpage

\appendix


\section{Incompressible Newtonian Stars with Two Phases}
\label{sec:newtonian stars}

We examine the feature extraction procedure laid out in Sec.~\ref{sec:new stats} within a simpler context: incompressible stars with two phases in Newtonian gravity.
Despite its simplicity, this demonstrates the main features of more realistic stars while greatly simplifying the mathematics.

We consider incompressible stars with a piecewise constant density $\rho$ as a function of the pressure $p$ separated by a transition pressure $p_T$
\begin{equation}
    \rho(p) = \left\{\begin{matrix} \rho_L & \text{if} & p \leq p_T \\ \rho_H & \text{if} & p > p_T\end{matrix} \right. \,.
\end{equation}
We combine this EoS with the Newtonian equations of stellar structure 
\begin{align}
    \frac{dm}{dr} = 4\pi r^2 \rho\,, \\
    \frac{dp}{dr} = - \frac{Gm\rho}{r^2}\,,
\end{align}
and a central pressure $p_c$, where $m$ is the enclosed mass up to radius $r$.

\begin{figure}
    \includegraphics[width=0.90\columnwidth, clip=True, trim=0.55cm 0.10cm 0.10cm 0.50cm]{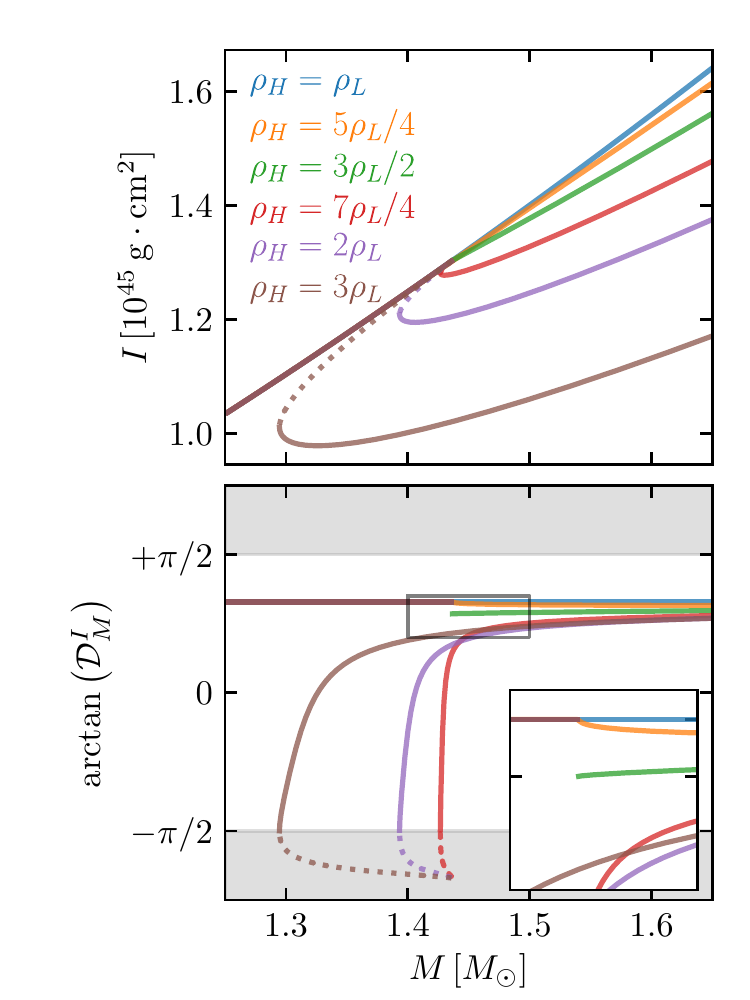}
    \caption{
        Stellar sequences for incompressible two-phase Newtonian stars with \result{$\rho_L=2\rhonuc=5.6\times10^{14}\mathrm{g}/\mathrm{cm}^3$}, \result{$p_T = 5\times10^{34} \mathrm{dyne}/\mathrm{cm}^2$}, and various values of $\rho_H$.
        We plot (\textit{top}) the $M$-$I$ relation and (\textit{bottom}) $\arctan(\moifeature)$~as a function of the stellar mass.
        Stable branches are shown with solid lines, and unstable branches are shown with dotted lines.
        The bottom panel inset focuses near the discontinuity for curves with; ticks on the y-axis correspond to the values in Eq.~\ref{eq:newtonian limits}.
    }
    \label{fig:newtonian stellar sequences}
\end{figure}

\begin{figure*}
    \hfill
    \begin{minipage}{0.20\textwidth}
        \begin{center}
            \vspace{1.0cm}
            $\Delta \arctan(\moifeature) = 0.0$ \\
            $\mathcal{R}_{c_s^2} = 1.0$ \\
            \vspace{2.75cm}
            $\Delta \arctan(\moifeature) = 0.15$ \\
            $\mathcal{R}_{c_s^2} = 1.0$ \\
            \vspace{2.75cm}    
            $\Delta \arctan(\moifeature) = 0.0$ \\
            $\mathcal{R}_{c_s^2} = 1.5$ \\
            \vspace{2.0cm}
        \end{center}
    \end{minipage}
    \hfill
    \begin{minipage}{0.64\textwidth}
        \begin{center}
            \includegraphics[width=1.0\textwidth, clip=True, trim=0.0cm 1.15cm 0.0cm 3.65cm]{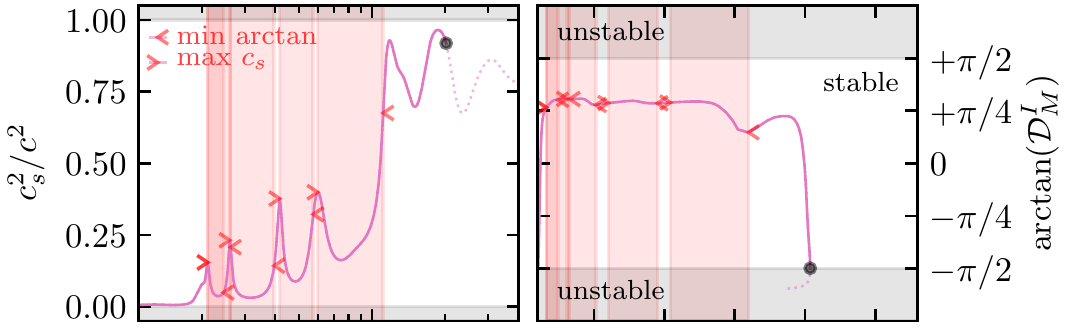} \\
            \vspace{0.05cm}
            \includegraphics[width=1.0\textwidth, clip=True, trim=0.0cm 1.15cm 0.0cm 3.65cm]{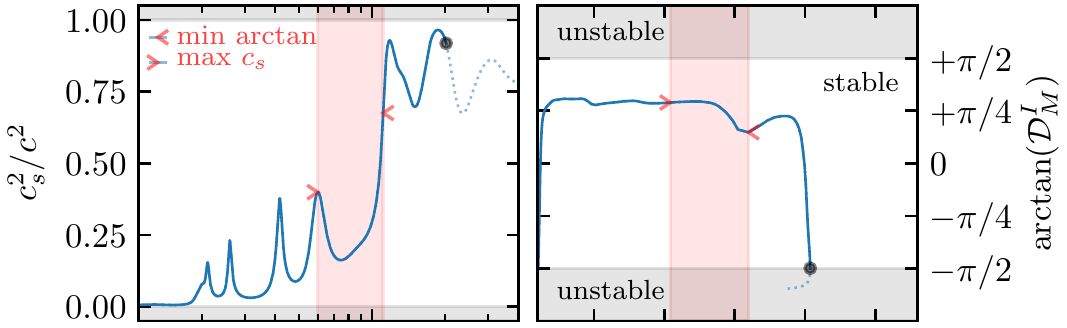} \\
            \vspace{0.05cm}
            \includegraphics[width=1.0\textwidth, clip=True, trim=0.0cm 0.00cm 0.0cm 3.65cm]{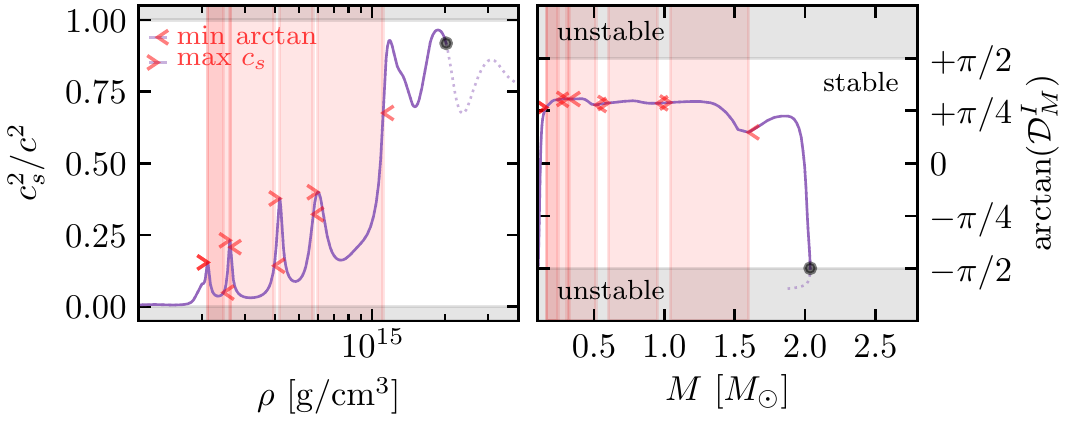} \\
        \end{center}
    \end{minipage}
    \hfill
    \caption{
        An additional example of the impact of thresholds within the feature extraction algorithm with an EoS realization with a relatively short correlation length.
        (\textit{top}) trivial thresholds; (\textit{middle}) threshold on the size of $\Delta\arctan(\moifeature)$; (\textit{bottom}) threshold on the amount $c_s^2$ must decrease (analogous to Fig.~\ref{fig:flowchart}).
        The rapid oscillations in $c_s^2$ are identified when selecting based on $\mathcal{R}_{c_s^2}$ but they are rejected when selecting based on $\Delta\arctan(\moifeature)$; their relatively small \latentenergy~do not produce significant changes in the $M$-$I$ relation.
    }
    \label{fig:GP examples - short correlation}
\end{figure*}

For $p_c \leq p_T$, the solution is trivial as the star is described by a single fluid:
\begin{align} 
        R & = \sqrt{\frac{3p_c}{2\pi G \rho_L^2}}\,, \\
        M & = \frac{4\pi}{3} \rho_L R^3\,, \\
        I & = \frac{2}{5}M R^2\,,
\end{align}
for the radius $R$, mass $M$ and moment of inertia $I$.
In this case, the star is always stable as $dM/dp_c > 0$ and $\moifeature=d\log I/d\log M = 5/3$ is constant.

For $p_c > p_T$, the star contains a core of high-density matter with radius
\begin{equation}
    R_c = \sqrt{\frac{3(p_c - p_T)}{2\pi G \rho_H^2}}\,.
\end{equation}
The entire star's macroscopic properties are then implicitly determined by
\begin{align}
    p_T & = \frac{4\pi G \rho_L (\rho_H - \rho_L)R_c^3}{3}\left(\frac{1}{R_c} - \frac{1}{R}\right) \nonumber \\
        & + \frac{2\pi G \rho_L^2}{3}\left(R^2 - R_c^2\right)\,, \\
    M & = \frac{4\pi}{3}\left[(\rho_H - \rho_L) R_c^3 + \rho_L R^3\right]\,, \\
    I & = \frac{8\pi}{15}\left[(\rho_H-\rho_L)R_c^5 + \rho_L R^5\right]\,,
\end{align}
In this case, the star can become unstable ($dM/dp_c < 0$) if $\rho_H$ is much larger than $\rho_L$.
Regardless of stability, \moifeature~is discontinuous whenever $\rho_H \geq \rho_\mathrm{thr}\equiv 3\rho_L/2$.
Fig.~\ref{fig:newtonian stellar sequences} shows that
\begin{equation}\label{eq:newtonian limits}
    \lim\limits_{p_c\rightarrow p_T^+} \frac{d\log I}{d\log M} = \left\{
    \begin{matrix}
        +5/3  & \text{if } \rho_H < \rho_\mathrm{thr} \\
        +5/4  & \text{if } \rho_H = \rho_\mathrm{thr} \\
        -5/3 & \text{if } \rho_H > \rho_\mathrm{thr}
    \end{matrix}
    \right. \,.
\end{equation}
Similar threshold behavior is encountered in other parameters combinations, for example the mass, radius or tidal deformability, as also shown for relativistic polytropic NSs with 1$^\mathrm{st}$-order phase transitions~\cite{Lindblom:1998dp}.


\section{The role of thresholds within feature extraction}
\label{sec:thresholds}

As part of the feature identification algorithm introduced in Sec.~\ref{sec:new stats}, we included a threshold on the amount the sound-speed must decrease within a candidate \moifeature~feature.
We now discuss the motivation for and impact of this and other thresholds in more detail.

We represent our uncertainty in the EoS as a random process for $c_s$ as a function of pressure with support for every possible causal and thermodynamically stable EoS.
We can therefore think of the behavior of our feature extraction algorithm in terms ``fluctuations'' in $c_s$ under different realizations of this random process.
Specifically, by selecting the running local maximum, we \textit{de facto} set a threshold on $c_s$ that subsequent local maxima must pass if they are to be associated with the start of a phase transition.
This means that small fluctuations in the height of subsequent local maxima, either above or below the previous running local maximum, can change the features extracted.
These changes can sometimes be dramatic, as the proxy for the onset density selected may jump to a much lower density.
By imposing a threshold on $R_{c_s^2}$, we make this type of selection explicit within the algorithm.
Although this does not remove the issue of small fluctuations qualitatively changing the estimated onset density, it at least provides a more concrete way to control the types of features selected.
Fig.~\ref{fig:flowchart} demonstrates the impact of a large threshold on $R_{c_s^2}$ for one EoS realization.

Although not used within our main analysis, we implement an additional threshold on the change in $\arctan(\moifeature)$ observed within the candidate phase transition.
That is, we define $\Delta \arctan(\moifeature)$ as the difference between the maximum $\arctan(\moifeature)$ for any density between the onset and end points and the local minimum in $\arctan(\moifeature)$ that defines the end point.
If this value is small, it will likely be difficult to detect such a feature from macroscopic properties of NSs.
One may wish to remove them at the time of extracting features.
In practice, though, we choose to record all features, regardless of how small $\Delta \arctan(\moifeature)$ is, and then filter them \textit{post hoc} by selecting subsets of features with different \latentenergy.

Fig.~\ref{fig:GP examples - short correlation} shows the impacts of threshold on both $R_{c_s^2}$ and $\Delta \arctan(\moifeature)$ for an EoS realization with rapid oscillations in $c_s$.
Our main results require $\Delta\arctan(\moifeature) \geq 0$ (satisfied axiomatically) and $R_{c_s^2} \geq 1.1$.


\section{Computational Challenges}
\label{sec:sampling headaches}

\begin{figure}
    \includegraphics[width=1.0\columnwidth]{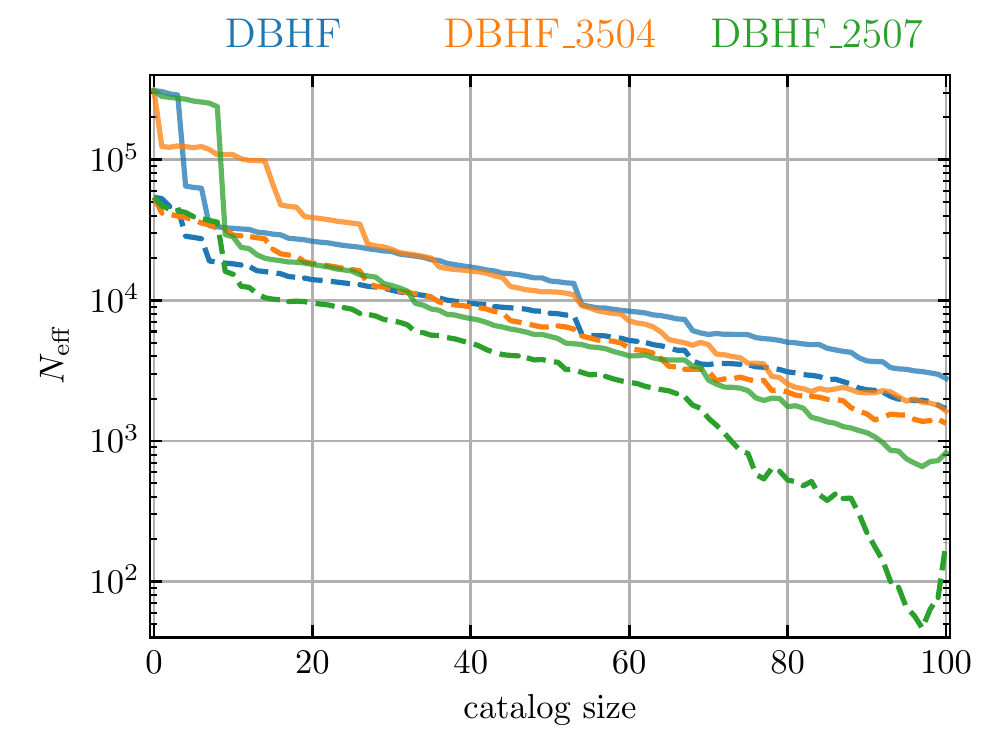}
    \caption{
        The effective number of EoS samples from the posterior process as a function of catalog size for (\textit{solid}) catalogs comprised of only mock GW observations and (\textit{dashed}) catalogs that include real pulsar mass measurements in addition to mock GW observations.
        For each of the three true EoS considered in Sec.~\ref{sec:prospects}, we find an approximately exponential decrease of the number of effective samples with the catalog size.
    }
    \label{fig:eos sample size}
\end{figure}

As discussed in Sec.~\ref{sec:prospects}, our current nonparametric sampling methods (i.e., direct Monte Carlo sampling) may not scale to catalogs of $\gtrsim 100$ detections.
This is perhaps not surprising.
That is, the total likelihood becomes increasingly peaked with more detections, and the majority of realizations from the nonparametric prior will have vanishingly small likelihoods.
As such, they do not contribute to the posterior.
With our current set of \result{$\sim \numpriorsamples$} prior samples, we retain \result{$\sim \astroneff$} effective samples in the posterior conditioned on real astrophysical data.
Heavy pulsar mass measurements alone rule out the largest portion of our prior, about $\psrneffpercent\%$.
See, e.g., Fig. 4 of~\citet{Essick:2020}.

The number of effective samples is substantially higher in our simulation campaigns if we do not include massive pulsars (Fig.~\ref{fig:eos sample size}).
Since our main goal is to explore how well GWs can constrain phase transitions, we only consider catalogs of simulated GW events in Sec.~\ref{sec:prospects} and do not include the heavy pulsars.

Although the existing set of EoS realizations from the nonparametric prior process will be sufficient for the catalog sizes expected over the next few years (current data and an additional $O(10)$ GW detections~\cite{Landry:2020vaw}), analyzing larger simulated catalogs might be challenging.
Fig.~\ref{fig:eos sample size} shows the number of effective EoS samples in the posterior as a function of the simulated GW catalog size and for different simulated EoS.
Solid lines only include simulated GW events; dashed lines include both heavy pulsars and simulated GW events.
Although there are differences between the injected EoS, we observe an approximately exponential decay in the number of effective posterior samples with the size of the catalog.
This implies we will need exponentially more draws from the current prior in order to analyze larger catalogs, which is computationally untenable in the long run.

However, given the expected rate of detections over the next few years, brute force may still be sufficient in the short run.
That is, given the low computational cost of producing additional EoS realizations, we may be able to draw more samples from the existing prior processes, solve the TOV equations, and compute the corresponding astrophysical weights fast enough to keep up.
With the current implementation, this takes $O(10)\,\mathrm{sec}/\mathrm{EoS}$, which is tractable compared to the expected rate of GW detections of $O(\mathrm{few})/\mathrm{year}$.

However, this approach will not work indefinitely.
We would be much better off spending (finite) computational resources in regions of the (infinite dimensional) vector-space of EoS with significant posterior support.
This is one motivation for sampling from the posterior using a Monte Carlo Markov Chain (MCMC) rather than direct Monte Carlo sampling.
Some authors in the broader GP literature have investigated implementations of GPs within MCMC schemes.
These typically involve evolving a handful of reference points used to model the GP's mean function along with the hyperparameters of the covariance kernel (see, for example,~\citet{Titsias:2011}).
This \textit{de facto} parametrizes the EoS prior with a handful of hyperparameters, at which point standard techniques for sampling from parametric distributions in hierarchical inference can be employed.
Other authors have suggested neural networks as a computationally efficient way to generate EoS proposal, but many (if not all) of these proposal are also \textit{de facto} parametric representations of the EoS itself or uncertainty in the EoS, which are then sampled with standard techniques~\cite{Fujimoto:2018, Fujimoto:2020, Fujimoto:2021zas, Han:2021kjx, Han:2022rug}.

An alternative method to focus computational efforts in high-likelihood region is to use the posterior from initial analyses with small catalogs to draw additional EoS proposals for future (larger) catalogs, similar to simulated annealing~\cite{Littenberg:2023xpl}.
The rate of detection is likely to be slow enough that new posteriors could be periodically developed (along with emulators to efficiently draw more samples) without the need for extensive automation.
As long as the noise at the time of each event is independent, this may be a computationally efficient path forward.
However, we leave exploration of such methods for future work.


\section{Additional Representations of Current Astrophysical constraints}
\label{sec:additional representations}

\begin{figure*}
    \includegraphics[width=1.0\textwidth, clip=False, trim=0.75cm 0.9cm 0.5cm 0.5cm]{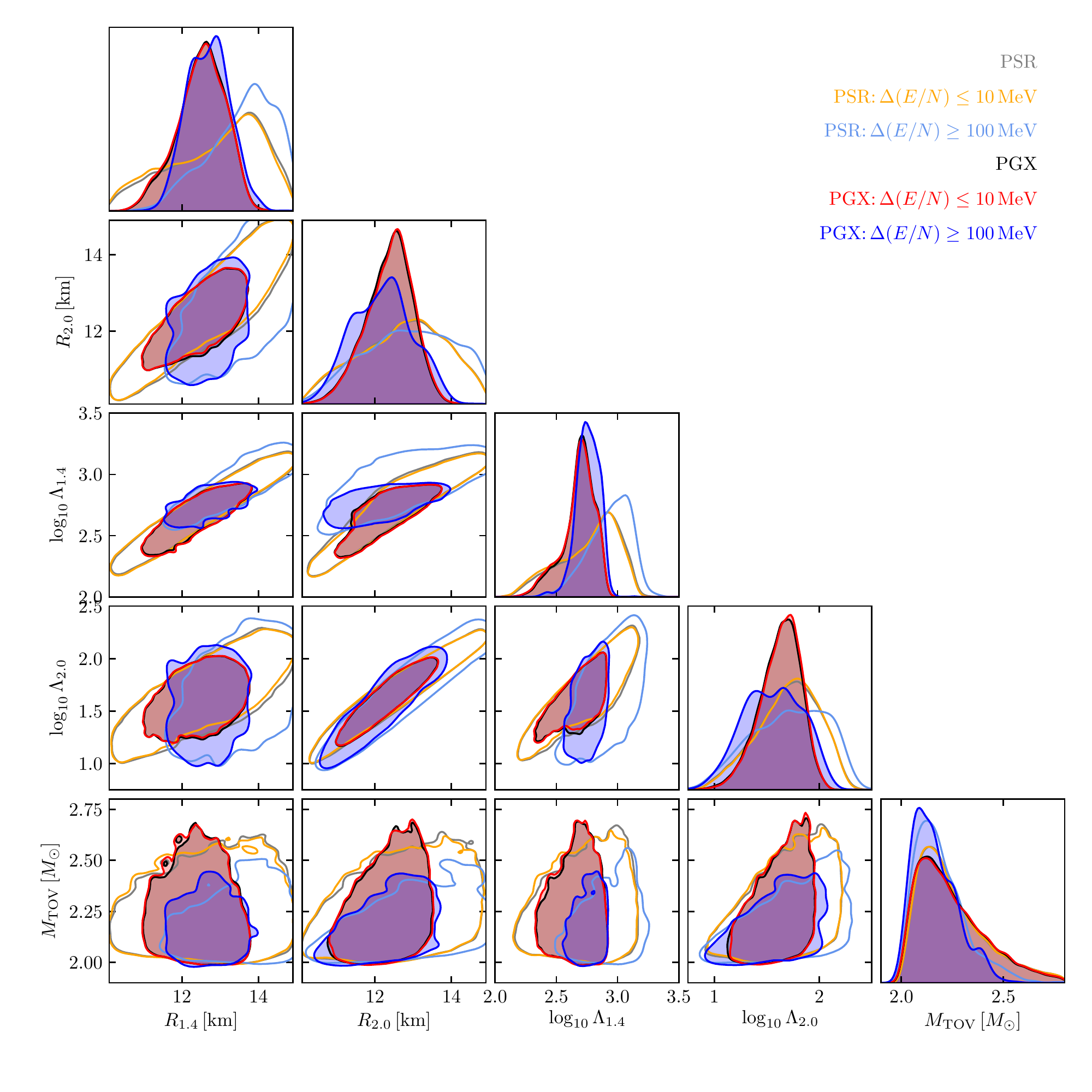}
    \caption{
        Distributions of radii and tidal deformabilities at reference masses as well as $M_{\mathrm{TOV}}$ conditioned on current data.
        These distributions \textit{de facto} exclude EoSs with $M_{\mathrm{TOV}} < 2\,\Msolar$ by requiring $\Lambda_{2.0} > 0$ (enforced through the logarithmic scale).
        As in Fig.~\ref{fig:current constraints M-R}, there are much weaker correlations between low-mass and high-mass observables.
    }
    \label{fig:current constraints M-Lambda}
\end{figure*}

Here we present additional representations of the constraints on phase transition phenomenology with current astrophysical data.
Similar to Fig.~\ref{fig:current constraints M-R}, Fig.~\ref{fig:current constraints M-Lambda} shows posteriors for macroscopic observables conditioned on EoSs with either small ($\latentenergy \leq 10\,\mathrm{MeV}$) or large ($\latentenergy \geq 100\,\mathrm{MeV}$) phase transitions for masses between 1.1--2.3$\,\Msolar$.
In general, we see that there are weaker correlations between macroscopic properties at low masses ($1.4\,\Msolar$) and high masses ($2.0\,\Msolar$) for EoSs with large phase transitions than for EoSs with small phase transitions, even though the marginal uncertainty for each is approximately the same.
Notable exceptions are that EoS with small \latentenergy~can support smaller $R_{1.4}$ and larger $M_\mathrm{TOV}$ than EoS with large \latentenergy.

Tables~\ref{tab:additional current maxL branch constraints}--\ref{tab:additional current Bayes features constraints} show additional detection statistics for different types of features conditioned on different subsets of the data, analogous to Table~\ref{tab:current constraints}.
We report different combinations of (P) pulsar mass measurements, (G) GW tidal measurements, and (X) X-ray pulse profiling with NICER.
Tables~\ref{tab:additional current maxL branch constraints} and~\ref{tab:additional current Bayes branch constraints} report the evidence for multiple stable branches.
Tables~\ref{tab:additional current maxL features constraints} and~\ref{tab:additional current Bayes features constraints} report the evidence for \moifeature~features.
Note that one can compute additional Bayes factors for different combinations of the data based on these numbers.
For example,
\begin{equation}
    \mathcal{B}(GX|P) = \frac{\mathcal{B}(GXP)}{\mathcal{B}(P)}
\end{equation}

\begin{table*}
    \centering
    \caption{
        Additional ratios of maximized likelihoods for the number of stable branches based on current astrophysical observations: (P) pulsar masses, (G) GW observations from LIGO/Virgo, and (X) X-ray timing from NICER.
    }
    \label{tab:additional current maxL branch constraints}
    {\renewcommand{\arraystretch}{1.5}
    \begin{tabular}{@{\extracolsep{0.2cm}} c ccccc}
        \hline\hline
        \multirow{2}{*}{$M\,[\MassUnit]$} & \multicolumn{5}{c}{Stable Branches} \\
        & $\max\mathcal{L}^{n\geq2}_{n=1}(\mathrm{P})$ & $\max\mathcal{L}^{n\geq2}_{n=1}(\mathrm{G})$ & $\max\mathcal{L}^{n\geq2}_{n=1}(\mathrm{X})$ & $\max\mathcal{L}^{n\geq2}_{n=1}(\mathrm{PG})$ & $\max\mathcal{L}^{n\geq2}_{n=1}(\mathrm{PGX})$ \\
        \hline\hline
        \LowMassBinMin-\LowMassBinMax
          & \MaxLikeRatioBranchLowMassPSR
          & \MaxLikeRatioBranchLowMassGW
          & \MaxLikeRatioBranchLowMassXray
          & \MaxLikeRatioBranchLowMassPSRGW
          & \MaxLikeRatioBranchLowMassPSRGWXray \\
        \MidMassBinMin-\MidMassBinMax
          & \MaxLikeRatioBranchMidMassPSR
          & \MaxLikeRatioBranchMidMassGW
          & \MaxLikeRatioBranchMidMassXray
          & \MaxLikeRatioBranchMidMassPSRGW
          & \MaxLikeRatioBranchMidMassPSRGWXray \\
        \HighMassBinMin-\HighMassBinMax
          & \MaxLikeRatioBranchHighMassPSR
          & \MaxLikeRatioBranchHighMassGW
          & \MaxLikeRatioBranchHighMassXray
          & \MaxLikeRatioBranchHighMassPSRGW
          & \MaxLikeRatioBranchHighMassPSRGWXray \\
        \hline
    \end{tabular}
    }
\end{table*}

\begin{table*}
    \centering
    \caption{
        Additional ratios of marginal likelihoods for the number of stable branches based on current observations.
    }
    \label{tab:additional current Bayes branch constraints}
    {\renewcommand{\arraystretch}{1.5}
    \begin{tabular}{@{\extracolsep{0.2cm}} c ccccccc}
        \hline\hline
        \multirow{2}{*}{\thead{$M$\\$[\MassUnit]$}} & \multicolumn{7}{c}{Stable Branches} \\
        & $\mathcal{B}^{n\geq2}_{n=1}(\mathrm{P})$ & $\mathcal{B}^{n\geq2}_{n=1}(\mathrm{G})$ & $\mathcal{B}^{n\geq2}_{n=1}(\mathrm{X})$ & $\mathcal{B}^{n\geq2}_{n=1}(\mathrm{PG})$ & $\mathcal{B}^{n\geq2}_{n=1}(\mathrm{PGX})$ & $\mathcal{B}^{n\geq2}_{n=1}(\mathrm{G|P})$ & $\mathcal{B}^{n\geq2}_{n=1}(\mathrm{GX|P})$\\
        \hline\hline
        \LowMassBinMin-\LowMassBinMax
         & \BayesBranchLowMassPSR
         & \BayesBranchLowMassGW
         & \BayesBranchLowMassXray
         & \BayesBranchLowMassPSRGW
         & \BayesBranchLowMassPSRGWXray
         & \BayesBranchLowMassGWGivenPSR
         & \BayesBranchLowMassGWXrayGivenPSR \\
        \MidMassBinMin-\MidMassBinMax
         & \BayesBranchMidMassPSR
         & \BayesBranchMidMassGW
         & \BayesBranchMidMassXray
         & \BayesBranchMidMassPSRGW
         & \BayesBranchMidMassPSRGWXray
         & \BayesBranchMidMassGWGivenPSR
         & \BayesBranchMidMassGWXrayGivenPSR \\  
        \HighMassBinMin-\HighMassBinMax
         & \BayesBranchHighMassPSR
         & \BayesBranchHighMassGW
         & \BayesBranchHighMassXray
         & \BayesBranchHighMassPSRGW
         & \BayesBranchHighMassPSRGWXray
         & \BayesBranchHighMassGWGivenPSR
         & \BayesBranchHighMassGWXrayGivenPSR \\  
        \hline
    \end{tabular}
    }
\end{table*}

\begin{table*}
    \centering
    \caption{
        Additional ratios of maximized likelihoods for the number of \moifeature~features based on current observations.
    }
    \label{tab:additional current maxL features constraints}
    {\renewcommand{\arraystretch}{1.5}
    \begin{tabular}{@{\extracolsep{0.2cm}} c c ccccc}
        \hline\hline
        \multirow{2}{*}{\thead{$M$\\$[\MassUnit]$}} & \multirow{2}{*}{\thead{$\min\latentenergy$\\$[\LatentEnergyUnit]$}} & \multicolumn{5}{c}{\moifeature~Features} \\
        & & $\max\mathcal{L}^{n\geq1}_{n=0}(\mathrm{P})$ & $\max\mathcal{L}^{n\geq1}_{n=0}(\mathrm{G})$ & $\max\mathcal{L}^{n\geq1}_{n=0}(\mathrm{X})$ & $\max\mathcal{L}^{n\geq1}_{n=0}(\mathrm{PG})$ & $\max\mathcal{L}^{n\geq1}_{n=0}(\mathrm{PGX})$ \\
        \hline\hline
        \multirow{3}{*}{\LowMassBinMin-\LowMassBinMax}
          & \MidLatentEnergy
            & \MaxLikeRatioMoILowMassMidLatentEnergyPSR
            & \MaxLikeRatioMoILowMassMidLatentEnergyGW
            & \MaxLikeRatioMoILowMassMidLatentEnergyXray
            & \MaxLikeRatioMoILowMassMidLatentEnergyPSRGW
            & \MaxLikeRatioMoILowMassMidLatentEnergyPSRGWXray \\
          & \HighLatentEnergy
            & \MaxLikeRatioMoILowMassHighLatentEnergyPSR
            & \MaxLikeRatioMoILowMassHighLatentEnergyGW
            & \MaxLikeRatioMoILowMassHighLatentEnergyXray
            & \MaxLikeRatioMoILowMassHighLatentEnergyPSRGW
            & \MaxLikeRatioMoILowMassHighLatentEnergyPSRGWXray \\
          & \HugeLatentEnergy
            & \MaxLikeRatioMoILowMassHugeLatentEnergyPSR
            & \MaxLikeRatioMoILowMassHugeLatentEnergyGW
            & \MaxLikeRatioMoILowMassHugeLatentEnergyXray
            & \MaxLikeRatioMoILowMassHugeLatentEnergyPSRGW
            & \MaxLikeRatioMoILowMassHugeLatentEnergyPSRGWXray \\
        \cline{2-7}
        \multirow{3}{*}{\MidMassBinMin-\MidMassBinMax}
          & \MidLatentEnergy
            & \MaxLikeRatioMoIMidMassMidLatentEnergyPSR
            & \MaxLikeRatioMoIMidMassMidLatentEnergyGW
            & \MaxLikeRatioMoIMidMassMidLatentEnergyXray
            & \MaxLikeRatioMoIMidMassMidLatentEnergyPSRGW
            & \MaxLikeRatioMoIMidMassMidLatentEnergyPSRGWXray \\
          & \HighLatentEnergy
            & \MaxLikeRatioMoIMidMassHighLatentEnergyPSR
            & \MaxLikeRatioMoIMidMassHighLatentEnergyGW
            & \MaxLikeRatioMoIMidMassHighLatentEnergyXray
            & \MaxLikeRatioMoIMidMassHighLatentEnergyPSRGW
            & \MaxLikeRatioMoIMidMassHighLatentEnergyPSRGWXray \\
          & \HugeLatentEnergy
            & \MaxLikeRatioMoIMidMassHugeLatentEnergyPSR
            & \MaxLikeRatioMoIMidMassHugeLatentEnergyGW
            & \MaxLikeRatioMoIMidMassHugeLatentEnergyXray
            & \MaxLikeRatioMoIMidMassHugeLatentEnergyPSRGW
            & \MaxLikeRatioMoIMidMassHugeLatentEnergyPSRGWXray \\
        \cline{2-7}
        \multirow{3}{*}{\HighMassBinMin-\HighMassBinMax}
          & \MidLatentEnergy
            & \MaxLikeRatioMoIHighMassMidLatentEnergyPSR
            & \MaxLikeRatioMoIHighMassMidLatentEnergyGW
            & \MaxLikeRatioMoIHighMassMidLatentEnergyXray
            & \MaxLikeRatioMoIHighMassMidLatentEnergyPSRGW
            & \MaxLikeRatioMoIHighMassMidLatentEnergyPSRGWXray \\
          & \HighLatentEnergy
            & \MaxLikeRatioMoIHighMassHighLatentEnergyPSR
            & \MaxLikeRatioMoIHighMassHighLatentEnergyGW
            & \MaxLikeRatioMoIHighMassHighLatentEnergyXray
            & \MaxLikeRatioMoIHighMassHighLatentEnergyPSRGW
            & \MaxLikeRatioMoIHighMassHighLatentEnergyPSRGWXray \\
          & \HugeLatentEnergy
            & \MaxLikeRatioMoIHighMassHugeLatentEnergyPSR
            & \MaxLikeRatioMoIHighMassHugeLatentEnergyGW
            & \MaxLikeRatioMoIHighMassHugeLatentEnergyXray
            & \MaxLikeRatioMoIHighMassHugeLatentEnergyPSRGW
            & \MaxLikeRatioMoIHighMassHugeLatentEnergyPSRGWXray \\
        \hline
    \end{tabular}
    }
\end{table*}

\begin{table*}
    \centering
    \caption{
        Additional ratios of marginal likelihoods for the number of \moifeature~features based on current astrophysical observations.
    }
    \label{tab:additional current Bayes features constraints}
    {\renewcommand{\arraystretch}{1.5}
    \begin{tabular}{@{\extracolsep{0.1cm}} c c ccccccc}
        \hline\hline
        \multirow{2}{*}{\thead{$M$\\$[\MassUnit]$}} & \multirow{2}{*}{\thead{$\min\latentenergy$\\$[\LatentEnergyUnit]$}} & \multicolumn{7}{c}{\moifeature~Features} \\
        & & $\mathcal{B}^{n\geq1}_{n=0}(\mathrm{P})$ & $\mathcal{B}^{n\geq1}_{n=0}(\mathrm{G})$ & $\mathcal{B}^{n\geq1}_{n=0}(\mathrm{X})$ & $\mathcal{B}^{n\geq1}_{n=0}(\mathrm{PG})$ & $\mathcal{B}^{n\geq1}_{n=0}(\mathrm{PGX})$ & $\mathcal{B}^{n\geq1}_{n=0}(\mathrm{G|P})$ & $\mathcal{B}^{n\geq1}_{n=0}(\mathrm{GX|P})$\\
        \hline\hline
        \multirow{3}{*}{\LowMassBinMin-\LowMassBinMax}
         & \MidLatentEnergy
          & \BayesMoILowMassMidLatentEnergyPSR
          & \BayesMoILowMassMidLatentEnergyGW
          & \BayesMoILowMassMidLatentEnergyXray
          & \BayesMoILowMassMidLatentEnergyPSRGW
          & \BayesMoILowMassMidLatentEnergyPSRGWXray
          & \BayesMoILowMassMidLatentEnergyGWGivenPSR
          & \BayesMoILowMassMidLatentEnergyGWXrayGivenPSR \\
         & \HighLatentEnergy
          & \BayesMoILowMassHighLatentEnergyPSR
          & \BayesMoILowMassHighLatentEnergyGW
          & \BayesMoILowMassHighLatentEnergyXray
          & \BayesMoILowMassHighLatentEnergyPSRGW
          & \BayesMoILowMassHighLatentEnergyPSRGWXray
          & \BayesMoILowMassHighLatentEnergyGWGivenPSR
          & \BayesMoILowMassHighLatentEnergyGWXrayGivenPSR \\
         & \HugeLatentEnergy
          & \BayesMoILowMassHugeLatentEnergyPSR
          & \BayesMoILowMassHugeLatentEnergyGW
          & \BayesMoILowMassHugeLatentEnergyXray
          & \BayesMoILowMassHugeLatentEnergyPSRGW
          & \BayesMoILowMassHugeLatentEnergyPSRGWXray
          & \BayesMoILowMassHugeLatentEnergyGWGivenPSR
          & \BayesMoILowMassHugeLatentEnergyGWXrayGivenPSR \\
        \cline{2-9}
        \multirow{3}{*}{\MidMassBinMin-\MidMassBinMax}
          & \MidLatentEnergy
          & \BayesMoIMidMassMidLatentEnergyPSR
          & \BayesMoIMidMassMidLatentEnergyGW
          & \BayesMoIMidMassMidLatentEnergyXray
          & \BayesMoIMidMassMidLatentEnergyPSRGW
          & \BayesMoIMidMassMidLatentEnergyPSRGWXray
          & \BayesMoIMidMassMidLatentEnergyGWGivenPSR
          & \BayesMoIMidMassMidLatentEnergyGWXrayGivenPSR \\
         & \HighLatentEnergy
          & \BayesMoIMidMassHighLatentEnergyPSR
          & \BayesMoIMidMassHighLatentEnergyGW
          & \BayesMoIMidMassHighLatentEnergyXray
          & \BayesMoIMidMassHighLatentEnergyPSRGW
          & \BayesMoIMidMassHighLatentEnergyPSRGWXray
          & \BayesMoIMidMassHighLatentEnergyGWGivenPSR
          & \BayesMoIMidMassHighLatentEnergyGWXrayGivenPSR \\
         & \HugeLatentEnergy
          & \BayesMoIMidMassHugeLatentEnergyPSR
          & \BayesMoIMidMassHugeLatentEnergyGW
          & \BayesMoIMidMassHugeLatentEnergyXray
          & \BayesMoIMidMassHugeLatentEnergyPSRGW
          & \BayesMoIMidMassHugeLatentEnergyPSRGWXray
          & \BayesMoIMidMassHugeLatentEnergyGWGivenPSR
          & \BayesMoIMidMassHugeLatentEnergyGWXrayGivenPSR \\ 
        \cline{2-9}
        \multirow{3}{*}{\HighMassBinMin-\HighMassBinMax}
          & \MidLatentEnergy
          & \BayesMoIHighMassMidLatentEnergyPSR
          & \BayesMoIHighMassMidLatentEnergyGW
          & \BayesMoIHighMassMidLatentEnergyXray
          & \BayesMoIHighMassMidLatentEnergyPSRGW
          & \BayesMoIHighMassMidLatentEnergyPSRGWXray
          & \BayesMoIHighMassMidLatentEnergyGWGivenPSR
          & \BayesMoIHighMassMidLatentEnergyGWXrayGivenPSR \\
         & \HighLatentEnergy
          & \BayesMoIHighMassHighLatentEnergyPSR
          & \BayesMoIHighMassHighLatentEnergyGW
          & \BayesMoIHighMassHighLatentEnergyXray
          & \BayesMoIHighMassHighLatentEnergyPSRGW
          & \BayesMoIHighMassHighLatentEnergyPSRGWXray
          & \BayesMoIHighMassHighLatentEnergyGWGivenPSR
          & \BayesMoIHighMassHighLatentEnergyGWXrayGivenPSR \\
         & \HugeLatentEnergy
          & \BayesMoIHighMassHugeLatentEnergyPSR
          & \BayesMoIHighMassHugeLatentEnergyGW
          & \BayesMoIHighMassHugeLatentEnergyXray
          & \BayesMoIHighMassHugeLatentEnergyPSRGW
          & \BayesMoIHighMassHugeLatentEnergyPSRGWXray
          & \BayesMoIHighMassHugeLatentEnergyGWGivenPSR
          & \BayesMoIHighMassHugeLatentEnergyGWXrayGivenPSR \\ 
        \hline
    \end{tabular}
    }
\end{table*}


\section{Additional Examples of Phase Transition Phenomenology}
\label{sec:additional examples}

\begin{figure*}
    \begin{minipage}{0.66\textwidth}
        \begin{center}
            \includegraphics[width=1.0\textwidth, clip=True, trim=0.0cm 0.0cm 0.0cm 0.0cm]{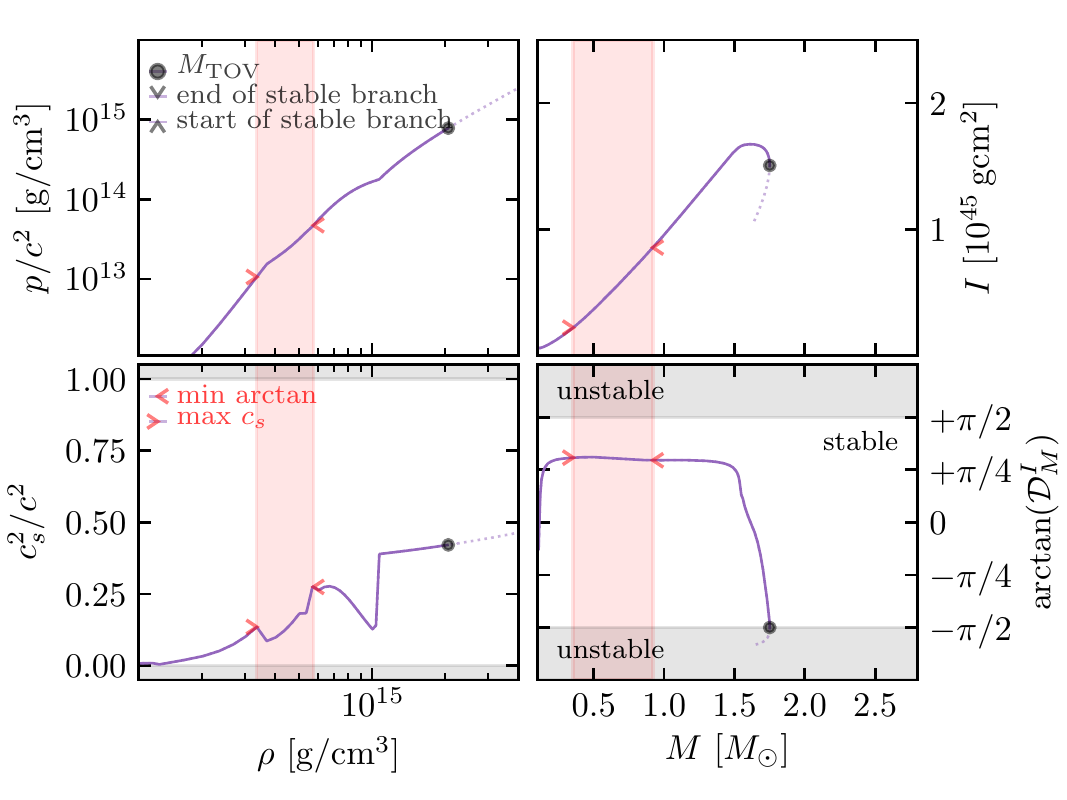}
        \end{center}
    \end{minipage}
    \hfill
    \begin{minipage}{0.33\textwidth}
        \begin{center}
            \includegraphics[width=1.0\textwidth, clip=True, trim=5.4cm 0.0cm 0.0cm 0.0cm]{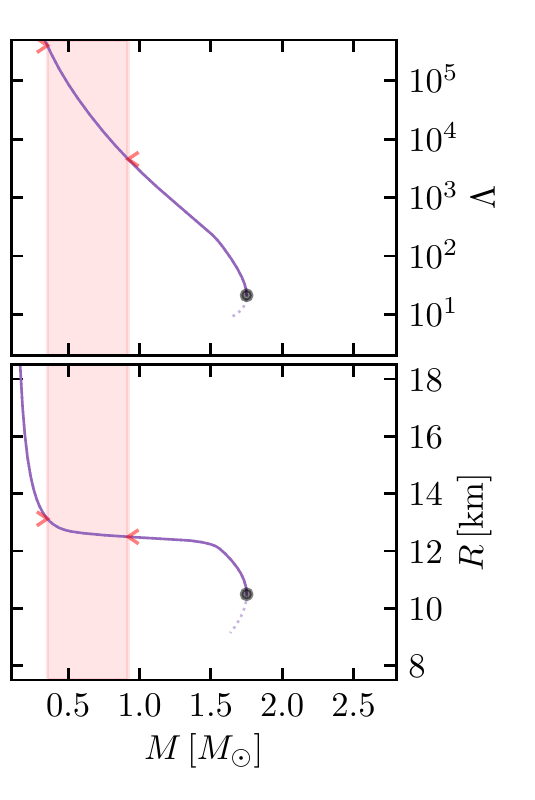}
        \end{center}
    \end{minipage}
    \caption{
        An additional example of an EoS with mixed phases (Gibbs construction) from~\citet{Han:2019}, analogous to Fig.~\ref{fig:Gibbs examples}.
    }
    \label{fig:Gibbs examples extra}
\end{figure*}

\begin{figure*}
    \begin{minipage}{0.66\textwidth}
        \begin{center}
            \includegraphics[width=1.0\textwidth, clip=True, trim=0.0cm 1.15cm 0.0cm 0.0cm]{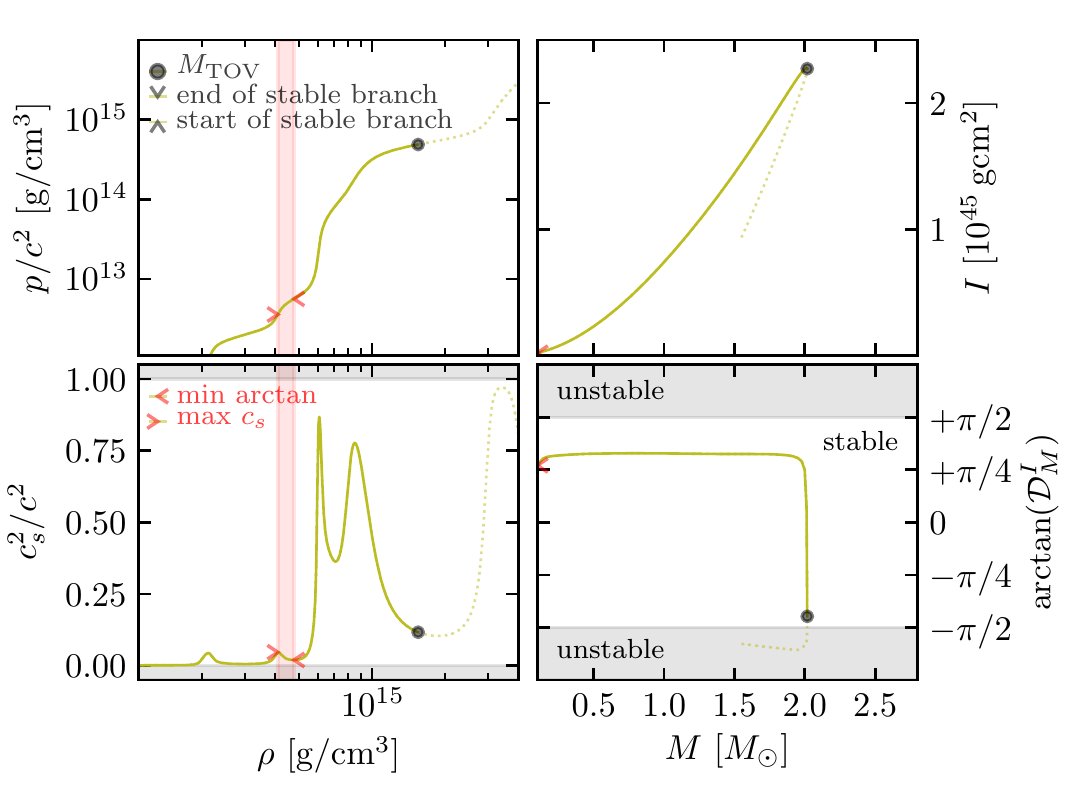}
            \includegraphics[width=1.0\textwidth, clip=True, trim=0.0cm 1.15cm 0.0cm 0.3cm]{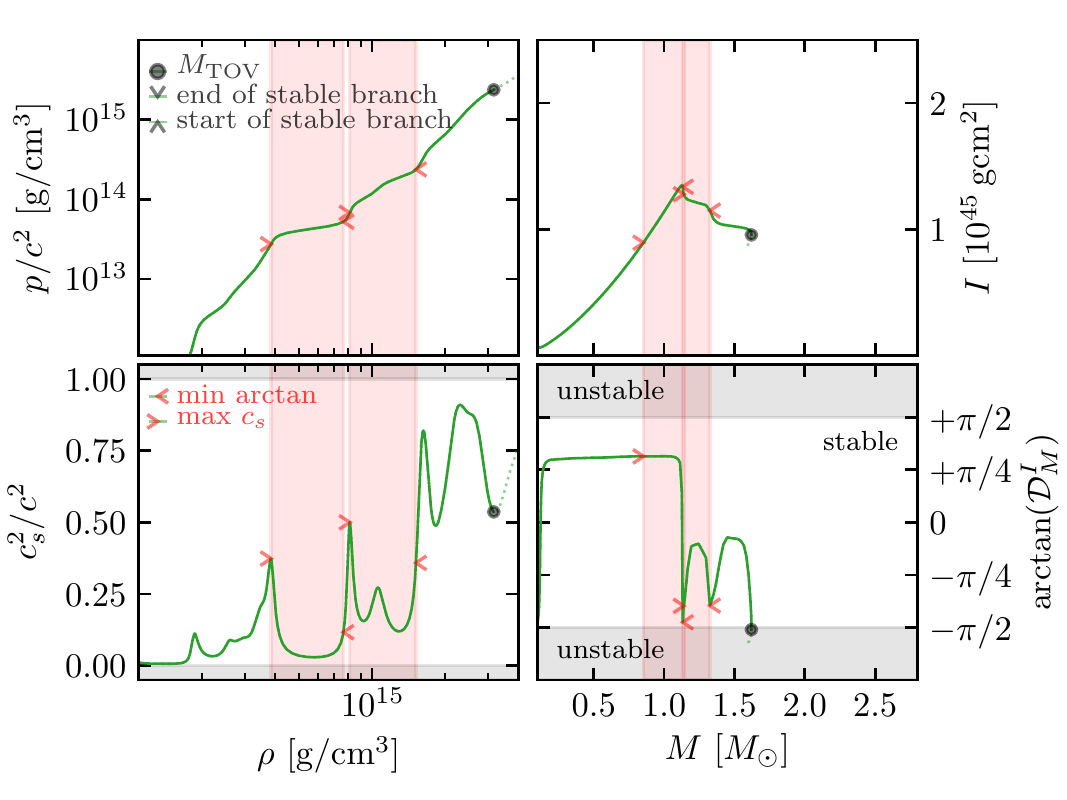}
            \includegraphics[width=1.0\textwidth, clip=True, trim=0.0cm 0.00cm 0.0cm 0.3cm]{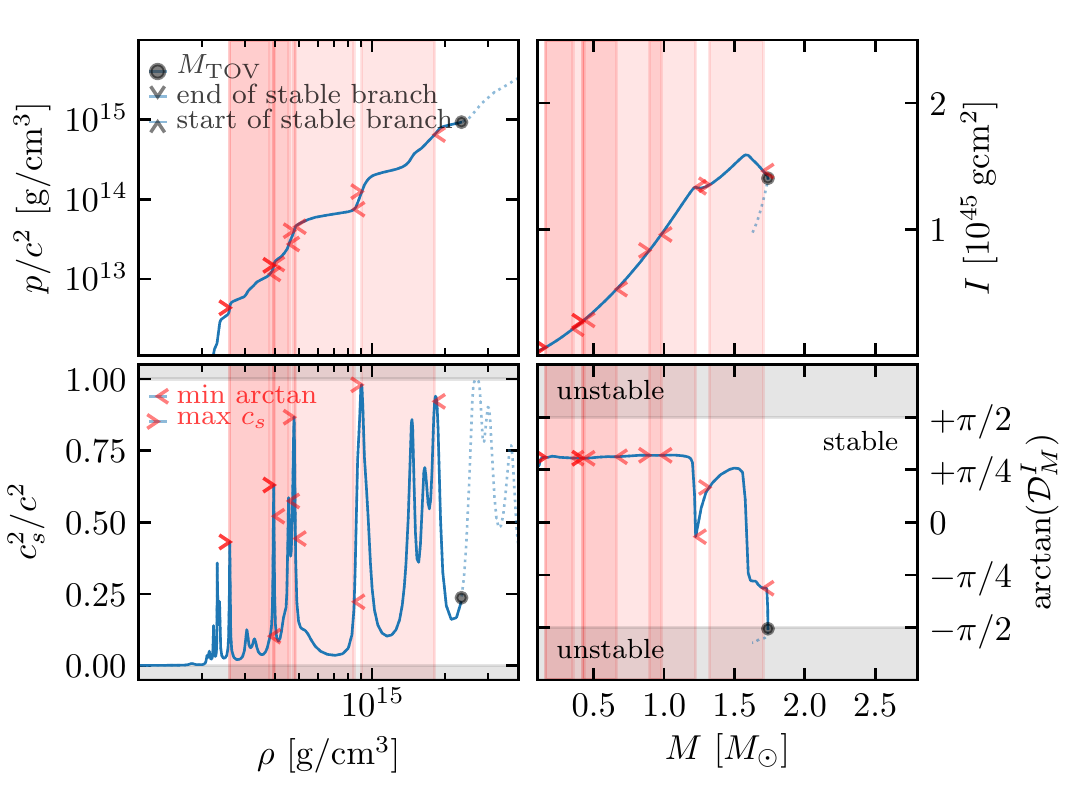}
        \end{center}
    \end{minipage}
    \hfill
    \begin{minipage}{0.33\textwidth}
        \begin{center}
            \includegraphics[width=1.0\textwidth, clip=True, trim=5.4cm 1.15cm 0.0cm 0.0cm]{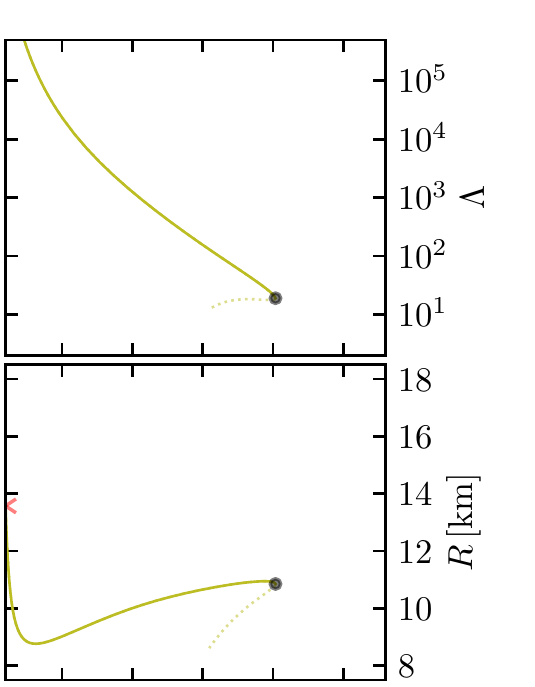}
            \includegraphics[width=1.0\textwidth, clip=True, trim=5.4cm 1.15cm 0.0cm 0.3cm]{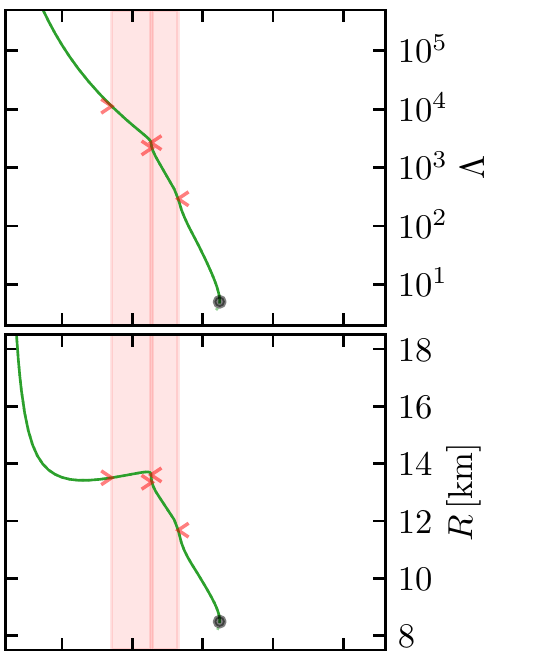}
            \includegraphics[width=1.0\textwidth, clip=True, trim=5.4cm 0.00cm 0.0cm 0.3cm]{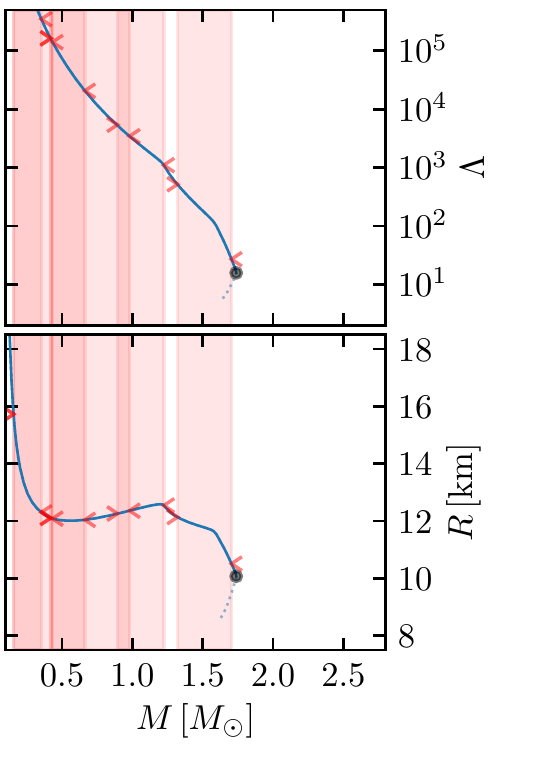}
        \end{center}
    \end{minipage}
    \caption{
        Several realizations from our nonparametric prior, each with a single stable branch but with different numbers of phase transitions.
    }
    \label{fig:wacky GP 1 branch}
\end{figure*}

\begin{figure*}
    \begin{minipage}{0.66\textwidth}
        \begin{center}
            \includegraphics[width=1.0\textwidth, clip=True, trim=0.0cm 1.15cm 0.0cm 0.0cm]{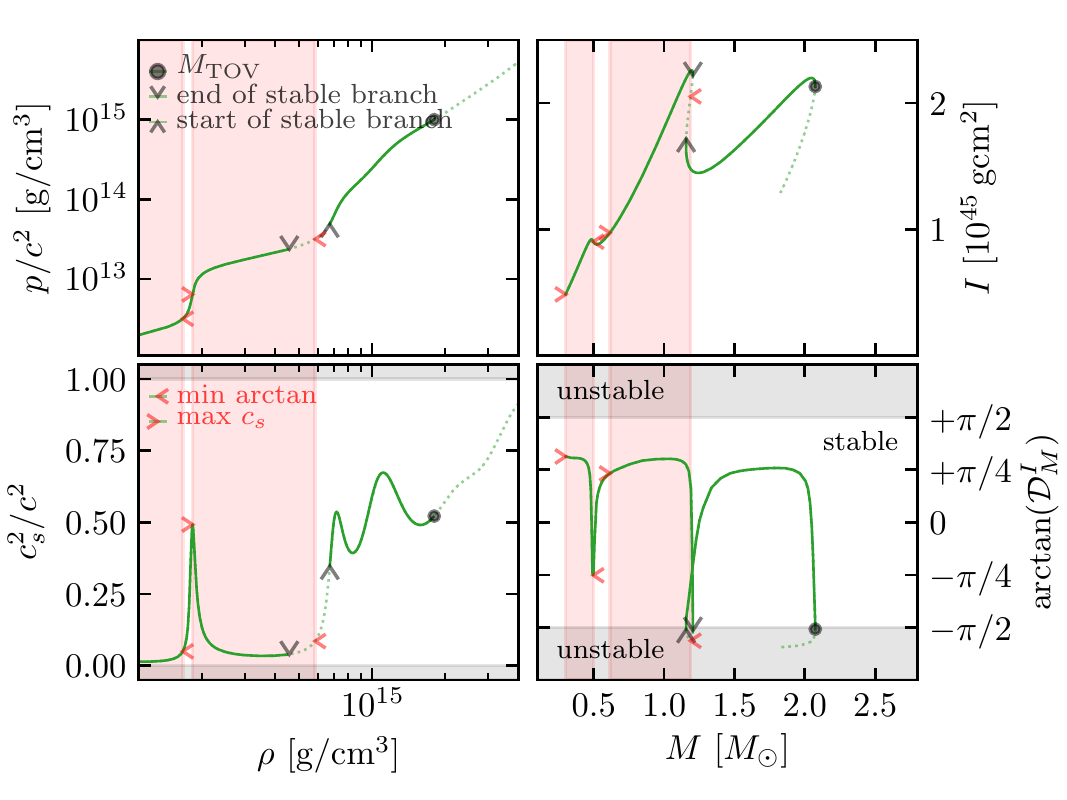}
            \includegraphics[width=1.0\textwidth, clip=True, trim=0.0cm 1.15cm 0.0cm 0.3cm]{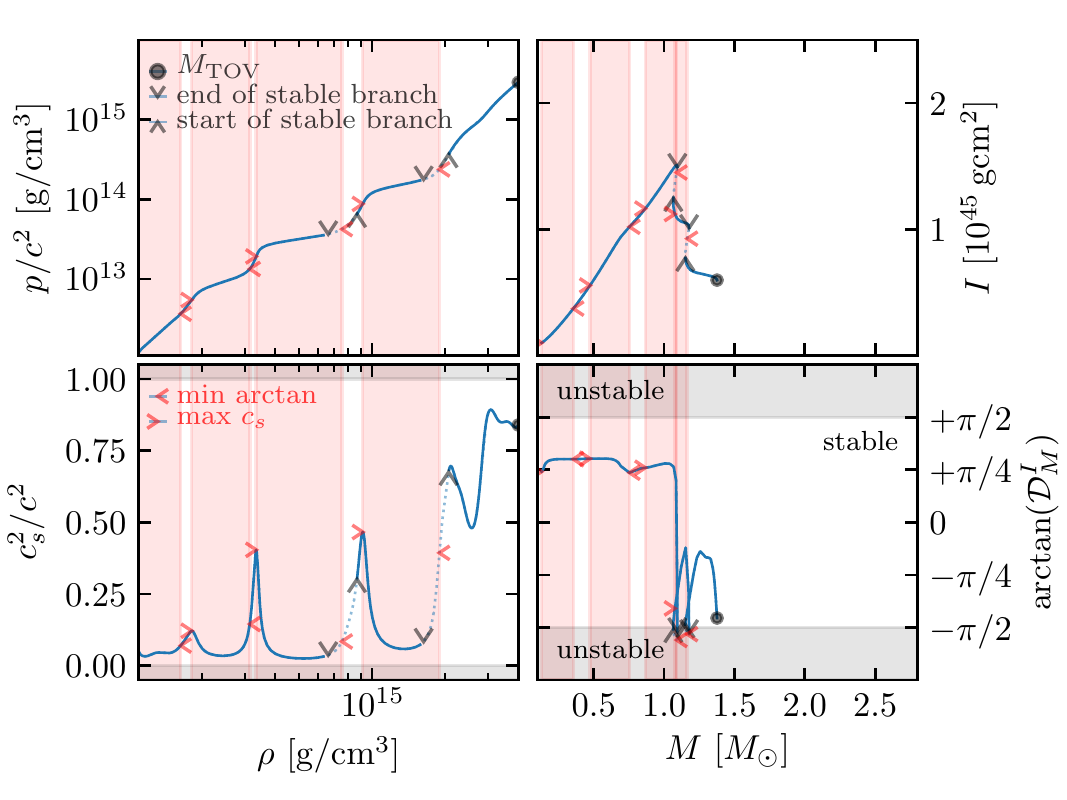}
            \includegraphics[width=1.0\textwidth, clip=True, trim=0.0cm 0.00cm 0.0cm 0.3cm]{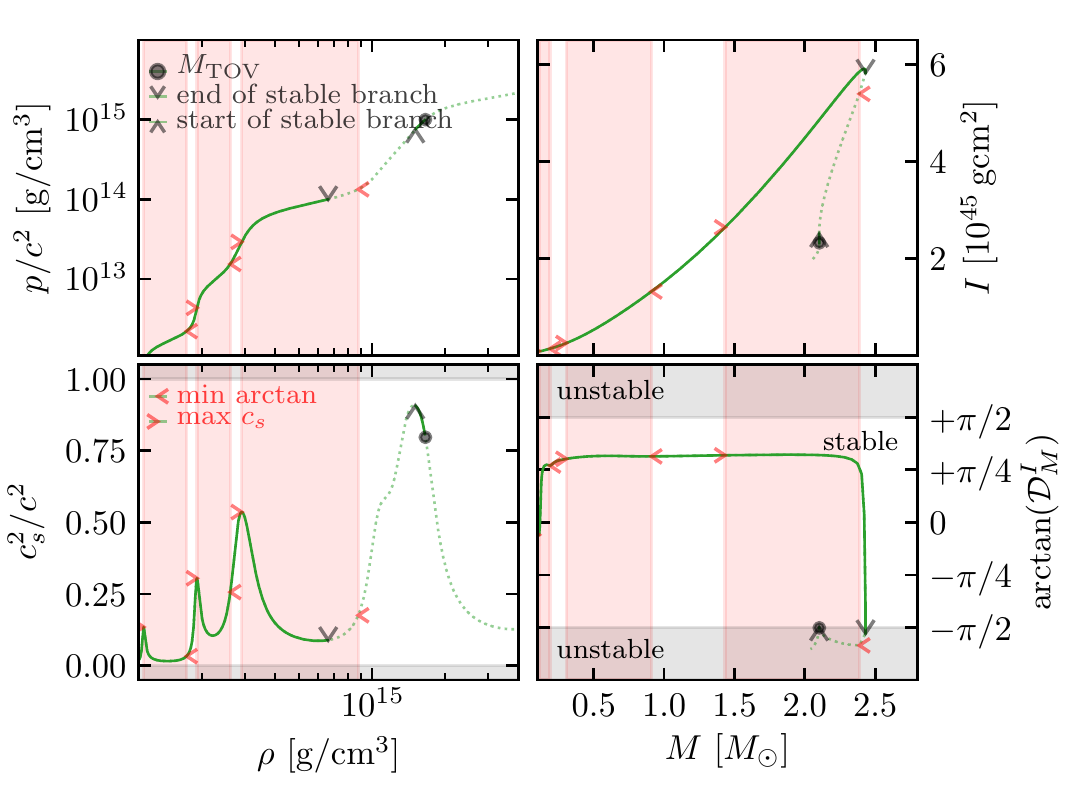}
        \end{center}
    \end{minipage}
    \hfill
    \begin{minipage}{0.33\textwidth}
        \begin{center}
            \includegraphics[width=1.0\textwidth, clip=True, trim=5.4cm 1.15cm 0.0cm 0.0cm]{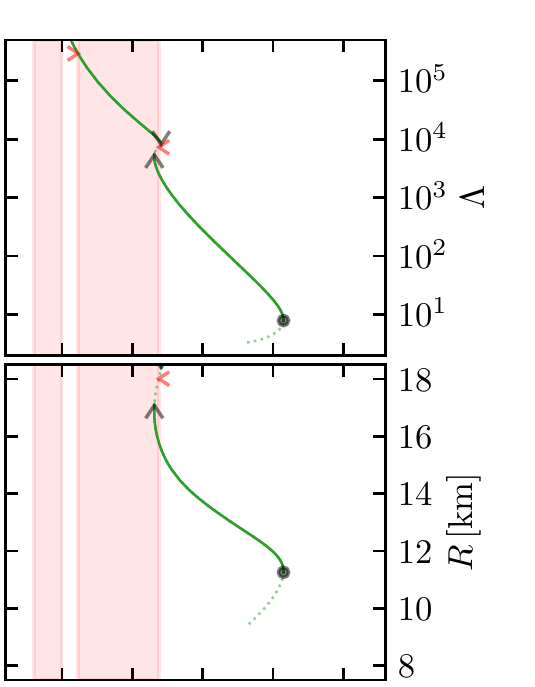}
            \includegraphics[width=1.0\textwidth, clip=True, trim=5.4cm 1.15cm 0.0cm 0.3cm]{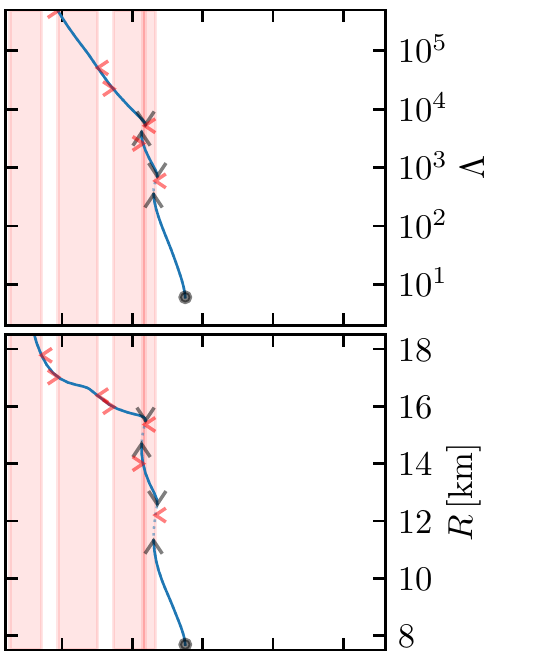}
            \includegraphics[width=1.0\textwidth, clip=True, trim=5.4cm 0.00cm 0.0cm 0.3cm]{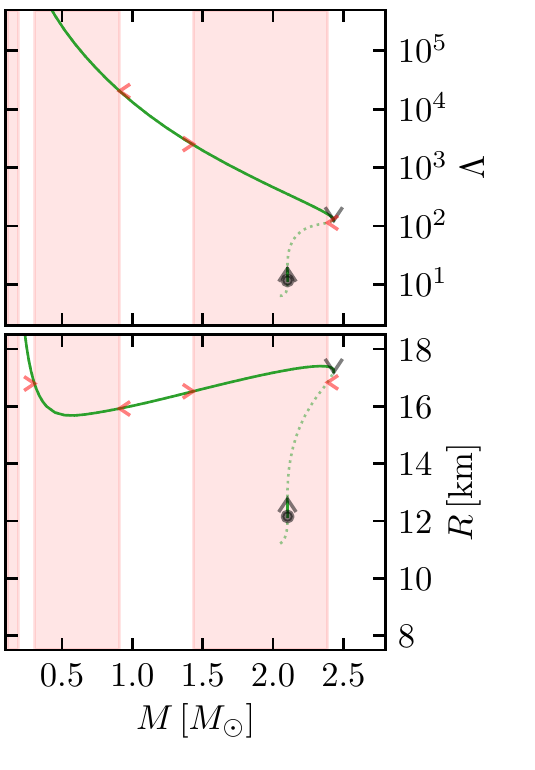}
        \end{center}
    \end{minipage}
    \caption{
        Additional realizations from our nonparametric prior, each with multiple stable branches.
        Typically, we always identify a phase transition associated with the loss of stability between stable branches, even if the stable branches are small (\textit{bottom row}).
    }
    \label{fig:wacky GP 2+ branches}
\end{figure*}

This appendix includes additional examples of phase transition phenomenology using both EoSs with known microphysical descriptions (Fig.~\ref{fig:Gibbs examples extra}) as well as realizations from our nonparametric prior (Figs.~\ref{fig:wacky GP 1 branch} and~\ref{fig:wacky GP 2+ branches}).

Fig.~\ref{fig:Gibbs examples extra} shows an EoS with mixed phases, analogous to Fig.~\ref{fig:Gibbs examples}.
The more complicated structure in $c_s$ demonstrates two shortcomings of the new feature introduced in Sec.~\ref{sec:new stats}.
The feature does not always identify the correct beginning and end of the phase transition; the microphysical model used to construct this transition has the mixed phase extend beyond the end of the identified region.
The true end of the phase transition occurs near $\rho \sim 10^{15}\,\mathrm{g}/\mathrm{cm}^3$ and $M \sim 1.5 \,\Msolar$.
Also, some features may be difficult to identify as they are overwhelmed by the final collapse to a BH, which often means there is no local minimum in $\arctan(\moifeature)$.
This is the case for the true end of this transition.

Figs.~\ref{fig:wacky GP 1 branch} and~\ref{fig:wacky GP 2+ branches} show a few realizations from our nonparametric prior with particularly complex behavior, such as multiple strong phase transitions leading to three disconnected stable branches.
These demonstrate that our \moifeature~feature identifies and classifies a broad range of behavior, some of which may not have been anticipated with parametric descriptions.
For example, \citet{Tan:2021ahl} and~\citet{Mroczek:2023eff} introduced a variety of parametric features in the sound-speed and attempted to classify which types of features led to observable effects within macroscopic relations.
Our procedure can identify relevant density scales associated with these behaviors and others \textit{without} access to the underlying parametric construction.

This flexibility is due to the fact that our nonparametric prior contains support for multiple different correlation length scales and marginal variances in the speed of sound, particularly compared to some others in the literature, e.g., Refs.~\cite{Mroczek:2023eff, Gorda:2023usm, J0740-Miller}.
This is achieved by marginalizing over covariance-kernel hyperparameters as described in~\citet{Essick:2019ldf} so that the overall prior process contains $O(150)$ different GPs, each of which generates different types of correlation behavior.

\end{document}